\documentclass[11pt]{article}
\usepackage{color,graphicx,amssymb,amsfonts}
\usepackage[mathscr]{eucal}
\title{On the Regularized Fermionic Projector of the Vacuum}
\author{Felix Finster}
\date{January 2008}

\newtheorem{Def}{Definition}[section]
\newtheorem{Thm}[Def]{Theorem}
\newtheorem{Prp}[Def]{Proposition}
\newtheorem{Lemma}[Def]{Lemma}
\newtheorem{Remark}[Def]{Remark}
\newcommand{\Proof}{{\em{Proof. }}}
\newcommand{\QED}{\ \hfill $\FBox$ \\[1em]}
\newcommand{\QEDrem}{\ \hfill $\blacklozenge$}
\newcommand{\spc}{\;\;\;\;\;\;\;\;\;\;}
\newcommand{\bra}{\mbox{$< \!\!$ \nolinebreak}}
\newcommand{\ket}{\mbox{\nolinebreak $>$}}

\newcommand{\R}{\mathbb{R}}
\newcommand{\1}{\mbox{\rm 1 \hspace{-1.05 em} 1}}

\newcommand{\N}{\mbox{\rm I \hspace{-.8 em} N}}

\newcommand{\Pdd}{\mbox{$\partial$ \hspace{-1.2 em} $/$}}
\newcommand{\gslsh}{\mbox{$g$ \hspace{-1.15 em} $/$}}
\newcommand{\slsh}{\mbox{ \hspace{-1.1 em} $/$}}
\newcommand{\Tr}{\mbox{Tr\/}}

\newcommand{\FBox}{\rule{2mm}{2.25mm}}
\newcommand{\OBox}{\raisebox{.6ex}{\fbox{}}\,}
\newcommand{\beq}{\begin{equation}}
\newcommand{\eeq}{\end{equation}}
\newcommand{\M}{{\mathcal{M}}}
\newcommand{\tM}{\tilde{\mathcal{M}}}
\newcommand{\hM}{\hat{\mathcal{M}}}
\newcommand{\Rflog}{\mbox{\rm{Rlog}}}
\newcommand{\Rf}{\mbox{\rm{R}}}
\newcommand{\hRflog}{\hat{\mbox{\rm{R}}}\mbox{\rm{log}}}
\newcommand{\hRf}{\hat{\mbox{\rm{R}}}}
\newcommand{\m}{{\mathfrak{m}}}

\begin{document}
\maketitle

\begin{abstract}
We construct families of fermionic projectors with spherically
symmetric regularization, which satisfy the condition
of a distributional ${\mathcal{M}} P$-product.
The method is to analyze regularization tails with a power-law
or logarithmic scaling
in composite expressions in the fermionic projector.
The resulting regularizations break the Lorentz symmetry and
give rise to a multilayer structure of
the fermionic projector near the light cone. Furthermore,
we construct regularizations which go beyond the
distributional ${\mathcal{M}} P$-product in that they yield
additional distributional contributions supported at the origin.
The remaining freedom for the regularization parameters and the
consequences for the normalization of the fermionic states are
discussed.
\end{abstract}

{\small{
\tableofcontents}}

\section{Introduction} \label{sec0}
\setcounter{equation}{0}
It is generally believed that the concept of a space-time continuum
(such as Minkowski space or a Lorentzian manifold) should be modified
for distances as small as the Planck length. Naively, general relativity
and quantum field theory become inconsistent on the Planck scale because
the energy fluctuations would lead to the formation of microscopic black holes.
This led to the intuitive picture that space-time
should have a structure with a complicated
topology and geometry~\cite{W}, sometimes
referred to as a ``quantum foam'' \cite{H}.
However, this intuitive picture has not yet been made rigorous, and indeed there is not even a
consensus on what the correct mathematical framework for physics on the Planck scale is.
The most prominent approaches are string theory~\cite{V} and loop quantum gravity~\cite{R}.
Another framework is non-commutative geometry~\cite{C}, where one works instead of
the commutative algebra of functions on a manifold with a non-commutative algebra
and replaces the classical action by the so-called spectral action~\cite{CC}.
The requirement of background independence led to the idea that space-time on the
Planck scale should be in some sense discrete (see for example~\cite{RW}).
The most notable discrete approaches are the causal sets~\cite{BLMS},
discrete gauge theory~\cite{Oeckl}, spin networks and spin foam models~\cite{Baez},
as well as group field theories describing a simplicial geometry~\cite{O}.

As an alternative to the above approaches, the principle of the fermionic projector~\cite{PFP} proposes a mathematical framework for Planck scale physics
where the physical equations are formulated via a variational
principle~\cite{PFP}, which is set up in a so-called {\em{discrete space-time}}
for a collection of projectors in an indefinite inner product space
(for an introduction to the discrete setting see Section~\ref{sec21} below).
In the so-called {\em{continuum limit}}
\cite[Chapter~4]{PFP}, this variational principle can be analyzed in Minkowski space,
and one can relate the resulting Euler-Lagrange equations to differential equations for
gauge and Dirac fields.
The analysis of the continuum limit is based on the assumption that the
{\em{vacuum}} corresponds to a {\em{Dirac sea configuration}}, in the sense
that the continuum limit of the fermionic projector of the vacuum should be composed
of free Dirac seas (see also Section~\ref{sec22} below). This assumption clearly
needs justification, and it is therefore an important task to show that Dirac
sea configurations really are stable minima of our variational principle.
In~\cite[\S5.6]{PFP} it is shown that, as a special feature of our particular
variational principle, all the composite expressions in the fermionic projector
which appear in the stability analysis can be defined as distributions.
However, this raises the subtle question of whether there really are regularizations
of the fermionic projector with the property that the corresponding composite
expressions are so ``well behaved''
that they converge in the continuum limit to the
distributions in~\cite[\S5.6]{PFP}. The goal of the present paper is to give the
definitive answer ``yes'' by constructing a family of spherically symmetric
regularizations with the desired properties.

Apart from justifying the considerations in~\cite[\S5.6]{PFP}, our analysis also settles
a few other important issues. First of all, we will specify the light-cone singularity
of~$\tM$, a distribution which in~\cite[\S5.6]{PFP} was defined only modulo
singular contributions on the light cone (see Section~\ref{sec2}).
This result is the basis of a stability analysis in the continuum as carried out in~\cite{FH}.
Furthermore, we shall see that the analysis here will
give no constraints for the so-called regularization parameters as introduced in~\cite[Chapter~4]{PFP} (see Remark~\ref{rem7}). This is important because
any additional relations between the regularization parameters would have a sensitive effect on the analysis of the continuum limit. Finally, our
analysis will give very strong conditions for the regularization. It seems
miraculous that these conditions
can all be fulfilled; this can be regarded as a further confirmation for our variational
principle and the concept of the continuum limit. Despite the fact that we consider only
the restrictive class of spherically symmetric regularizations, our analysis seems to
reveal a few general properties of admissible regularizations, most notably a
{\em{multilayer structure}} near the light cone (see Section~\ref{sec8}).
If one believes that the regularized fermionic projector describes nature,
we thus get concrete hints on how the vacuum should look like on the Planck scale.

Unfortunately, some of the calculations presented in this paper are lengthy and quite tedious.
In order to make the paper as convenient to read as possible, we always
explain the ideas of the calculations in a non-technical way before entering the
details. The calculations in Sections~\ref{sec4}--\ref{appB} were carried out with the
help of a computer algebra program; the corresponding Mathematica worksheets
are available from the author on request.

\section{Preliminaries, Statement of the Main Results} \label{sec1}
\setcounter{equation}{0}
In this section we give a basic introduction to our variational principle in the discrete setting
and explain the connection between discrete space-time and Minkowski space.
We then formulate our main results.

\subsection{A Variational Principle in Discrete Space-Time} \label{sec21}
We let~$(H, \bra .|. \ket)$ be a finite-dimensional complex inner product space.
Thus~$\bra .|. \ket$ is linear in its second and anti-linear in its first argument, and it is symmetric,
\[ \overline{\bra \Psi \:|\: \Phi \ket} \;=\; \bra \Phi \:|\: \Psi \ket \quad
\spc {\mbox{for all~$\Psi,\Phi \in H$}} \,, \]
and non-degenerate,
\[ \bra \Psi \:|\: \Phi \ket \;=\; 0 \;\;\; {\mbox{for all $\Phi \in H$}}
 \quad \Longrightarrow \quad
\Psi \;=\; 0 \:. \]
In contrast to a scalar product, $\bra .|. \ket$ need {\em{not}} be positive.

A {\em{projector}}~$A$ in~$H$ is defined just as in Hilbert spaces as a linear
operator which is idempotent and self-adjoint,
\[ A^2 = A \spc {\mbox{and}} \spc \bra A\Psi \:|\: \Phi \ket = \bra \Psi \:|\: A\Phi \ket \quad
{\mbox{for all $\Psi, \Phi \in H$}}\:. \]
Let~$M$ be a finite set. To every point~$x \in M$ we associate a projector
$E_x$. We assume that these projectors are orthogonal and
complete in the sense that
\beq \label{oc}
E_x\:E_y \;=\; \delta_{xy}\:E_x \spc {\mbox{and}} \spc
\sum_{x \in M} E_x \;=\; \1\:.
\eeq
Furthermore, we assume that the images~$E_x(H) \subset H$ of these
projectors are non-de\-ge\-ne\-rate subspaces of~$H$, which
all have the same signature~$(n,n)$. The parameter~$n$ is referred
to as the {\em{spin dimension}}. The points~$x \in M$ are
called {\em{discrete space-time points}}, and the corresponding
projectors~$E_x$ are the {\em{space-time projectors}}. The
structure~$(H, \bra .|. \ket, (E_x)_{x \in M})$ is
called {\em{discrete space-time}}.

We next introduce the so-called {\em{fermionic projector}} $P$ as
a projector in~$H$ whose image~$P(H) \subset H$ is
{\em{negative definite}}.
The vectors in the image of~$P$ have the interpretation as the
quantum states of the particles of our system. Thus the rank of~$P$
gives the {\em{number of particles}} $f := \dim P(H)$.
The name ``fermionic projector'' is motivated from the correspondence to
Minkowski space, where our particles should go over to Dirac particles,
being fermions (see Section~\ref{sec22}).
We call the obtained system~$(H, \bra .|. \ket, (E_x)_{x \in M}, P)$
a {\em{fermion system in discrete space-time}}.
For a discussion of the underlying physical principles see~\cite{F2}
or~\cite[Chapter~4]{PFP}).

In order to introduce an interaction of the fermions, we now set up a variational principle. 
For any~$u \in H$, we refer to the projection~$E_x u \in E_x(H)$ as the
{\em{localization}} of~$u$ at~$x$. We also use the short notation~$u(x)=E_x u$ and
sometimes call~$u(x)$ the {\em{wave function}} corresponding to the vector~$u$.
Furthermore, we introduce the short notation
\beq \label{notation}
P(x,y) \;=\; E_x\,P\,E_y \spc x,y \in M \:.
\eeq
This operator product maps~$E_y(H) \subset H$ to~$E_x(H)$, and it is often
useful to regard it as a mapping only between these subspaces,
\[ P(x,y)\;:\; E_y(H) \: \rightarrow\: E_x(H)\:. \]
Using the properties of the space-time projectors~(\ref{oc}), we find
\[ (Pu)(x) \;=\; E_x\: Pu \;=\; \sum_{y \in M} E_x\,P\,E_y\:u
\;=\; \sum_{y \in M} (E_x\,P\,E_y)\:(E_y\,u) \:, \]
and thus
\beq \label{diskernel}
(Pu)(x) \;=\; \sum_{y \in M} P(x,y)\: u(y)\:.
\eeq
This relation resembles the representation of an operator with an integral kernel,
and thus we refer to~$P(x,y)$ as the {\em{discrete kernel}} of the fermionic projector.
Next we introduce the {\em{closed chain}} $A_{xy}$ as the product
\beq \label{chain0}
A_{xy} \;:=\; P(x,y)\: P(y,x) \;=\; E_x \:P\: E_y \:P\: E_x \:;
\eeq
it maps~$E_x(H)$ to itself.
Let~$\lambda_1, \ldots, \lambda_{2n}$ be the roots of the characteristic polynomial
of~$A_{xy}$, counted with multiplicities. We define the {\em{Lagrangian}} by
\beq \label{Lcrit}
{\mathcal{L}}[A] \;=\;
\frac{1}{4n} \sum_{i,j=1}^{2n} \left( |\lambda_i| - |\lambda_j| \right)^2
\eeq
and form the {\em{action}} by summing over the space-time points,
\beq \label{Sdef}
{\mathcal{S}}[P] \;=\; \sum_{x,y \in M} {\mathcal{L}}[A_{xy}]\:.
\eeq
Our variational principle is to minimize~(\ref{Sdef}), keeping the number of particles~$f$ as
well as discrete space-time fixed. 
This variational principle was first introduced in~\cite{PFP}. In~\cite{F1}
it is analyzed mathematically in a more general context (it is there referred to as 
the auxiliary variational principle in the critical case).

We next derive the corresponding {\em{Euler-Lagrange equations}}
(for details see~\cite[\S3.5 and \S5.2]{PFP}).
Suppose that~$P$ is a critical point of the action~(\ref{Sdef}).
We consider a variation~$P(\tau)$ of projectors with~$P(0)=P$.
Denoting the gradient of the Lagrangian by~$\mathcal{M}$,
\beq \label{fvar}
{\mathcal{M}}[A]^\alpha_\beta \;:=\;
\frac{\partial {\mathcal{L}}[A]}{\partial A^\beta_\alpha}\:, \qquad 
{\mbox{with }} \alpha, \beta
\in \{1,\ldots, 2n\}\:,
\eeq
we can write the variation of the Lagrangian as a trace on~$E_x(H)$,
\[ \delta {\mathcal{L}}[A_{xy}] \;=\; \frac{d}{d \tau}  {\mathcal{L}}[A_{xy}(\tau)]
\Big|_{\tau=0} \;=\; \Tr \left( E_x \,{\mathcal{M}}[A_{xy}] \:\delta A_{xy} \right) . \]
Using the Leibniz rule
\[ \delta A_{xy} \;=\; \delta P(x,y) \:P(y,x) \:+\: P(x,y) \:\delta P(y,x) \]
together with the fact that the trace is cyclic, after summing over the space-time points
we find
\[ \sum_{x,y \in M} \delta {\mathcal{L}}_\mu[A_{xy}] \;=\; \sum_{x,y \in M} 
4 \: \Tr \left( E_x\, Q_\mu(x,y) \:\delta P(y,x) \right)\:, \]
where we set
\[ Q(x,y) \;=\; \frac{1}{4} \Big( {\mathcal{M}}[A_{xy}]\:P(x,y)
\:+\: P(x,y) \:{\mathcal{M}}[A_{yx}] \Big) \:. \]
It follows from general properties of the spectral decompositions of~$A_{xy}$ and~$A_{yx}$
(see \cite[Lemma~5.2.1]{PFP}) that the two summands on the right coincide,
and thus we can write~$Q(x,y)$ as the product
\beq \label{Qxydef}
Q(x,y) \;=\; \frac{1}{2} \:{\mathcal{M}}[A_{xy}]\:P(x,y) \:.
\eeq
Thus the first variation of the action can be written in the compact form
\beq \label{dS1}
\delta {\mathcal{S}}[P] \;=\; 4 \,\Tr \left( Q\: \delta P \right) ,
\eeq
where~$Q$ is the operator in~$H$ with kernel~(\ref{Qxydef}).
This equation can be simplified using
that the operators~$P(\tau)$ are all projectors of fixed rank. Namely,
there is a family of unitary operators~$U(\tau)$ with~$U(\tau)=\1$ and
\[ P(\tau) \;=\; U(\tau)\, P\, U(\tau)^{-1}\:. \]
Hence~$\delta P = i[B, P]$, where~$B=-iU'(0)$ is the infinitesimal generator of the
family~$U(\tau)$. Using this relation in~(\ref{dS1}) and again using that the trace is
cyclic, we find $\delta {\mathcal{S}}[P] = 4i \,\Tr \left( [P,Q]\: B \right)$.
Since~$B$ is an arbitrary self-adjoint operator, we conclude that
\beq \label{EL}
[P,Q] \;=\; 0\:.
\eeq
This commutator equation with~$Q$ given by~(\ref{Qxydef}) are the
Euler-Lagrange equations corresponding to our variational principle.

Before moving on to the space-time continuum, we briefly mention a few
results on fermion systems in discrete space-time. In~\cite{F3} it is shown under
under general assumptions that the {{permutation symmetry}} of the space-time points
is {\em{spontaneously broken}} by the fermionic projector. This implies that the
fermionic projector induces nontrivial relations between the space-time points.
In particular, one can introduce a {\em{discrete causal structure}}~\cite{F2}.
In~\cite{DFS} the spontaneous symmetry breaking and the emergence of a
discrete causal structure are illustrated in simple examples.

\subsection{Connection to Minkowski Space, the Distributional~$\M  \!\cdot\! P$-Product}
\label{sec22}
It is conjectured that, in a suitable limit where the number of particles and space-time points
tends to infinity, the emergent discrete causal structure gives rise to the local and causal
structure of Minkowski space. For a discussion of this conjecture we refer to the
recent survey article~\cite{F4}. Here we focus on a particular aspect of this problem,
namely to the question of how the Euler-Lagrange equations~(\ref{EL}) can be introduced
for vacuum Dirac sea configurations in Minkowski space. The crucial point
is to make sense of the gradient of the Lagrangian~(\ref{fvar}) and of
the product of~$\M$ with~$P$ in the definition of~$Q$, (\ref{Qxydef}).

Before we can specify what needs to be done, we need to get a connection between
discrete space-time and Minkowski space.
The simplest method for obtaining a correspondence to relativistic quantum mechanics in
Minkowski space is to replace the discrete space-time points~$M$ by the space-time
continuum~$\R^4$ and the sums over~$M$ by space-time integrals. For a
vector~$\Psi \in H$, the corresponding localization~$E_x \Psi$
should be a four-component Dirac
wave function, and the scalar product $\bra \Psi(x) \,|\, \Phi(x) \ket$ on~$E_x(H)$
should correspond to the usual Lorentz invariant scalar product on Dirac
spinors~$\overline{\Psi} \Phi$
with~$\overline{\Psi} = \Psi^\dagger \gamma^0$ the adjoint spinor. Since this last
scalar product is indefinite of signature~$(2,2)$, we are led to choosing~$n=2$.
In view of~(\ref{diskernel}), the discrete kernel should in the continuum go over to the
integral kernel of an operator~$P$ on the Dirac wave functions,
\[ (P \Psi)(x) \;=\; \int P(x,y)\, \Psi(y)\: d^4y \:. \]
The image of~$P$ should be spanned by the occupied fermionic states. We take Dirac's
concept literally that in the vacuum all negative-energy states are occupied by fermions
forming the so-called {\em{Dirac sea}}. Thus we are led to describe the vacuum by the
integral over the lower mass shell
\[ P(x,y) \;=\; \int \frac{d^4k}{(2 \pi)^4}\: (k \slsh+m)\:
\delta(k^2-m^2)\: \Theta(-k^0)\: e^{-ik(x-y)} \]
(here~$\Theta$ is the Heaviside function). In order to take into account
the three generations of elementary particles (such as the quarks $u$, $s$, $t$ in
the standard model), we take the sum of three Dirac seas,
\beq \label{A}
P(x,y) \;=\; \sum_{\beta=1}^3 \rho_\beta \int \frac{d^4k}{(2 \pi)^4}\: (k \slsh+m_\beta)\:
\delta(k^2-m_\beta^2)\: \Theta(-k^0)\: e^{-ik(x-y)}\:.
\eeq
Compared to the situation in~\cite[\S2.2]{PFP}, the ansatz~(\ref{A}) is more general
in that it involves additional weight factors~$\rho_\beta > 0$.
We treat these factors as a priori given positive constants. For a discussion of the
physical significance and the implications of the weights~$\rho_\beta$ see Appendix~\ref{appA}.
More generally, in~\cite[\S5.1]{PFP} a realistic system of fermions is built up
by taking a direct sum of operators acting on so-called sectors.
In this more general situation, our variational principle splits into
separate variational principles in the individual sectors. Thus we may
restrict attention to one sector, and considering a {\em{massive sector}}
again gives a fermionic projector of the form~(\ref{A}).

The Fourier integrals in~(\ref{A}) are clearly well defined in the distributional
sense. Carrying them out using Bessel functions (see~\cite[\S2.5]{PFP} and
Section~\ref{sec2} below), one sees
that~$P(x,y)$ is even a smooth function away from the light cone (i.e.\ for~$(y-x)^2
\neq 0$), but that it has poles and singularities on the light cone. Before we can set
up our variational principle, these singularities must be removed by a regularization procedure.
To us, the regularization has a physical significance in that a suitably regularized fermionic
projector should describe the physical fermionic projector in discrete space-time.
Since we do not have any information on what the physically correct regularization is,
we use the {\em{method of variable regularization}} and consider a sufficiently large class of
regularizations. More precisely, as in~\cite[\S4.1]{PFP} we assume that the regularizations
are {\em{homogeneous}} and have a {\em{vector-scalar structure}}. This means that the
regularized fermionic projector, which we denote by~$P^\varepsilon$, can be written
as a Fourier integral
\beq P^\varepsilon(x,y) \;=\; \int \frac{d^4k}{(2 \pi)^4}\: \hat{P}^\varepsilon(k)\:
e^{-ik(x-y)}\:, \label{B}
\eeq
where~$\hat{P}^\varepsilon$ is a distribution of the form
\beq \label{C}
\hat{P}^\varepsilon(k) \;=\; \hat{g}_j(k)\: \gamma^j + \hat{h}(k)
\eeq
with real-valued distributions~$\hat{g}_j$ and~$\hat{h}$.
Here the parameter~$\varepsilon>0$ denotes the length scale of the regularization.
Thus, expressed in momentum space, the distributions~$\hat{g}_j$
and~$\hat{h}$ should
decay at infinity on the scale~$k \sim \varepsilon^{-1}$. Furthermore, we restrict
attention to {\em{spherically symmetric regularizations}} which are composed of
{\em{surface states}}. The last assumption means that, similar to the situation in~(\ref{A}),
the distributions~$\hat{g}_j$ and~$\hat{h}$ should be supported on three hypersurfaces
(for details see~(\ref{Pansatz}) and Section~\ref{sec3}).
We consider a family of regularized fermionic projectors~$(P^\varepsilon)_{\varepsilon>0}$
with the above properties. As~$\varepsilon$ tends to zero, the regularized fermionic
projectors should go over to the unregularized fermionic projector,
\beq
\lim_{\varepsilon \searrow 0} P^\varepsilon(x,y) \;=\; P(x,y)
\spc {\mbox{as a distribution.}} \label{Z}
\eeq

Having defined the fermionic projector, we can introduce the closed chain in analogy
to~(\ref{chain0}) by
\beq \label{chain}
A^\varepsilon_{xy} \;=\; P^\varepsilon(x,y)\: P^\varepsilon(y,x)\:.
\eeq
For ease in notation, we will often omit the subscript `$xy$'.
In our setting of one sector and a vector-scalar structure, the roots of the characteristic
polynomials of~$A^\varepsilon$ have a particularly simple structure.
\begin{Lemma} \label{lemma11} For the fermionic projector~(\ref{B}, \ref{C}), the characteristic polynomial of the closed chain~$A^\varepsilon_{xy}$ has two roots~$\lambda_\pm$, each of multiplicity two.
Either the~$\lambda_\pm$ form a complex conjugate pair, $\overline{\lambda_+}=\lambda_-$,
or else the~$\lambda_\pm$ are both real and have the same sign.
\end{Lemma}
{\Proof} According to~(\ref{B}, \ref{C}) and the fact that the distributions~$\hat{g}_j$
and~$\hat{h}$ are real-valued, we can write the fermionic projector in position space as
\[ P^\varepsilon(x,y) \;=\; g_j(x,y)\: \gamma^j + h(x,y) \:,\spc
P^\varepsilon(y,x) \;=\; \overline{g_j(x,y)}\: \gamma^j + \overline{h(x,y)} \:. \]
Thus, omitting the arguments~$x$ and~$y$,
\[ A^\varepsilon \;=\; (\gslsh + h)(\overline{\gslsh} + \overline{h})\:. \]
A short calculation using the anti-commutation relations of the Dirac matrices
shows that the characteristic polynomial of~$A^\varepsilon$ has the two roots
\[ \lambda_\pm \;=\; g \overline{g} + h \overline{h} \pm \sqrt{(g \overline{g})^2 -
g^2\: \overline{g}^2 + (g \overline{h} + h \overline{g})^2 } \]
(for more details see~\cite[\S5.3]{PFP}; we use the short notations
$g \overline{g} \equiv g_j \overline{g^j}$ and~$g^2 \equiv g_j g^j$,
$\overline{g}^2 \equiv \overline{g_j g^j}$). If the discriminant is negative,
the~$\lambda_\pm$ form a complex conjugate pair. If conversely the discriminant is positive,
the~$\lambda_\pm$ are both real. In order to show that they have the same sign, we compute
their product,
\begin{eqnarray*}
\lambda_+ \lambda_- &=& (g \overline{g} + h \overline{h})^2 - \left[ (g \overline{g})^2
- g^2\: \overline{g}^2 + (g \overline{h} + h \overline{g})^2 \right] \\
&=& 2\:(g \overline{g})\: |h|^2 + |h|^4 + g^2\: \overline{g}^2 -
(g \overline{h} + h \overline{g})^2 \\
&=& |h|^4 + g^2\: \overline{g}^2 - g^2\: \overline{h}^2 - h^2\: \overline{g}^2 \\
&=& (g^2-h^2)(\overline{g}^2 - \overline{h}^2) \;\geq\; 0\:.
\end{eqnarray*}

\vspace*{-0.79cm}
\QED

Using this lemma, we can simplify the Lagrangian~(\ref{Lcrit}) to
\beq \label{F}
{\mathcal{L}}[A] \;=\; \left\{ \begin{array}{cl} (\lambda_+ - \lambda_-)^2 &
{\mbox{if $\lambda_\pm \in \R$}} \\
0 & {\mbox{if $\lambda_\pm \not \in \R$}}\:.
\end{array} \right.
\eeq
Furthermore, we can easily compute~$\M$ as defined by~(\ref{fvar}).
Namely, in the case~$\lambda_\pm \not \in \R$, the roots of the characteristic polynomial
by continuity will be non-real on an open neighborhood of~$A$. However, the degeneracy
will in general not be preserved, and thus the spectrum of~$A$ will consist of two complex
conjugate pairs~$\lambda_1, \overline{\lambda_1}$ and~$\lambda_2, \overline{\lambda_2}$
(possibly with~$\lambda_1=\lambda_2$). The Lagrangian~(\ref{Lcrit}) then becomes
\beq \label{Lcp}
{\mathcal{L}}[A] \;=\; \left( |\lambda_1| - |\lambda_2| \right)^2 .
\eeq
According to standard perturbation theory with degeneracies, the spectrum is Lipschitz
continuous in~$A$. Hence~(\ref{Lcp}) is differentiable at the point~$\lambda_1=\lambda_2$,
and its gradient vanishes. If conversely
$\lambda_\pm \in \R$ and $\lambda_+ \neq \lambda_-$, these properties will again be
preserved in a neighborhood of~$A$. In this neighborhood, we can write~(\ref{F}) as
\[ {\mathcal{L}}[A] \;=\; \Tr(A^2) - \frac{1}{4}\: \Tr(A)^2\:, \]
and varying this Lagrangian gives two times the trace-free part of~$A$,
\[ \M[A] \;=\; 2 A - \frac{1}{2}\, \Tr(A)\,\1\:. \]
In the remaining case~$\lambda_+=\lambda_- \in \R$, the Lagrangian~(\ref{F}) is
continuously differentiable and has vanishing gradient.
We thus obtain
\beq \label{0}
\M[A] \;=\; \left\{ \begin{array}{cl} \displaystyle
2 A - \frac{1}{2}\, \Tr(A)\,\1 & {\mbox{if $\lambda_\pm \in \R$}} \\[.8em]
0 & {\mbox{if $\lambda_\pm \not \in \R$}}\:.
\end{array} \right.
\eeq

In the region away from the light cone, where~$P(x,y)$ is a smooth function, the
matrix~$\M[A_{xy}]$ is well defined even without regularization. We
postpone the detailed calculations to Section~\ref{sec2}; here it suffices to derive
the general structure of~$\M[A_{xy}]$ from the following simple consideration.
For given space-time points~$x$ and~$y$ with~$(y-x)^2 \neq 0$, we know from the Lorentz symmetry that the unregularized fermionic projector~(\ref{A}) is of the form
\[ P(x,y) \;=\; \alpha\, \xi\slsh + \beta\:\1 \:,\spc
P(y,x) \;=\; \overline{\alpha}\, \xi\slsh + \overline{\beta}\:\1 \]
for some complex parameters~$\alpha$ and~$\beta$,
where we set~$\xi \equiv y-x$. As a consequence,
\beq \label{1}
A_{xy} \;=\; a\, \xi\slsh + b\, \1 \;=\; A_{yx}
\eeq
with real parameters~$a = \alpha \overline{\beta} + \beta \overline{\alpha}$
and~$b=|\alpha|^2 \xi^2 + |\beta|^2$. Suppose that the vector~$\xi$ is spacelike.
Then there is a Lorentz ``rotation'' which transforms $\xi$ to $-\xi$. Using
the Lorentz symmetry of~$A_{xy}$,
we find that~$A_{yx}$ is obtained from~$A_{xy}$ simply by the
replacement~$\xi \rightarrow -\xi$,
\[ A_{yx} \;=\; -a \xi\slsh + b \1\:. \]
This is consistent with the right equation in~(\ref{1}) only if~$a$ vanishes. We
conclude that~$A_{xy}$ is of the form
\[ A_{xy} \;=\; \left\{ \begin{array}{cl} a \xi\slsh + b\1 &
{\mbox{if $\xi$ is timelike}} \\
b \1 & {\mbox{if $\xi$ is spacelike}}\:.
\end{array} \right. \]
In the case when~$\xi$ is timelike, our last argument does not apply because a Lorentz
transformation which maps~$\xi$ to $-\xi$ involves a reversal of the time direction,
but the fermionic projector~(\ref{A}) has no time reflection symmetry as it distinguishes
the lower from the upper mass shells. However, if~$\xi$ is timelike, the identity
$\xi\slsh^2 = \xi^2>0$ shows that the matrix~$\xi\slsh$ has real eigenvalues.
We thus obtain from~(\ref{0}) that
\[ \M[A_{xy}] \;=\; \left\{ \begin{array}{cl} 2 a \xi\slsh &
{\mbox{if $\xi$ is timelike}} \\
0 & {\mbox{if $\xi$ is spacelike}}\:.
\end{array} \right. \]
Moreover, it is clear from~(\ref{1}) that~$\M[A_{xy}] = \M[A_{yx}]$
and, again using Lorentz symmetry, we conclude that there is a real function~$f$ such
that
\beq \label{3}
\M[A_{xy}] \;=\; \left\{ \begin{array}{cl} \xi\slsh\: \epsilon(\xi^0)\:f(\xi^2) &
{\mbox{if $\xi$ is timelike}} \\
0 & {\mbox{if $\xi$ is spacelike}} \end{array} \right.
\eeq
($\epsilon$ denotes the step function~$\epsilon(x)=1$ if~$x>0$ and~$\epsilon(x)=-1$ otherwise).
It is important to observe that~$\M[A_{xy}]$ vanishes for spacelike $y-x$
and in this way encodes the {\em{causal structure}} of Minkowski space.

Clearly, we cannot use the above argument on the light cone $\xi^2=0$,
where~$\M[A_{xy}]$ is ill-defined. Indeed, the function~$f(\xi^2)$ has a non-integrable
pole as~$\xi^2 \searrow 0$ (for details see Section~\ref{sec2}), and
therefore~$\M[A_{xy}]$ has a serious singularity on the light cone. If we
regularized by replacing~$P$ in the above construction by~$P^\varepsilon$, the
resulting~$\M[A^\varepsilon_{xy}]$ would be well defined for all~$\xi$.
Qualitatively speaking, the singularity on the light cone would be ``smeared out'' on the scale~$\varepsilon$, whereas away from the light
cone~$\M[A^\varepsilon_{xy}]$ would be well-approximated by the unregularized~$\M[A_{xy}]$
(at least for the regularizations considered in this paper, which all satisfy the
condition introduced in~\cite[\S5.6]{PFP} that~$P^\varepsilon$ should be macroscopic
away from the light cone). However, it is far from obvious what happens in the
limit~$\varepsilon \searrow 0$, and this is precisely the question which we shall address
here. We will see that in this limit, $\M[A_{xy}]$ will in
general develop singularities on the light cone, which diverge even in the distributional
sense. However, we will show that there are {\em{special}} regularizations
where~$\M[A_{xy}]$ does indeed converge as a distribution. Before we can state our
result, we must briefly recall the procedure in~\cite[\S5.6]{PFP}.
In order to make sense of the pole of~$\M[A_{xy}]$ across the light cone, a
distribution~$\tM(\xi)$ is introduced which coincides
with~$\M[A_{xy}]$ away from the light cone. $\tM(\xi)$ can be represented as the
distributional derivative of a function~$F$ which has an integrable pole on
the light cone and can thus be regarded as a regular distribution.
More precisely,
\beq \tM(\xi) \;=\; \Pdd_\xi \,\OBox_\xi \,F(\xi) \:, \label{5} \eeq
where~$F$ is causal, Lorentz invariant and odd under time reversals, i.e.
\beq F(\xi) \;=\; \Theta(\xi^2)\:\epsilon(\xi^0)\: g(\xi^2) \label{6} \eeq
for a suitable real function~$g(\xi^2)$ which has an integrable pole as~$\xi^2 \searrow 0$.
Let us consider the Fourier transform of~$\tM(\xi)$, which we denote for
convenience by $\hM(k)$ (omitting the tilde cannot lead to confusion because
the Fourier transform of~$\M[A_{xy}]$ is ill-defined due to the singularities
on the light cone).
The differential operators in~(\ref{5}) clearly correspond to multiplication
operators in momentum space,
\beq \hM(k) \;=\; -i k\slsh\, k^2\: \hat{F}(k)\:. \label{7} \eeq
As in~\cite{PFP}, we denote the {\em{mass cone}} by~${\mathcal{C}} = \{ k \:|\: k^2 > 0\}$
and define the upper and lower mass cone by~${\mathcal{C}}^\vee = \{k \in {\mathcal{C}} \:|\:
k^0 > 0\}$ and~${\mathcal{C}}^\wedge = \{k \in {\mathcal{C}} \:|\: k^0 < 0\}$, respectively.
The following argument shows that~$\hat{F}$ vanishes identically outside the closed
mass cone. Suppose that~$k \not \in \overline{\mathcal{C}}$. Due to Lorentz symmetry, we
can assume that~$k$ is purely spatial, $k=(0, \vec{k})$. When we now take the Fourier
transform of~(\ref{6}), the integrand of the resulting time integral is odd because of the
step function~$\epsilon(\xi^0)$, and thus the whole integral vanishes. In view of~(\ref{7}),
we come to the important conclusion that also~$\hM$ is
{\em{supported inside the mass cone}},
\beq \label{7a}
{\mbox{supp}}\, \hM \;\subset\; \overline{\mathcal{C}}\:.
\eeq

\begin{figure}[tb]
\begin{center}
{\includegraphics{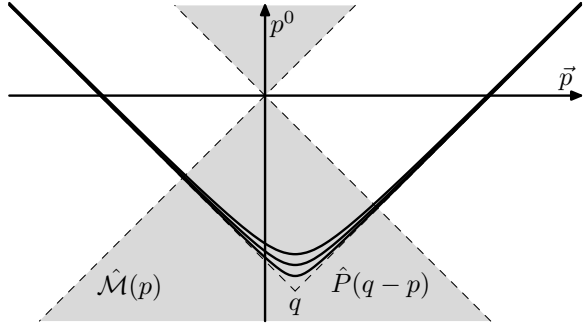}}
\caption{The convolution~${\hat{\mathcal{M}}} * \hat{P}$.}
\label{fig1}
\end{center}
\end{figure}

This support property makes it possible to make sense of the pointwise product in~(\ref{Qxydef})
even without regularization. Namely, setting
\[ Q(x,y) \;=\; \frac{1}{2}\: \tM(\xi)\: P(x,y) \:, \]
we rewrite the product as a convolution in momentum space,
\beq \label{ci}
\hat{Q}(q) \;=\; \frac{1}{2}\: (\hM * \hat{P})(q)
\;=\; \frac{1}{2} \int \frac{d^4p}{(2 \pi)^4}\: \hM(p)\:
\hat{P}(q-p)\:.
\eeq
If~$q$ is inside the lower mass cone, the last integrand has compact support
(see Figure~\ref{fig1}), and the integral is finite.
If however~$q \not \in \overline{{\mathcal{C}}^\wedge}$, the convolution
integral extends over an unbounded region and is thus ill-defined.

In~\cite[\S5.6]{PFP}, the regularization of the product in~(\ref{Qxydef}) is
treated using an ad-hoc assumption, which in our setting can be stated as follows.
\begin{Def} \label{def1} The regularized fermionic projector~$(P^\varepsilon)_{\varepsilon>0}$
satisfies the assumption of a {\bf{distributional $\M P$-product}} if the following conditions
are satisfied:
\begin{description}
\item[(i)] There is a distribution~$\tM(\xi)$ of the form~(\ref{5}, \ref{6})
such that $\lim_{\varepsilon \searrow 0} \M[A^\varepsilon_{xy}] =
\tM(\xi)$ in the distributional sense.
\item[(ii)] For every~$k$ for which~$\lim_{\varepsilon \searrow 0} \hat{Q}^\varepsilon(k)$
exists, the convolution integral~(\ref{ci}) is well defined
and $\lim_{\varepsilon \searrow 0} \hat{Q}^\varepsilon(k) = \hat{Q}(k)$.
\end{description}
\end{Def}

\subsection{Statement of the Main Results}
The main result of the present paper is to justify the assumption of a
distributional $\M P$-product in the following sense.
\begin{Thm} \label{thm1}
Suppose that~${\tilde{\mathcal{M}}}(\xi)$ is a distribution
which away from the light-cone coincides with~${\mathcal{M}}[A_{xy}]$,
(\ref{0}), and can be represented in the form~(\ref{5}, \ref{6})
with~$F$ a regular distribution.
Then there is a family of regularized fermionic projectors~$(P^\varepsilon)_{\varepsilon>0}$
with the following properties:
\begin{description}
\item[(1)] The~$P^\varepsilon$ satisfy~(\ref{Z}), are homogeneous~(\ref{B}) and have
vector-scalar structure~(\ref{C}). Furthermore, they are spherically symmetric and are
composed of surface states.
\item[(2)] $\lim_{\varepsilon \searrow 0} \M[A^\varepsilon_{xy}] =
\tM(\xi)$ in the distributional sense.
\item[(3)] For every~$k \in {\mathcal{C}}^\wedge$,
\[ \lim_{\varepsilon \searrow 0} \hat{Q}^\varepsilon(k) \;=\; \hat{Q}(k) \:, \]
with~$\hat{Q}(k)$ as defined by~(\ref{ci}).
\item[(4)] For every~$k \not\in \overline{{\mathcal{C}}^\wedge}$,
$\hat{Q}^\varepsilon(k)$ diverges in the sense that
\[ \lim_{\varepsilon \searrow 0}
\inf \sigma \!\left(\hat{Q}^\varepsilon(k) \right) \;=\; +\infty\:. \]
\end{description}
\end{Thm}
The constructions used for proving this theorem leave us with the freedom to modify~$Q$
by distributional contributions supported at the origin~$\xi=0$. Expressed in momentum
space, we prove the following generalization of Theorem~\ref{thm1}.

\begin{Thm} \label{thmbeyond}
Under the assumptions of Theorem~\ref{thm1}, for any real parameters~$c_2, c_3, c_4$
there is a family of regularized fermionic projectors~$(P^\varepsilon)_{\varepsilon>0}$,
which has the properties~(1), (2) and~(4) in the statement of Theorem~\ref{thm1}
and instead of~(3) satisfies the condition
\begin{description}
\item[(3')] For every~$k \in {\mathcal{C}}^\wedge$,
\[ \lim_{\varepsilon \searrow 0} \hat{Q}^\varepsilon(k)
\;=\;\frac{1}{2}\: (\hM * \hat{P})(k) + c_2 + c_3 \,k\slsh + c_4 \,k^2\:. \]
\end{description}
\end{Thm}

The proofs of these theorems are constructive and will give us detailed information on the
admissible regularizations. This will be discussed in Section~\ref{sec8}.

\section{The Singularities of~$\tM$ in Polar Coordinates} \label{sec2}
\setcounter{equation}{0}
We begin the explicit calculations by determining the function~$f$ in~(\ref{3}).
Clearly, we may assume that the vector~$\xi$ is timelike and future directed. Then the Fourier
integral~(\ref{B}) can be given explicitly in terms of Bessel functions, which
in turn have a series expansion,
\begin{eqnarray}
P(x,y) &=& \sum_{\beta=1}^3 \rho_\beta\:
(i \Pdd_x + m_\beta) \left(a_\beta(\xi^2) + i b_\beta(\xi^2)
\right) \label{Pdef} \\
a_\beta(\xi^2) &=& \frac{1}{2} \int \frac{d^4k}{(2 \pi)^4}\:
\delta(k^2-m_\beta^2)\: e^{-ik(x-y)} \;=\;
\frac{m_\beta^2}{16 \pi^2} \: \frac{Y_1(m_\beta
\sqrt{\xi^2})}{m_\beta \sqrt{\xi^2}} \nonumber \\
&=& -\frac{1}{8 \pi^3\: \xi^2} \:+\: \frac{m_\beta^2}{32 \pi^3}
\left(\log(m_\beta^2 \xi^2) + c \right) \:+\:
{\mathcal{O}}(\xi^2 \log \xi^2) \label{bessel1} \\
b_\beta(\xi^2) &=& \frac{i}{2} \int \frac{d^4k}{(2 \pi)^4}\:
\delta(k^2-m_\beta^2)\: \epsilon(k^0)\: e^{-ik(x-y)} \;=\;
\frac{m_\beta^2}{16 \pi^2} \: \frac{J_1(m_\beta
\sqrt{\xi^2})}{m_\beta \sqrt{\xi^2}} \nonumber \\
&=& \frac{m_\beta^2}{32 \pi^2} - \frac{m_\beta^4}{256 \pi^2}\:\xi^2 \:+\:
{\mathcal{O}}(\xi^4) \:, \label{bessel2}
\end{eqnarray}
where~$c=2C - 2 \log2 - 1$ and~$C$ is Euler's constant.
Using that the functions~$a_\beta$ and~$b_\beta$ are real, a short calculation
yields that
\[ A_{xy} - \frac{1}{4}\, \Tr(A_{xy})\, \1 \;=\;
-2 \xi\slsh \sum_{\alpha, \beta=1}^3 \left( a'_\alpha\: m_\beta b_\beta +
m_\alpha b_\alpha\: a'_\beta
-b_\alpha'\: m_\beta a_\beta - m_\alpha a_\alpha\: b'_\beta \right) , \]
and comparing with~(\ref{0}, \ref{3}), we obtain the simple formula
\beq \label{fdef}
f \;=\; -8 \sum_{\alpha, \beta=1}^3 \left(
a_\alpha'\: m_\beta b_\beta - m_\alpha a_\alpha\: b'_\beta \right) .
\eeq
Substituting the above asymptotic expansions of the Bessel functions, we find that~$f$ really has a non-integrable pole. More precisely,
\beq \label{fex}
f(z) \;=\; \frac{\m_3}{z^2} + \frac{\m_5}{z} \:+\: {\mathcal{O}}(\log z)\:,
\eeq
where we set~$z \equiv \xi^2$ and introduced the abbreviations
\begin{eqnarray*}
\m_3 &=& -\frac{1}{64\, \pi^5}
\sum_{\alpha, \beta=1}^g \rho_\alpha\, \rho_\beta\:(m_\alpha^3+m_\beta^3) \\
\m_5 &=& \frac{1}{512\, \pi^5} \sum_{\alpha, \beta=1}^g  \rho_\alpha\, \rho_\beta\:
(m_\alpha-m_\beta)^2\, (m_\alpha+m_\beta)^3 \:.
\end{eqnarray*}

Next we compute the function~$g$ in~(\ref{6}). The Dirac operator and the Laplacian
of a Lorentz invariant function~$h(\xi^2)$ are computed by
\[ \Pdd h(z) \;=\; 2 \xi\slsh\, h'(z)\:,\spc
\OBox h(z) \;=\; \frac{4}{z} \left(z^2\: h'(z) \right)'. \]
Using these formulas, it is immediately verified that the function~$g(z)$, defined for
given real constants~$c_0$, $c_1$ and any~$z>0$ by
\beq \label{gdef}
g(z) \;=\; \frac{1}{8} \int_1^z \frac{d\tau}{\tau^2} \int_0^\tau \sigma\, d\sigma
\int_1^\sigma f \;+\; \frac{c_0}{8} \:+\: \frac{c_1}{16}\:z \;,
\eeq
in the interior of the light cone is a solution of the equation
\[ \xi\slsh\, f(z) \;=\; \Pdd \:\OBox g(z)\:. \]
Furthermore, using~(\ref{fex}) in~(\ref{gdef}) one sees that~$g$ has
only a logarithmic pole as~$z \searrow 0$. Therefore, we can
use~(\ref{6}) to define~$F$ as a regular distribution, and taking its distributional
derivative~(\ref{5}) gives the distribution~$\tM(\xi)$, which coincides
away from the light cone with the function~$\M[A_{xy}]$. The parameters~$c_0$
and~$c_1$ can be interpreted as integration constants. From the calculations
\begin{eqnarray*}
\Pdd \,\OBox \left( \Theta(\xi^2)\: \epsilon(\xi^0) \right) &=&
4 \,\Pdd \left( \delta(\xi^2)\: \epsilon(\xi^0) \right) \;=\; 8 \,\xi\slsh\:
\delta'(\xi^2)\: \epsilon(\xi^0) \\
\Pdd \,\OBox \left( \xi^2\,\Theta(\xi^2)\: \epsilon(\xi^0) \right) &=&
8 \,\Pdd \left( \Theta(\xi^2)\: \epsilon(\xi^0) \right) \;=\; 16 \,\xi\slsh\:
\delta(\xi^2)\: \epsilon(\xi^0)
\end{eqnarray*}
one sees that they give rise to contributions to~$\tM$ supported on the light
cone,
\[ \frac{\partial}{\partial c_0} \tM(x,y) \;=\; \xi\slsh\:
\delta'(\xi^2)\: \epsilon(\xi^0)\:, \spc
\frac{\partial}{\partial c_1} \tM(x,y) \;=\; \xi\slsh\:
\delta(\xi^2)\: \epsilon(\xi^0)\:. \]

Our next lemma shows that the distribution~$\tM(\xi)$ is regular at
the {\em{origin}} $\xi=0$. We let~$\eta \in C^\infty_0(\R^4)$ be a test function
which is identically equal to one in a neighborhood of the origin. We set for
any~$\delta>0$
\beq \label{tf}
\eta_\delta(\xi) \;=\; \eta\!\left(\frac{\xi}{\delta}\right) .
\eeq

\begin{Lemma} \label{lemma21}
$\lim_{\delta \searrow 0} \eta_\delta(\xi)\: \tM(\xi) = 0$
in the distributional sense.
\end{Lemma}
{\Proof} According to~(\ref{5}), we have for any test function~$h \in C^\infty_0(\R^4)$,
\begin{eqnarray*}
\int \eta_\delta(\xi)\: \tM(\xi)\: h(\xi)\: d^4\xi
&=& -\int F(\xi)\; \OBox_\xi \,\Pdd_\xi \Big(h(\xi)\: \eta_\delta(\xi) \Big) d^4\xi \\
&=& -\delta \int_{{\mbox{\scriptsize{supp}}}\, \eta} F(\delta \zeta)\;
\OBox_\zeta \,\Pdd_\zeta \Big(h(\delta \zeta)\: \eta(\zeta) \Big) d^4\zeta \:,
\end{eqnarray*}
where in the last step we introduced the new variable~$\zeta = \xi/\delta$. Estimating
the integrand on the support of~$\eta$ by
\[ |F(\delta \zeta)| \;\leq\; c \:|F(\zeta)|\:\log \delta \:,\spc
\left| \,\OBox_\zeta \Pdd_\zeta \!\left(h(\delta \zeta)\: \eta(\zeta) \right) \right|
\;\leq\; c\:, \]
we can bound the integral by a constant times~$\log \delta$.
\QED

It remains to analyze the singularities of~$\tM(\xi)$ on the light cone,
away from the origin. Due to the symmetry under the transformation~$\xi \rightarrow -\xi$,
we may restrict attention to the {\em{future light cone}}. Thus choosing polar
coordinates~$(t=\xi^0, r=|\vec{\xi}|, \vartheta, \varphi)$, we consider the region
$t \approx r>0$, and thus it is convenient to introduce the ``small'' time
variable~$s=t-r$. Furthermore, due to spherical symmetry, we can decompose~$\tM$
into a time and a radial component as follows,
\beq \tM \;=\; \tM^0 \gamma^0 - \tM^r \gamma^r\:, \label{Mde}
\eeq
where we set
\[ \gamma^r \;=\; \frac{\vec{\xi} \vec{\gamma}}{r}\:, \]
and~$\tM^{t/r}(s,r)$ are two real-valued, spherically
symmetric distributions.

In view of the calculations in the following sections, it is most convenient to describe
the singularity on the light cone by evaluating for fixed~$r$ weakly in~$s$. More precisely,
we consider for any~$0<s_0<r$ the integrals
\[ \int_{-s_0}^{s_0} s^n\: \tM^{t/r}(s,r)\: ds\:,\spc
n=0,1,\ldots , \]
which can be thought of as measuring the ``$n^{\mbox{\scriptsize{th}}}$ moment'' of the
singularity. Here the lower limit of the integral is irrelevant because the integrand
vanishes identically on the interval~$(-s_0, 0)$; we only need to ensure to integrate
over an open interval containing the origin. The upper limit of the integral is
also not interesting because for different values of~$s_0$, the corresponding integrals
coincide up to a contribution which is completely determined by the well-known behavior
of~$\tM$ away from the light cone. Therefore, we would like to take the
limit~$s_0 \searrow 0$. However, since the distributions~$\tM^{t/r}$ have
non-integrable poles at~$s=0$, the above integrals need not converge as~$s_0 \searrow 0$.
But we can take the limit~$s_0 \searrow 0$ after subtracting indefinite integrals of
the poles. More precisely, from~(\ref{3}) and~(\ref{fex}) we have for small~$s>0$ the expansions
\begin{eqnarray*}
\tM^0(s,r) &=& \frac{\m_3}{4r s^2} + \frac{\m_5}{2 s}
+ {\mathcal{O}}(\log s) \\
\tM^r(s,r) &=& \frac{\m_3}{4r s^2} + \frac{\m_5}{2 s} - \frac{\m_3}{4r^2 s}
+ {\mathcal{O}}(\log s)\:.
\end{eqnarray*}
Thus we introduce the quantities
\begin{eqnarray}
I^0(r) &=& \lim_{s_0 \searrow 0} \left\{ \int_{-s_0}^{s_0} \tM^0(s,r)\:ds
\:+\: \frac{\m_3}{4r s_0} - \frac{\m_5}{2}\: \log s_0 \right\} \label{Itdef} \\
I^r(r) &=& \lim_{s_0 \searrow 0} \left\{ \int_{-s_0}^{s_0} \tM^r(s,r)\:ds
\:+\: \frac{\m_3}{4r s_0} - \frac{\m_5}{2}\: \log s_0 + \frac{\m_3}{4r^2}\: \log s_0
\right\} \label{Irdef} \\
J^{t/r}(r) &=& \lim_{s_0 \searrow 0} \left\{ \int_{-s_0}^{s_0} s\:
\tM^{t/r}(s,r)\:ds \:-\: \frac{\m_3}{4r} \:\log s_0 \right\}
\label{Jtrdef} \\
K^{t/r}_n(r) &=& \lim_{s_0 \searrow 0} \int_{-s_0}^{s_0} s^n\: \tM^{t/r}(s,r)\:ds\:,
\spc n \geq 2\:. \label{hmdef}
\end{eqnarray}
These real-valued functions are well defined for all~$r>0$. They have the nice property
of being independent of all smooth contributions to~$\tM(\xi)$ as well
as of the integrable poles of~$\tM(\xi)$. This makes it possible
in what follows to work only with the expansion of~$f$ as given
by~(\ref{fex}), thus avoiding the complicated formulas involving
Bessel functions.

\begin{Lemma} \label{lemma22}
The functions~$K^{t/r}_n$ all vanish. Furthermore,
\begin{eqnarray}
I^0(r) &=& \frac{\m_3}{8}\: \frac{1}{r^2} \:+\: \frac{\m_5}{2}\: \log 2r \:+\: \frac{\m_3+c_1}{2} \label{2a} \\
J^0(r) &=& \frac{\m_3}{4}\: \frac{\log 2r}{r} \:+\: \frac{\m_3 -2 c_0}{8r} \label{2b} \\
I^r &=& I^0 \:-\: \frac{1}{r}\: J^0\:,\spc
J^r \;=\; J^0\:. \label{2c}
\end{eqnarray}
\end{Lemma}
{\Proof} The identities for the radial component follow from those for the time component
simply using the Lorentz symmetry of~$\tM$. Namely, assume for the moment that the lemma
holds for the time component. According to~(\ref{5}) and~(\ref{Mde}), we can
write~$\tM^{t/r}$ as derivatives of a Lorentz invariant distribution
$G := \OBox F$,
\[ \tM^0 \;=\; \partial_t G(t^2-r^2)\:,\spc
\tM^r \;=\; -\partial_r G(t^2-r^2) \:. \]
Differentiating with the Leibniz rule, one sees that~$r \tM^0 = t \tM^r$, and thus
\beq \label{2ctr}
\tM^r \;=\; \frac{r}{r+s}\: \tM^0 \;=\; \left(1 - \frac{s}{r} +
{\mathcal{O}}(s^2) \right) \tM^0\:.
\eeq
Integrating over~$s$ yields~(\ref{2c}) as well as the vanishing of~(\ref{hmdef})
for the radial component.

It remains to consider the time component. Using~(\ref{fex}) in~(\ref{gdef}), we obtain
for~$g$ the expansion
\[ g(z) \;=\; -\frac{\m_3}{8}\: \log z + \frac{\m_5}{16}\: z \log z
+ \frac{4 c_0-2 \m_3+3 \m_5}{32} + \frac{2 c_1+2 \m_3-3 \m_5}{32}\:z
+ {\mathcal{O}}(z^2 \log z) \:. \]
One possible way to proceed would be to regularize the poles of~$g$, to compute the corresponding
regularized~$\tM$ via~(\ref{5}, \ref{6}), to evaluate weakly in time and finally to remove
the regularization (this method is well-suited for computer algebra). Here we describe a
different method, which is a bit more elegant. We choose a test function~$\eta \in
C^\infty_0(\R)$ which vanishes in a neighborhood of the origin. Due to spherical symmetry,
we can omit the angular integrals. Then, applying the definition of weak derivatives
to~(\ref{5}), we obtain
\[ \int_0^\infty r^2\: \eta(r)\: dr \int_{-s_0}^{s_0} ds\:s^n\: \tM^0(t,r)
\;=\; -\int_0^\infty r^2\: dr \int_0^\infty ds\: g(t^2-r^2)\: \OBox \partial_t \Big(
\eta(r)\: s^n\: \Theta(s_0-s) \Big) , \]
where the derivatives act on~$\Theta(s_0-s)$ in the distributional sense, giving
$\delta$- and $\delta'$-con\-tri\-bu\-tions supported on the hypersurface $s=s_0$ where~$g$ is
smooth. The derivatives of~$\eta$ can be rewritten as directional derivatives
$\partial_t + \partial_r$ tangential to the light cone,
$\eta^{(k)}(r) = (\partial_t + \partial_r)^k \eta(r)$. These directional derivatives may
be integrated by parts because the resulting tangential derivatives of~$g$ are again
integrable. The resulting expression is of the general form
\[ \int_0^\infty r^2\: \eta(r)\: dr \int_0^\infty ds \:\cdots \]
and justifies that~$\tM^0$ can indeed be evaluated for fixed~$r$ weakly in~$t$. Furthermore,
going through the detailed calculation, one finds that
\begin{eqnarray*}
\lefteqn{ \int_{-s_0}^{s_0} s^n\: \tM^0(s,r)\:dr \;=\; \left. -2n r^{-1} s^{n-1}\: g
\:-\: 4\,(n-2)\: s^n\: g' \:+\: 4 \,(t+r)\, s^{n+1}\: g'' \right|_{s=s_0} } \\
&&+ \int_0^{s_0} \left( 2 n \,(n-1)\: r^{-1} \: s^{n-2}\: g
\:+\: 4 n\,(n-1+r^{-1} s) s^{n-1}\: g'
\:+\: 4n\: s^{1+n}\: g'' \right) ds \:.
\end{eqnarray*}
The last integral vanishes in the limit~$s_0 \searrow 0$ for any $n \geq 0$.
Furthermore, in the case~$n \geq 2$ it is obvious that the boundary terms at~$s_0$ tend to zero
as~$s_0 \searrow 0$. In the cases~$n=0,1$, the boundary terms diverge
as $s_0$ tends to zero. But these divergences are exactly canceled by the
counterterms in~(\ref{Itdef}) and~(\ref{Jtrdef}). Thus we can compute the
limit~$s_0 \searrow 0$ to obtain the result.
\QED

\section{The Momentum Cone Conditions} \label{sec3a}
\setcounter{equation}{0}
In this section we explain a particular difficulty in proving
Theorem~\ref{thm1}~(3). Apart from illustrating the
statement of our main theorem, our analysis will lead
to conditions which will be important for the constructions of the
subsequent sections.
Let us assume for the moment that Theorem~\ref{thm1}~(2) holds,
i.e.\ suppose that we already
know that $\M[A^\varepsilon_{xy}]$ converges as a
distribution to~$\tM(\xi)$. This statement entails that, in the
limit~$\varepsilon \searrow 0$,
$\M[A^\varepsilon_{xy}]$ develops singularities on the
light cone as characterized by Lemma~\ref{lemma22}. If~$P^\varepsilon(x,y)$
converged locally
uniformly to a smooth function, we could immediately
conclude that the product~$\M[A^\varepsilon_{xy}]\, P^\varepsilon(x,y)$
converges in the distributional sense to~$\tM(\xi)\, P(x,y)$.
Unfortunately, the situation is more difficult because
$P(x,y)$ also has singularities on the light cone
(see~(\ref{Pdef}--\ref{bessel2})). As a consequence,
we must expect that in the limit~$\varepsilon \searrow 0$, the function
$\M[A^\varepsilon_{xy}]\, P^\varepsilon(x,y)$ 
will develop singularities on the
light cone, which do not even make sense as distributions.
The main difficulty in proving Theorem~\ref{thm1}~(3) is to
show that, despite the divergences of the
product~$\M[A^\varepsilon_{xy}]\, P^\varepsilon(x,y)$ on the light cone,
its Fourier transform has a well defined limit inside the lower mass
cone, which coincides with~(\ref{ci}).

The nature of this problem becomes clearer when working purely in
momentum space. Similar to~(\ref{ci}), we can rewrite the product in
position space as a convolution in momentum space,
\beq \label{ci2}
\int \frac{d^4p}{(2 \pi)^4}\: \hM^\varepsilon(p)\: \hat{P}^\varepsilon(q-p)\:.
\eeq
We let~$q$ be a fixed vector inside the lower mass cone.
In contrast to the situation in Figure~\ref{fig1},
$\hM^\varepsilon$ will not be supported inside the mass cone,
because the regularization will in general give rise also to a contribution
outside the mass cone.
Furthermore, for~$p^0$ on the order of the Planck energy, the support
of~$\hat{P}^\varepsilon$ may no longer lie on the mass
hyperbolas, but might have a completely different form.
For~$p$ in any compact set, these regularization effects are
not a problem, because the distributional convergences~$\hM^\varepsilon \rightarrow \hM$ and~$P^\varepsilon \rightarrow P$
ensure that for bounded~$p$ and~$q$ the effects of the regularization
tend to zero as~$\varepsilon \searrow 0$. However, the problem is
to show that the contribution due to the regularization for large~$p^0$ to the
convolution integral~(\ref{ci2}) tends to zero
as~$\varepsilon \searrow 0$.

\begin{figure}[tb]
\begin{center}
\begin{picture}(0,0)%
\includegraphics{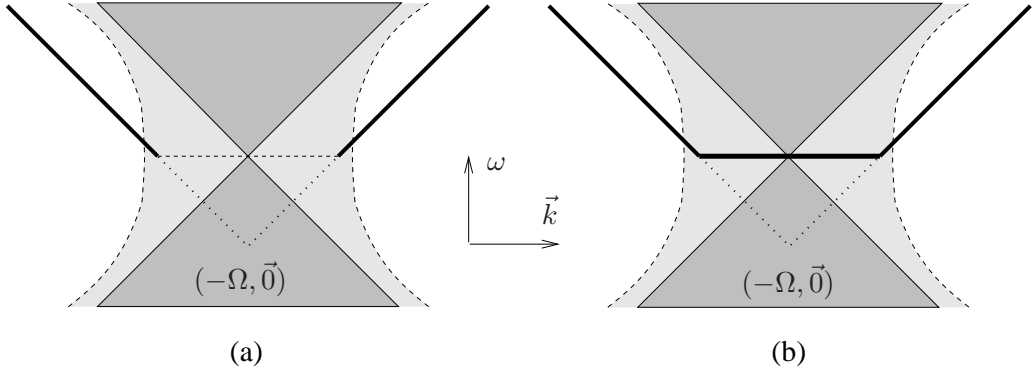}%
\end{picture}%
\setlength{\unitlength}{1657sp}%
\begingroup\makeatletter\ifx\SetFigFont\undefined%
\gdef\SetFigFont#1#2#3#4#5{%
  \reset@font\fontsize{#1}{#2pt}%
  \fontfamily{#3}\fontseries{#4}\fontshape{#5}%
  \selectfont}%
\fi\endgroup%
\begin{picture}(15408,5507)(397,-8657)
\put(7628,-5664){\makebox(0,0)[lb]{\smash{{\SetFigFont{11}{13.2}{\familydefault}{\mddefault}{\updefault}$\omega$}}}}
\put(8447,-6474){\makebox(0,0)[lb]{\smash{{\SetFigFont{11}{13.2}{\familydefault}{\mddefault}{\updefault}$\vec{k}$}}}}
\put(11431,-7486){\makebox(0,0)[lb]{\smash{{\SetFigFont{11}{13.2}{\familydefault}{\mddefault}{\updefault}$(-\Omega,\vec{0})$}}}}
\put(3241,-7441){\makebox(0,0)[lb]{\smash{{\SetFigFont{11}{13.2}{\familydefault}{\mddefault}{\updefault}$(-\Omega,\vec{0})$}}}}
\end{picture}%
\caption{The convolution~${\hat{\mathcal{M}}}^\varepsilon * \hat{H}$
for different choices of~$\hat{H}$.}
\label{figcon}
\end{center}
\end{figure}

In order to develop a method for analyzing an integral of
the form~(\ref{ci2}) for large~$p^0$, we first make a few
simplifications. First, we shall {\em{regularize only one factor}} and
consider the product~$\M[A^\varepsilon_{xy}]\, P(x,y)$
of the regularized~$\M$ with the unregularized~$P$.
Thus~$P$ is singular on the light cone, whereas~$\M[A^\varepsilon_{xy}]$
is, due to the regularization, a smooth function.
Furthermore, we assume that~$\M[A^\varepsilon_{xy}]$
{\em{vanishes in a neighborhood of the light cone}}. This assumption
seems to make the problem trivial, because as a consequence
the product~$\M[A^\varepsilon_{xy}]\, P(x,y)$ is even a smooth function.
Nevertheless, this case is interesting because
the singularities on the light cone will reappear as soon as we take
the limit~$\varepsilon \searrow 0$. Moreover, we replace the mass shells of~$P$ for
large~$p^0$ by one {\em{cone}}, whereas for~$p$ in a compact set
we are free to choose~$P$ in a convenient way.
Finally, we choose a reference frame such that~$q=(-\Omega, \vec{0})$
with~$\Omega>0$. We are thus led to the question of whether the integrals
\beq \label{ci4}
B^\varepsilon \;:=\;
\int \frac{d^4p}{(2 \pi)^4}\: \hM^\varepsilon(p)\: \hat{H}(q-p)
\eeq
vanish in the limit~$\varepsilon \searrow 0$, where~$\hat{H}$ is a
distribution whose support can be chosen for example as in
Figure~\ref{figcon}~(a) or~(b).
More specifically, in order to model the scalar and vector components
of~$P$, we shall choose~$\hat{H}$ equal to
\beq
\hat{H}_{\mbox{\tiny{scal}}}(k) \;=\; \delta(k^2)\: \Theta(-k^0-\Omega)
\spc {\mbox{(see Figure~\ref{figcon}~(a))}} \label{H1}
\eeq
and
\begin{eqnarray}
\hat{H}_{\mbox{\tiny{vect}}}(k) &=& k \slsh \,\delta(k^2)\: \Theta(-k^0 - \Omega)
- \frac{1}{2} \gamma^0\, \delta(-k^0 - \Omega)\,\Theta(k^2) \nonumber \\
&=& \frac{1}{2} \:\Pdd_k \left( \Theta(k^2)\: \Theta(-k^0 - \Omega) \right)
\spc {\mbox{(see Figure~\ref{figcon}~(b))}} \:,
\label{H2}
\end{eqnarray}
respectively. This choice of~$\hat{H}$ has the advantage that
it will be possible to easily compute the Fourier integral of~$\hat{H}$.

Despite all the simplifications made, the analysis of~(\ref{ci4})
for~$H$ according to~(\ref{H1}) and~(\ref{H2}) will be helpful
in order to understand under which conditions on~$\M[A^\varepsilon_{xy}]$
we can expect that Theorem~\ref{thm1}~(3) holds.
We first bring the integral~(\ref{ci4}) into a more explicit form
in position space.
\begin{Lemma} \label{lemma4n1}
Suppose that the function~$\M[A^\varepsilon_{xy}]$ vanishes in a
neighborhood of the light cone. Then for~$\hat{H}$ according
to~(\ref{H1}) and~(\ref{H2}), the corresponding convolution
integral~(\ref{ci4}) is given by
\begin{eqnarray}
B^\varepsilon_{\mbox{\tiny{\rm{scal}}}} &=& \int \frac{d^4 \xi}{16 \pi^3}
\: \M[A^\varepsilon_{xy}] \left(\frac{e^{-i \Omega r}}{r\, (t+r)} -
\frac{e^{i \Omega r}}{r\, (t-r)} \right) \label{Bs} \\
B^\varepsilon_{\mbox{\tiny{\rm{vect}}}} &=& \int \frac{d^4 \xi}{16 \pi^3}
(\M[A^\varepsilon_{xy}]\, i \xi\slsh)
\:\frac{1}{r} \frac{\partial}{\partial r}
\left(\frac{e^{-i \Omega r}}{r\, (t+r)} -
\frac{e^{i \Omega r}}{r\, (t-r)} \right) , \label{Bv}
\end{eqnarray}
respectively, where again~$\xi = (t, \vec{x})$ and~$r=|\vec{x}|$.
\end{Lemma}
{\Proof} We first consider the function~$\hat{H}_{\mbox{\tiny{scal}}}$.
Its Fourier transform can be computed in polar coordinates,
\begin{eqnarray}
H_{\mbox{\tiny{scal}}}(\xi) &=& \frac{1}{(2 \pi)^3} \int_{-\infty}^{-\Omega} d\omega\, e^{i \omega t} \int_0^\infty p^2 \,dp
\: \delta(\omega^2 - p^2)
\int_{-1}^1 d\cos \vartheta\, e^{-i p r \cos \vartheta} \nonumber \\
&=& \frac{i}{8 \pi^3\,r}
\int_{-\infty}^{-\Omega} d\omega\, e^{i \omega t} \int_0^\infty p \,dp
\: \delta(\omega^2 - p^2) \left(e^{-i p r} - e^{i p r} \right) \nonumber \\
&=& \frac{i}{16 \pi^3\, r} \int_{-\infty}^{-\Omega} d\omega\,
\left( e^{i \omega (t+r)} - e^{i \omega (t-r)} \right) \label{hscalf} \\
&=& \frac{1}{16 \pi^3} \:\frac{1}{r} \left[
e^{-i \Omega (t+r)} \left(\frac{\mbox{PP}}{t+r} + i \pi \delta(t+r) \right)
- e^{-i \Omega (t-r)}\left(\frac{\mbox{PP}}{t-r} + i \pi \delta(t-r)
\right) \right] . \nonumber
\end{eqnarray}
Using Plancherel's theorem, we can rewrite~(\ref{ci4}) in position space as
\beq \label{ci5}
B^\varepsilon \;=\; \int e^{i \Omega t}
\M[A^\varepsilon_{xy}]\, H(\xi)\: d^4 \xi\:.
\eeq
Substituting the above formula for~$H_{\mbox{\tiny{scal}}}(\xi)$
and using that~$M[A^\varepsilon_{xy}]$ vanishes in a neighborhood of the
light cone, we obtain~(\ref{Bs}). The identity~(\ref{Bv}) follows similarly
from the calculation
\begin{eqnarray*}
H_{\mbox{\tiny{vect}}}(\xi) &=& \int \frac{d^4k}{32 \pi^4}
\:\Pdd_k \left( \Theta(k^2)\: \Theta(-k^0 - \Omega) \right) e^{i k \xi} \\
&=& -i \xi\slsh \int \frac{d^4k}{32 \pi^4}\:
\Theta(k^2)\: \Theta(-k^0 - \Omega) \:e^{i k \xi} \\
&=& \frac{\xi\slsh}{16 \pi^3\,r} \int_{-\infty}^{-\Omega} d\omega
\, e^{i \omega t} \int_0^{|\omega|} p \,dp
\left(e^{-i p r} - e^{i p r} \right) \\
&=& \frac{i \xi\slsh}{16 \pi^3\,r} \frac{\partial}{\partial r}
\int_{-\infty}^{-\Omega} d\omega \, e^{i \omega t} \int_0^{|\omega|}
\left(e^{-i p r} + e^{i p r} \right) dp \\
&=& -\frac{\xi\slsh}{16 \pi^3\,r} \frac{\partial}{\partial r}
\left\{ \frac{1}{r}
\int_{-\infty}^{-\Omega} d\omega \left( e^{i \omega (t+r)}
- e^{i \omega (t-r)} \right) \right\} \\
&=& \frac{i \xi\slsh}{16 \pi^3\,r} \frac{\partial}{\partial r}
\left\{ \frac{1}{r}
\left. e^{-i \Omega u} \left(\frac{\mbox{PP}}{u} + i \pi \delta(u)
\right) \right|^{u=t+r}_{u=t-r} \right\} .
\end{eqnarray*}

\vspace*{-0.79cm}
\QED
Let us discuss the result of this lemma, first in the case of a
{\em{Lorentz invariant regularization}}, which for convenience we write
in the form
\beq \label{3X}
\M[A^\varepsilon_{xy}] \;=\; \frac{1}{2}\, (\Pdd g_\varepsilon(\xi^2))\,
\epsilon(\xi^0)
\;=\; \xi\slsh \, g'_\varepsilon(\xi^2)\, \epsilon(\xi^0)
\eeq
with a smooth real function~$g_\varepsilon$ which vanishes if~$\xi^2<0$.
Then the argument leading to~(\ref{7a}) still holds and thus~${\mbox{supp}}\,
\hM[A^\varepsilon] \subset \overline{\mathcal{C}}$.
As a consequence, a support argument (see Figure~\ref{figcon}) shows
that the integral~(\ref{ci4}) vanishes. It is instructive to reproduce
this fact using the formulas of Lemma~\ref{lemma4n1}.
We first collect the terms for a fixed phase factor~$e^{-i \Omega r}$
in~(\ref{Bs}). Due to spherical symmetry, the radial
component~$\sim \gamma^r$ of~(\ref{3X}) drops out.
Using the symmetry under time reflections, we can
write the time integral of the remaining component~$\sim \gamma^0$ as
\begin{eqnarray*}
e^{-i \Omega r}
\int_{-\infty}^\infty \epsilon(t)\; g'_\varepsilon(\xi^2)\: \frac{t}{r(t+r)}\:dt
&=& -e^{-i \Omega r} \int_0^\infty
\frac{g'_\varepsilon(\xi^2)}{\xi^2} \:2 t \,dt
\;=\; -e^{-i \Omega r} \int_{-r^2}^\infty
\frac{g'_\varepsilon(z)}{z} \:dz\:,
\end{eqnarray*}
where in the last step we introduced the ``Lorentz invariant''
integration variable~$z(t)=t^2-r^2$.
Since~$g'_\varepsilon$ vanishes for negative~$z$, it is obvious that
the last integral is equal to a constant~$c$, independent of~$r$
and the angular variables.
Similarly, for the contribution~$\sim \gamma^0$ involving the
phase factor~$e^{i \Omega r}$, the time integral is computed to be
\[ -e^{i \Omega r} \int_{-\infty}^\infty \epsilon(t)\; g'_\varepsilon(\xi^2)\: \frac{t}{r(t-r)}\:dt \;=\; -e^{i \Omega r} \int_{-r^2}^\infty \, \frac{g'_\varepsilon(z)}{z} \:dz \:, \]
and the last integral again equals~$c$. Adding the obtained terms
involving the phase factors~$e^{-i \Omega r}$ and~$e^{i \Omega r}$,
we can carry out the radial integral to get
\beq \label{svanish}
-c \int_0^\infty r^2 \left( e^{-i \Omega r} + e^{i \Omega r} \right)
dr \;=\; -c \int_{-\infty}^\infty r^2\, e^{-i \Omega r}\, dr \;=\;
c \:2 \pi \delta''(\Omega) \;=\; 0\:,
\eeq
where in the last step we used that~$\Omega \neq 0$. This explains
why~$B^\varepsilon_{\mbox{\tiny{scal}}}$ vanishes.
For~$B^\varepsilon_{\mbox{\tiny{vect}}}$, we likewise collect
the terms involving the phase factor~$e^{-i \Omega r}$
in~(\ref{Bv}). Then we can carry out the time integral as follows,
\begin{eqnarray}
\lefteqn{ \int_{-\infty}^\infty i \xi^2\, g'_\varepsilon(\xi^2)\, \epsilon(t)
\frac{1}{r} \frac{\partial}{\partial r} \left(
\frac{e^{-i \Omega r}}{r\, (t+r)} \right) dt } \nonumber \\
&=& e^{-i \Omega r}
\int_{-\infty}^\infty i \xi^2\, g'_\varepsilon(\xi^2)\: \epsilon(t)
\left[ -\frac{i \Omega r + 1}{r^3 (t+r)}
- \frac{1}{r^2 (t+r)^2} \right] dt \nonumber \\
&=& e^{-i \Omega r}
\int_{0}^\infty i \xi^2\, g'_\varepsilon(\xi^2)
\left[ -\frac{(i \Omega r + 1)\, 2t}{r^3 \, \xi^2}
+ \frac{4t}{r \, \xi^4} \right] dt \nonumber \\
&=& e^{-i \Omega r} \,\frac{\Omega r - i}{r^3} \int_{-r^2}^\infty
g'_\varepsilon(z)\, dz \:+\: e^{-i \Omega r}\, \frac{2 i}{r}
\int_{-r^2}^\infty
\frac{g'_\varepsilon(z)}{z}\, dz \:. \label{2ints}
\end{eqnarray}
The first integral in the last line can be evaluated and is seen
to vanish (note that, in view of the representation~(\ref{5}, \ref{6}),
we know that~$g(\infty)=0$). The second integral is again equal to a
constant~$c$, independent of~$r$ and the angular variables.
As a consequence, it vanishes similar
to~(\ref{svanish}) after adding the corresponding contribution
involving the phase factor~$e^{i \Omega r}$ and carrying out the
radial integral
\[ 2 i c \int_0^\infty r^2 \left( \frac{e^{-i \Omega r}}{r} -
\frac{e^{i \Omega r}}{r} \right) dr \;=\;
2 i c \int_{-\infty}^\infty r\, e^{-i \Omega r}\, dr \;=\;
-2 c\, 2 \pi \delta'(\Omega) \;=\; 0\:. \]
To summarize, if~$\M[A^\varepsilon_{xy}]$ is Lorentz invariant, the
time integrals can be written in an invariant form using the
measure~$dz = 2 t \, dt$. Then the radial integral becomes the
Fourier transform of a polynomial, giving a distribution
supported at~$\Omega=0$.

If conversely~$\M[A^\varepsilon_{xy}]$ is {\em{not}} Lorentz invariant,
it is impossible to rewrite the time integrals in the above invariant
form. As a consequence, they will in general depend on~$r$ in a complicated
way, and the $r$-integral will no longer be supported at~$\Omega=0$.
One might even conjecture that the vanishing of~(\ref{Bs}) and~(\ref{Bv})
implies the Lorentz invariance of~$\M[A^\varepsilon_{xy}]$. However, this
conjecture is false, as one sees easily in examples such as
\[ \M[A^\varepsilon_{xy}] \;=\; \Pdd \left( r^{2n}\, g_\varepsilon(t^2-r^2) \right)
\epsilon(t)\:, \]
which obviously violate Lorentz invariance, but where the resulting time
integrals are nevertheless polynomials in~$r$.

Let us work out systematically under which conditions the integrals~(\ref{Bs})
and~(\ref{Bv}) vanish. In order to avoid technical complications, we shall
assume that~$\M[A^\varepsilon_{xy}]$ is {\em{strictly causal}} in the
sense that it is supported inside the interior light cone.
Our analysis is simplified considerably by our assumption
that~$\M[A^\varepsilon_{xy}]$ converges as a distribution to~$\tM$. Namely,
whenever we have an expression of the form~$\M[A^\varepsilon_{xy}]$ times a
smooth function, this expression converges as~$\varepsilon \searrow 0$ in the distributional
sense, and we could easily compute its limit. Therefore, we may disregard such
so-called {\em{distributional expressions}}. 
In order to explain how to compute modulo distributional
expressions, let us simplify the expression~(\ref{Bs}) for~$B^\varepsilon_{\mbox{\tiny{scal}}}$.
Using the convention
\beq \label{Mconv}
\M(t,-r,\vartheta, \varphi) \;=\; \M(t,r,\vartheta+\pi, \varphi)\:,
\eeq
we may extend the function~$\M[A^\varepsilon_{xy}]$ to negative values of~$r$
(thus for fixed~$\vartheta$ and~$\varphi$, the curve~$r \in \R$
is a straight line through the origin). Then we can rewrite the radial
integral as
\beq \label{rfactor}
\int_0^\infty r^2\, dr \left( \frac{e^{-i \Omega r}}{r\, (t+r)} -
\frac{e^{i \Omega r}}{r\, (t-r)} \right) \M[A^\varepsilon_{xy}]
\;=\; \int_{-\infty}^\infty e^{-i \Omega r}\:
\frac{\M[A^\varepsilon_{xy}]}{t+r}\;r\, dr \:.
\eeq
Now our task is to analyze the time integral of~$\M[A^\varepsilon_{xy}]/(t+r)$.
Since~$\M[A^\varepsilon_{xy}]$ is strictly causal, we only need to consider
the two regions~$t>|r|$ and~$t<-|r|$. We can even restrict attention to the
region where the factor~$(t+r)^{-1}$ has a pole,
\[ \int_{-\infty}^\infty \frac{\M[A^\varepsilon_{xy}]}{t+r}\, dt
\;=\; \mp \int_0^{\mp \infty} \frac{\M[A^\varepsilon_{xy}]}{t+r}\, dt
\:+\: {\mbox{(distributional)}}\:, \]
where we distinguished the two cases~$r>0$ and~$r<0$, respectively.
Furthermore, we may add distributional terms; in particular,
\begin{eqnarray*}
\int_0^{\mp \infty} \frac{\M[A^\varepsilon_{xy}]}{t+r}\, dt
&=& \int_0^{\mp \infty} \M[A^\varepsilon_{xy}]
\left( \frac{1}{t+r} - \frac{1}{t-r} \right) dt
\:+\: {\mbox{(distributional)}} \\
&=& -2r \int_0^{\mp \infty} \frac{\M[A^\varepsilon_{xy}]}{t^2-r^2}
\: dt \:+\: {\mbox{(distributional)}} \:.
\end{eqnarray*}
In this way, we have rewritten~(\ref{Bs}) as
\begin{eqnarray}
16 \pi^3 B^\varepsilon_{\mbox{\tiny{scal}}} &=& 2
\int_0^{2 \pi} d\varphi \int_{-1}^1 d\cos \vartheta
\int_{-\infty}^\infty dr\, e^{-i \Omega r} \, r^2\,
\epsilon(r)
\int_0^{-\epsilon(r)\, \infty} \frac{\M[A^\varepsilon_{xy}]}{t^2-r^2}
\: dt \nonumber \\
&&+ {\mbox{(distributional)}}\:. \label{Bsr}
\end{eqnarray}
In the omitted distributional terms, we may replace~$\M[A^\varepsilon_{xy}]$
by a Lorentz invariant regularization of the form~(\ref{3X}).
As explained above, for such a Lorentz invariant regularization,
the time integral in~(\ref{Bsr}) is independent of~$r$, and thus
the radial integral gives zero (note that the factor~$\xi\slsh$
in~(\ref{3X}) contains a factor~$t$ which can be combined with
the $dt$ in~(\ref{Bsr}) to the Lorentz invariant measure~$dz$).
Hence for the Lorentz invariant distribution, the integrals
in~(\ref{Bsr}) and the distributional terms both tend to zero
as~$\varepsilon \searrow 0$. Again using that the
limit~$\varepsilon \searrow 0$ of the distributional terms is independent
of the regularization, we conclude that~$B^\varepsilon_{\mbox{\tiny{scal}}}$
vanishes as~$\varepsilon \searrow 0$
if and only if the integrals in~(\ref{Bsr}) tend to zero in this limit.
This leads us to impose that the time integral in~(\ref{Bsr}) should be
a polynomial in~$r$.

\begin{Def} \label{defsmcc}
Suppose that~$(P^\varepsilon)_{\varepsilon>0}$ is a family
of regularized fermionic projectors, such that
$P^\varepsilon \rightarrow P$ and~$\M[A^\varepsilon_{xy}] \rightarrow
\tilde{\M}$ as distributions.
The functions~$\M[A^\varepsilon_{xy}]$ satisfy the
{\bf{scalar momentum cone conditions}} if there is a finite number of
constants~$(c_i)_{i=0,1,\ldots}$ (which may depend on the angular variables,
but are independent of~$r$) such that for every~$r>0$,
\beq \label{smcc}
\int_0^{\pm \infty} \frac{\M[A^\varepsilon_{xy}](\pm r)}{t^2-r^2}
\, dt
\:\pm\: \frac{c_0(\varepsilon)}{r^2} + \frac{c_1(\varepsilon)}{r} \pm
c_2(\varepsilon) + c_3(\varepsilon)\, r + \cdots
\;\stackrel{\varepsilon \searrow 0}{\longrightarrow}\; 0\:,
\eeq
where~$\M$ is defined for negative~$r$ by~(\ref{Mconv}).
\end{Def}

The function~$B^\varepsilon_{\mbox{\tiny{vect}}}$, (\ref{Bv}), can be
handled in a similar way, where it is most convenient to integrate
the additional radial derivative by parts.
We first rewrite the radial integral as follows,
\begin{eqnarray}
\lefteqn{ \int_0^\infty r^2 dr (\M[A^\varepsilon_{xy}]\, i \xi\slsh)(r)
\:\frac{1}{r} \frac{\partial}{\partial r}
\left(\frac{e^{-i \Omega r}}{r\, (t+r)} -
\frac{e^{i \Omega r}}{r\, (t-r)} \right) } \nonumber \\
&=& \int_{-\infty} ^\infty
r\, dr \:(\M[A^\varepsilon_{xy}]\, i \xi\slsh)(r)
\:\frac{\partial}{\partial r} \left(
\frac{e^{-i \Omega r}}{r\, (t+r)} \right) \nonumber \\
&=& -\int_{-\infty} ^\infty
\frac{\partial}{\partial r} \Big( r\, (\M[A^\varepsilon_{xy}]\, i \xi\slsh)(r) \Big)
\: \frac{e^{-i \Omega r}}{t+r}\:\frac{dr}{r}\:. \label{Bvr}
\end{eqnarray}
Computing modulo distributional expressions, the resulting time integral can
be written in the two cases~$r>0$ and~$r<0$ as
\begin{eqnarray*}
\lefteqn{ \int_{-\infty}^\infty
\frac{\partial}{\partial r} \Big( r\, (\M[A^\varepsilon_{xy}]\, i \xi\slsh)(r) \Big)
\: \frac{1}{t+r}\: dt } \\
&=& \mp \int_0^{\mp \infty} \frac{\partial}{\partial r} \Big( r\, (\M[A^\varepsilon_{xy}]\, i \xi\slsh)(r) \Big)
\: \frac{1}{t+r}\: dt \:+\: {\mbox{(distributional)}} \\
&=& \mp 2 \int_0^{\mp \infty} \frac{\partial}{\partial r} \Big( r\, (\M[A^\varepsilon_{xy}]\, i \xi\slsh)(r) \Big)
\: \frac{1}{t^2-r^2}\:t\,dt \:+\: {\mbox{(distributional)}} .
\end{eqnarray*}
A short calculation shows that for the Lorentz invariant regularization~(\ref{3X}), the
last integral is a polynomial in~$r$ of degree two, and thus the radial integral vanishes.
We thus end up with the following condition.
\begin{Def} \label{defvmcc}
Suppose that~$(P^\varepsilon_{xy})_{\varepsilon>0}$ is a family
of regularized fermionic projectors, such that
$P^\varepsilon \rightarrow P$ and~$\M[A^\varepsilon_{xy}] \rightarrow
\tilde{\M}$ as distributions.
The functions~$\M[A^\varepsilon_{xy}]$ satisfy the
{\bf{vector momentum cone conditions}} if there is a finite number of
constants~$(c_i)_{i=0,1,\ldots}$ (which may depend on the angular variables,
but are independent of~$r$) such that for every~$r>0$,
\beq \label{vmcc}
\int_0^{\pm \infty} \frac{\partial}{\partial r} \Big( r\,
\M[A^\varepsilon_{xy}](\pm r) \, i \xi\slsh \Big)
\: \frac{t\, dt}{t^2-r^2}
\:+\: c_0(\varepsilon)\,r \pm c_1(\varepsilon)\, r^2 +
c_2(\varepsilon)\, r^3 + \cdots
\;\stackrel{\varepsilon \searrow 0}{\longrightarrow}\; 0\:.
\eeq
\end{Def}

We close this section with a few remarks. For clarity, we mention
that, since~$P(x,y)$ is the sum of a scalar and vector component,
we do not need to satisfy the conditions of Definitions~\ref{defsmcc}
and~\ref{defvmcc} separately; it suffices to satisfy a certain
linear combination of these conditions.

\begin{figure}[tb]
\begin{center}
\begin{picture}(0,0)%
\includegraphics{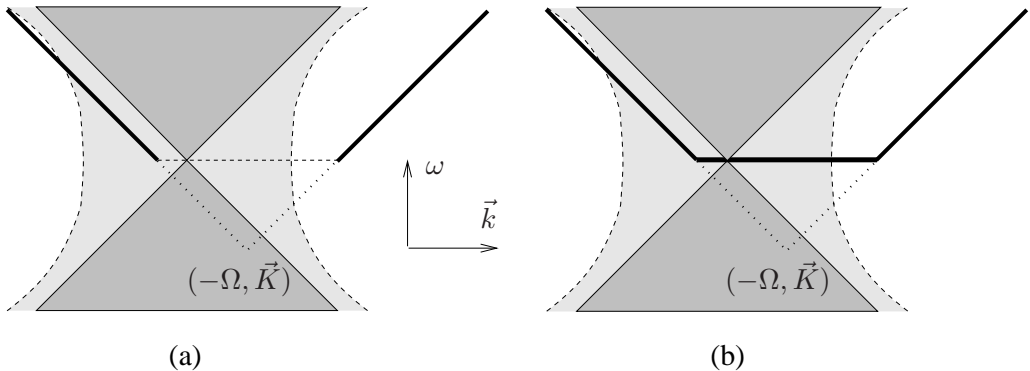}%
\end{picture}%
\setlength{\unitlength}{1657sp}%
\begingroup\makeatletter\ifx\SetFigFont\undefined%
\gdef\SetFigFont#1#2#3#4#5{%
  \reset@font\fontsize{#1}{#2pt}%
  \fontfamily{#3}\fontseries{#4}\fontshape{#5}%
  \selectfont}%
\fi\endgroup%
\begin{picture}(15355,5507)(1310,-8657)
\put(8447,-6474){\makebox(0,0)[lb]{\smash{{\SetFigFont{11}{13.2}{\familydefault}{\mddefault}{\updefault}$\vec{k}$}}}}
\put(7628,-5664){\makebox(0,0)[lb]{\smash{{\SetFigFont{11}{13.2}{\familydefault}{\mddefault}{\updefault}$\omega$}}}}
\put(4051,-7396){\makebox(0,0)[lb]{\smash{{\SetFigFont{11}{13.2}{\familydefault}{\mddefault}{\updefault}{$(-\Omega,\vec{K})$}%
}}}}
\put(12106,-7396){\makebox(0,0)[lb]{\smash{{\SetFigFont{11}{13.2}{\familydefault}{\mddefault}{\updefault}{$(-\Omega,\vec{K})$}%
}}}}
\end{picture}%
\caption{The convolution integral~${\hat{\mathcal{M}}}^\varepsilon
* \hat{H}$ in a general reference frame.}
\label{figcon2}
\end{center}
\end{figure}

In the previous analysis we worked in a particular reference frame, where
the $t$-coordinate pointed in the direction of the vector~$q$.
Nevertheless, the momentum cone conditions are the same in
any other reference frame, as the following consideration shows.
In a general reference frame, the vector~$q$ lies in the interior of the
lower mass cone and can thus be written as~$q=(-\Omega, \vec{K})$
with~$|\vec{K}|<\Omega$. On a compact set, 
$\hat{H}_{\mbox{\tiny{scal}}}$ and~$\hat{H}_{\mbox{\tiny{vect}}}$ are
again arbitrary. It is most convenient to choose
\begin{eqnarray*}
\hat{H}_{\mbox{\tiny{scal}}}(k) &=& \delta \Big(
(k^0)^2 - (\vec{k} - \vec{K})^2 \Big)
\: \Theta(-k^0-\Omega) \\
\hat{H}_{\mbox{\tiny{vect}}}(k) &=&
\frac{1}{2} \:\Pdd_k \left( \Theta \Big(
(k^0)^2 - (\vec{k} - \vec{K})^2 \Big)
\: \Theta(-k^0 - \Omega) \right) ,
\end{eqnarray*}
because these distributions differ from~(\ref{H1}, \ref{H2}) simply
by a shift in the spatial direction~$\vec{K}$.
Then the corresponding convolution integrals~(\ref{ci4}) are
obtained from the formulas of Lemma~\ref{lemma4n1} simply
by inserting the phase factor~$e^{-i \vec{K} \vec{\xi}}$ into the integrands
in~(\ref{Bs}, \ref{Bv}) (see also Figure~\ref{figcon2}).
As the additional phase~$\vec{K} \vec{\xi}$ is time independent and
linear in~$r$, it has no effect on the subsequent arguments.
In particular, the formulas~(\ref{Bsr}, \ref{Bvr}) remain valid if
we only replace~$\Omega$ by~$\tilde{\Omega} = \Omega + |\vec{K}|
\cos \alpha > 0$, where~$\alpha$ is the angle between~$\vec{K}$ and~$\vec{\xi}$.
Since~$\alpha$ is fixed in our argument, we again obtain the conditions
of Definitions~\ref{defsmcc} and~\ref{defvmcc}.

We finally recall that the momentum cone conditions were derived
under the assumption that~$P$ is not regularized. Furthermore,
we made the simplification of working with a cone instead of mass hyperbolas
(compare Figures~\ref{fig1} and~\ref{figcon2}).
These assumptions and simplifications still need to be justified.
Indeed, we will have to slightly modify the momentum cone conditions in
order to take these additional effects into account.
We postpone these more technical aspects to Sections~\ref{sec6z}
and~\ref{sec6}.

\section{The Spherically Symmetric Regularization} \label{sec3}
\setcounter{equation}{0}
For the regularized fermionic projector of the vacuum we take the general
spherically symmetric ansatz
\begin{eqnarray}
P^\varepsilon(x,y) &=& \int \frac{d^4k}{(2 \pi)^4}\; \sum_{\beta=1}^3\:\Theta(-\omega)\:
\delta \!\left(|\vec{k}| - K_\beta(\omega) \right)\:
\frac{1}{2 |\vec{k}|} \nonumber \\
&&\times 16 \pi^3 \left( k\slsh \:g_\beta(\omega)
+ \gamma^0\: f_\beta(\omega) + h_\beta(\omega) \right) e^{-ik(x-y)} \:, \label{Pansatz}
\end{eqnarray}
where~$\omega \equiv k^0$, and~$K_\beta$, $g_\beta$, $f_\beta$ and $h_\beta$
are positive functions. The fermionic
projector without regularization~(\ref{A}) is recovered by setting
\begin{eqnarray}
K_\beta(\omega) &=& \sqrt{\omega^2-m_\beta^2} \label{mex} \\
g_\beta(\omega) &=& \frac{1}{16 \pi^3} \:,\quad
f_\beta(\omega) \;=\; 0 \:,\quad
h_\beta(\omega) \;=\; \frac{m_\beta}{16 \pi^3} \:. \label{mex2}
\end{eqnarray}
For the regularization, we will always choose functions~$g_\beta$, $f_\beta$
and~$h_\beta$ which decay in~$\omega$ on the scale~$\varepsilon^{-1}$. Clearly,
the ansatz~(\ref{Pansatz}) is a special case of~(\ref{B}, \ref{C}); in particular, the
fermionic projector is homogeneous and has a vector-scalar structure. Furthermore,
$P^\varepsilon(k)$ is a sum of distributions supported on the hypersurfaces
$|\vec{k}| = K_\beta(\omega)$. This property is called that the
fermionic projector is composed of {\em{surface states}}. Another property of
potential interest is that the surface states should be {\em{half occupied}}
(see~\cite[\S4.4]{PFP}); in our setting this condition is equivalent to the
following equation,
\beq \label{hoss}
\left( \omega\: g_\beta + f_\beta \right)^2 - (K_\beta\: g_\beta)^2 \;=\;
h_\beta^2\:.
\eeq
The condition of half occupied surface states will be discussed in
Remark~\ref{remhalfnorm}.

Let us bring the ansatz~(\ref{Pansatz}) into a more convenient form.
First, we can use the spherical symmetry as well as the~$\delta$-distribution
$\delta(|\vec{k}|-K_\beta(\omega))$ to reduce to one-dimensional Fourier transforms.
\begin{Lemma} \label{lemma1}
The fermionic projector~(\ref{Pansatz}) has the representation
\begin{eqnarray*}
P^\varepsilon(x,y) &=& \sum_{\beta=1}^3 \frac{i}{r} \int_{\Xi_\beta} d\omega\: h_\beta(\omega) \:e^{i \omega t}
\left( e^{-i K_\beta(\omega) \:r} - e^{i K_\beta(\omega) \:r} \right) \\
&&+ \gamma^0 \sum_{\beta=1}^3 \frac{i}{r} \int_{\Xi_\beta} d\omega\: (f_\beta(\omega)+\omega \:g_\beta(\omega))\:
e^{i \omega t} \left( e^{-i K_\beta(\omega) \:r}
- e^{i K_\beta(\omega) \:r} \right) \\
&&- \gamma^r \sum_{\beta=1}^3 \frac{1}{r^2} \int_{\Xi_\beta} d\omega\: g_\beta(\omega) \:e^{i \omega t} \left( e^{-i K_\beta(\omega) \:r}
- e^{i K_\beta(\omega) \:r} \right) \\
&&- \gamma^r \sum_{\beta=1}^3 \frac{i}{r} \int_{\Xi_\beta} d\omega\: K_\beta(\omega)\: g_\beta(\omega) \:e^{i \omega t} \left( e^{-i K_\beta(\omega) \:r}
+ e^{i K_\beta(\omega) \:r} \right) \:,
\end{eqnarray*}
where~$\Xi_\beta$ are the sets
\[ \Xi_\beta \;:=\; \{ \omega<0 {\mbox{ with }} K_\beta(\omega)>0 \}\:. \]
\end{Lemma}
{\Proof} For notational convenience, we consider one Dirac sea and omit
the index~$\beta$. Choosing polar coordinates
\[ k \;=\; (\omega, p \cos \vartheta, p \sin \vartheta\:\cos \varphi,
p \sin \vartheta\:\sin \varphi) \:,\spc
y-x \;=\; (t,r, 0, 0) \]
(thus $\vartheta$ is the angle between the vectors $\vec{k}$ and
$\vec{y}-\vec{x}$), we obtain
\begin{eqnarray*}
\lefteqn{ \int \frac{d^4k}{(2 \pi)^4}\: \Theta(-\omega)\:
\delta \!\left(|\vec{k}| - K(\omega) \right)\:
\frac{1}{2 |\vec{k}|}\: 16 \pi^3\: h(\omega) \: e^{-ik(x-y)} } \\
&=& \int_{-\infty}^0 d\omega\: h(\omega)
\:e^{i \omega t}
\int_0^\infty p\: dp \:\delta(p - K(\omega)) \int_{-1}^1
d\cos \vartheta \: e^{-i p r \cos \vartheta} \\
&=& \frac{i}{r} \int_{-\infty}^0 d\omega\: h(\omega)
\:e^{i \omega t}\:
\int_0^\infty dp \:\delta(p - K(\omega))
\left( e^{-i p r} - e^{i p r} \right) \\
&=& \frac{i}{r} \int_\Xi d\omega\: h(\omega) \:e^{i \omega t} \left( e^{-i K(\omega) \:r} - e^{i K(\omega) \:r} \right) \:.
\end{eqnarray*}
Similarly,
\begin{eqnarray*}
\lefteqn{ \int \frac{d^4k}{(2 \pi)^4}\: \Theta(-\omega)\:
\delta \!\left(|\vec{k}| - K(\omega) \right)\:
\frac{1}{2 |\vec{k}|}\: 16 \pi^3\: \gamma^0\:f(\omega)\: e^{-ik(x-y)} } \\
&=& \gamma^0\: \frac{i}{r} \int_\Xi d\omega\: f(\omega) \:e^{i \omega t} \left( e^{-i K(\omega) \:r}
- e^{i K(\omega) \:r} \right)
\end{eqnarray*}
and
\begin{eqnarray*}
\lefteqn{ \int \frac{d^4k}{(2 \pi)^4}\: \Theta(-\omega)\:
\delta \!\left(|\vec{k}| - K(\omega) \right)\:
\frac{1}{2 |\vec{k}|}\: 16 \pi^3\: k\slsh\: g(\omega) \: e^{-ik(x-y)} } \\
&=& \left(-i \gamma^0 \partial_t - i \gamma^r \partial_r \right)
\int \frac{d^4k}{(2 \pi)^4}\: \Theta(-\omega)\:
\delta \!\left(|\vec{k}| - K(\omega) \right)\:
\frac{1}{2 |\vec{k}|}\: 16 \pi^3\: g(\omega) \: e^{-ik(x-y)} \\
&=& \left(-i \gamma^0 \partial_t - i \gamma^r \partial_r \right)
\frac{i}{r} \int_\Xi d\omega\: g(\omega) \:e^{i \omega t} \left( e^{-i K(\omega) \:r} - e^{i K(\omega) \:r} \right) \\
&=& \gamma^0 \:\frac{i}{r} \int_\Xi d\omega\: \omega\: g(\omega) \:e^{i \omega t} \left( e^{-i K(\omega) \:r} - e^{i K(\omega) \:r} \right) \\
&&-\gamma^r \:\frac{1}{r^2} \int_\Xi d\omega\: g(\omega) \:e^{i \omega t} \left( e^{-i K(\omega) \:r} - e^{i K(\omega) \:r}
\right) \\
&&-\gamma^r \:\frac{i}{r} \int_\Xi d\omega\: K(\omega)\: g(\omega) \:e^{i \omega t} \left( e^{-i K(\omega) \:r} + e^{i K(\omega) \:r}
\right) \:.
\end{eqnarray*}
Adding these terms gives the result.
\QED
In order to further simplify the representation for~$P^\varepsilon(x,y)$, we observe that without
a regularization, the expansion of~(\ref{mex})
\[ \sqrt{\omega^2 - m_\beta^2} \;=\; |\omega| \left( 1 + {\mathcal{O}}
\bigg(
\frac{m_\beta^2}{\omega^2} \bigg) \right) \]
shows that the functions~$K_\beta(\omega)$ and~$|\omega|$ come asymptotically close
for large~$|\omega|$ and coincide in the massless case~$m_\beta=0$. Since the
regularization effects come into play only for large energies, we can
expect that even if a regularization is present, the function
\beq \label{alphadef}
\alpha_\beta \;:=\; |\omega| - K_\beta(\omega)
\eeq
should be small for large~$|\omega|$. Therefore, it is useful to expand in
powers of~$\alpha$ as follows,
\beq \label{massex}
e^{i K_\alpha(\omega) r} \;=\; e^{-i \omega r}\: e^{-i \alpha_\beta(\omega) r}
\;=\; e^{-i \omega r} \sum_{k=0}^\infty \frac{(-i \alpha_\beta(\omega)\,
 r)^k}{k!}
\eeq
(note that~$\omega$ is always negative). Motivated by the fact that~$\alpha_\beta$
vanishes in the unregularized massless case, we refer to~(\ref{massex}) as the
{\em{mass expansion}}. It is unpleasant that Lemma~\ref{lemma1} involves factors
$e^{+i K_\beta r}$ and $e^{-i K_\beta r}$, making the resulting formulas
rather messy.
It is therefore convenient to introduce the following short notation.
Taking the complex conjugate of the representation of Lemma~\ref{lemma1}, one sees
that the fermionic projector is invariant under the transformations
\[ P^\varepsilon(x,y) \;\to\; \overline{P^\varepsilon(x,y)} \:,\qquad
t \;\to\; -t \:,\qquad \gamma^r \;\to\; -\gamma^r\:, \]
which we call a {\em{PCT-transformation}} (in analogy to the usual parity, charge
and time transformation; actually, in our context this invariance is nothing
but the Hermiticity condition $P(x,y)^* = P(y,x)$ expressed in polar coordinates).
A PCT-transformation converts the terms
involving the factors~$e^{-i K_\beta r}$ and~$e^{i K_\beta r}$ into
each other. Therefore, we can in what follows consider only
the terms involving~$e^{i K_\beta t}$ and refer to the terms
involving~$e^{-i K_\beta t}$ by ``PCT''.
\begin{Lemma} \label{lemma2} {\bf{(mass expansion)}}
The fermionic projector~(\ref{Pansatz}) has the representation
\begin{eqnarray*}
P^\varepsilon(x,y) &=& \sum_{\beta=1}^3\: \frac{1}{r} \sum_{k=0}^\infty \frac{(-i)^{k+1}\: r^k}{k!} \int_{\Xi_\beta}
d\omega\: h_\beta \:\alpha_\beta^k \;e^{i \omega s} \\
&&+ \gamma^0 \:\sum_{\beta=1}^3\: \frac{1}{r} \sum_{k=0}^\infty \frac{(-i)^{k+1}\: r^k}{k!}
\int_{\Xi_\beta} d\omega \left(f + \omega\: g \right)
\:\alpha_\beta^k \;e^{i \omega s} \\
&&-\gamma^r \:\sum_{\beta=1}^3\: \frac{1}{r} \sum_{k=0}^\infty \frac{(-i)^{k+1}\: r^k}{k!}
\int_{\Xi_\beta} d\omega\: \left(\omega + i \:\frac{k-1}{r} \right)
\: g\:\alpha_\beta^k \;e^{i \omega s} \:+\: {\mbox{(PCT)}}\:.
\end{eqnarray*}
\end{Lemma}
{\Proof} From Lemma~\ref{lemma1} we get
\begin{eqnarray*}
P^\varepsilon(x,y) &=&-\sum_{\beta=1}^3 \frac{i}{r} \int_{\Xi_\beta} d\omega\: h_\beta \:e^{i \omega s} \:e^{-i \alpha_\beta r} \\
&&- \gamma^0 \sum_{\beta=1}^3 \frac{i}{r} \int_{\Xi_\beta} d\omega\: (f_\beta+\omega \:g_\beta)\:e^{i \omega s} \:e^{-i \alpha_\beta r} \\
&&+\gamma^r \sum_{\beta=1}^3 \frac{1}{r^2} \int_{\Xi_\beta} d\omega\: g_\beta \:e^{i \omega s} \: e^{-i \alpha_\beta r} \\
&&-\gamma^r \sum_{\beta=1}^3 \frac{i}{r} \int_{\Xi_\beta} d\omega\: (-\omega-\alpha_\beta)\: g_\beta \:e^{i \omega s} \: e^{-i \alpha_\beta r}
\:+\: {\mbox{(PCT)}}
\end{eqnarray*}
and expand in powers of~$\alpha_\beta$. \QED
Before going on, we briefly describe how our spherically symmetric regularization is related to the general formulas for the regularized fermionic projector
as derived in~\cite[Chapter~4]{PFP}. In order to get a connection to the
general form of the vector component as considered in~\cite[{\S}4.4]{PFP}, we consider the fermionic projector near the future light cone~$t=r$. Then the ``PCT''-terms involve the
oscillatory phase factors~$e^{i\omega(t+r)}$, and the corresponding
contributions to the fermionic projector are smooth. Such terms were
not considered in~\cite[{\S}4]{PFP}, and therefore we also leave them
out here. Setting $s=t-r$ and
\[ \gamma^s \;=\; \frac{1}{2} \left( \gamma^0 - \gamma^r \right) ,\spc
\gamma^l \;=\; \frac{1}{2} \left( \gamma^0 + \gamma^r \right) , \]
the fermionic projector takes the form
\begin{eqnarray}
\lefteqn{ P(x,y) \;=\; {\mbox{(smooth contributions)}} } \nonumber \\
&&+\sum_{\beta=1}^3\: \frac{1}{ir} \sum_{k=0}^\infty
\frac{(-i r)^k}{k!} \int_{\Xi_\beta}
d\omega\: h_\beta \:\alpha_\beta^k \;e^{i \omega s} \label{a1} \\
&&-\gamma^s \:\sum_{\beta=1}^3\: \frac{1}{i r} \sum_{k=0}^\infty
\frac{(-i r)^k}{k!} \int_{\Xi_\beta} d\omega \left(-2 \omega\:g
- f -\frac{k}{k+1}\:g\:\alpha_\beta \right)
\alpha_\beta^k \;e^{i \omega s} \label{a2} \\
&&+\gamma^s \:\sum_{\beta=1}^3\: \frac{1}{(i r)^2}
\int_{\Xi_\beta} d\omega\:g  \;e^{i \omega s} \label{a3} \\
&&+\gamma^l \:\sum_{\beta=1}^3\: \frac{1}{(ir)^2} \sum_{k=0}^\infty
\frac{(-i r)^k}{k!}
\int_{\Xi_\beta} d\omega \left( (k-1) \:g \:\alpha_\beta^k
-k \: f\: \alpha_\beta^{k-1} \right)
 \;e^{i \omega s}\:. \label{a4}
\end{eqnarray}
The expansion of the scalar component~(\ref{a1}) is very similar
to that in~\cite[{\S}4.3]{PFP}. The only difference is that instead
of~$l$ we are
working here with the ``large'' coordinate~$r$,
which is more convenient in the spherically symmetric situation.
Setting~$r=l+s$ and expanding in powers of~$l$, one sees that~(\ref{a1})
amounts to a special form of the regularization expansion.
For the vector component one should notice that the
terms~$- f -\frac{k}{k+1}\:g\:\alpha$ in~(\ref{a2}) are of higher
order in~$\alpha_{\mbox{\scriptsize{max}}}/E_P$ and were thus left out in~\cite[{\S}4.3]{PFP}. The term~(\ref{a3}) can be identified with
the summand~$n=1$ of the regularization expansion. Comparing~(\ref{a4})
with the formulas in~\cite[{\S}4.3]{PFP}, one sees that the function~$f$
describes the {\em{shear}} of the surface states.
To summarize, the expansion~(\ref{a1}--\ref{a4}) is compatible with
the general expansion in~\cite{PFP}, but in our spherical situation
the regularization expansion has a special form.

We can now describe our general strategy for constructing admissible regularizations.
\label{discussion}
Without a regularization, the mass expansion corresponds in position space to an expansion
in powers of~$s$ (as one sees for example by expanding the Bessel functions
as in~(\ref{bessel1}, \ref{bessel2})). Likewise, in the representation of Lemma~\ref{lemma2}
with regularization, the terms for~$k=0$ will be much larger on the light cone than
the terms for~$k=1$, which will in turn be much larger than the terms for~$k=2$,
et cetera. It is a crucial observation that without a regularization and away from
the light cone~(\ref{3}, \ref{fex}), the {\em{leading term in the mass expansion is cubic}},
because the contributions~$\sim m^0$, $\sim m$ and $\sim m^2$ all drop
out in~(\ref{fdef}), as a consequence of the special form of the Fourier transform of
the Dirac sea configuration. Once we put in a regularization, these lower order
terms in the mass
expansion will in general no longer drop out, and this leads to the surprising effect that in
$\M[A^\varepsilon_{xy}]$ the regularization terms will be typically much larger on the light
cone than what is needed for compensating the singularity of~(\ref{3}, \ref{fex}).
Without this effect, it would be very hard if not impossible to construct admissible
regularization. But exploiting this fact, we can proceed as follows. We choose the
regularization functions $g_\beta$, $f_\beta$ and $h_\beta$ such that their Fourier
transforms $g_\beta(s)$, $f_\beta(s)$ and $h_\beta(s)$ are supported in the
interval~$(-\varepsilon, \varepsilon)$, except for small {\em{regularization tails}}
which are more spread out and decay polynomially $\sim s^{-\gamma}$ or logarithmically
$\sim s^{-\gamma} \log^p s$. We arrange that the regularization terms drop out
in~$\M[A^\varepsilon_{xy}]$, except for the regularization tails. We then have
a lot of freedom to choose the exponents~$\gamma$ and the amplitudes of
the regularization tails, and this will indeed make it possible to compensate for
the singularity of~(\ref{3}, \ref{fex}).

Let us specify which class of regularizations we shall consider.
It is desirable that the Fourier transform of the regularization functions can be
computed explicitly, because we can then work with closed formulas for the regularized
fermionic projector. A function which has a particularly simple Fourier transform is the
exponential,
\[ \int_{-\infty}^0 e^{i \omega s}\: e^{\varepsilon \omega}\: d\omega \;=\;
\int_0^\infty e^{-i \omega s}\: e^{-\varepsilon \omega}\: d\omega \;=\;
-\frac{i}{s-i\varepsilon} \;=\; -\frac{i}{s} + \frac{\varepsilon}{s^2}
+ \frac{i \varepsilon^2}{s^3} + \cdots . \]
In the special case~$\varepsilon=0$ we simply get the Fourier transform of the
Heaviside function. Furthermore, the terms involving the regularization decay as desired on
the scale~$\varepsilon$. However, unfortunately, the leading regularization term
decays at a fixed rate~$\sim s^{-2}$. We would like to be more flexible; in particular, we
would like to arrange that the regularization term decays at the faster rate
${\mathcal{O}}(s^{-2-n})$ for some~$n>0$. The key for getting more general decay rates is
the observation that the term~$\varepsilon/s^2$ arises because the exponential
$e^{\varepsilon \omega}$ has a non-vanishing derivative at~$\omega=0$. This motivates
our general method of multiplying the function~$e^{\varepsilon \omega}$ by the truncated
Taylor series of~$e^{-\varepsilon \omega}$. For the resulting product, the first~$n$
derivatives vanish at~$\omega=0$. Such so-called {\em{polynomial smoothing}}
really give faster decay of the regularization terms, as the next lemma shows.
\begin{Prp} \label{prpreg1}
{\bf{(polynomially smoothed exponential regularization)}} \\
For all~$\omega \in \R$ and~$\varepsilon>0$,
\[ \int_0^\infty e^{-i \omega s}\: e^{-\varepsilon \omega} \left( \sum_{l=0}^n
\frac{(\varepsilon \omega)^l}{l!} \right) d\omega
\;=\; -\frac{i}{s} + \frac{i}{s} \left( \frac{\varepsilon}{\varepsilon+is} \right)^{n+1} . \]
\end{Prp}
{\Proof} Rewriting the polynomial in~$\omega$ with $\varepsilon$-derivatives, we can evaluate the integral,
\begin{eqnarray*}
\int_0^\infty e^{-i\omega s - \varepsilon \omega} \left(
\sum_{l=0}^n \frac{(\varepsilon \omega)^l}{l!} \right) d\omega
&=& \sum_{l=0}^n \frac{1}{l!}\left(-\varepsilon\: \frac{d}{d\varepsilon} \right)^l \int_0^\infty e^{-i \omega s - \varepsilon \omega}\: d\omega \\
\;=\; \sum_{l=0}^n \frac{1}{l!}\left(-\varepsilon\: \frac{d}{d\varepsilon} \right)^l \frac{1}{\varepsilon+is}
&=& \frac{1}{\varepsilon+is} \sum_{l=0}^n \left( \frac{\varepsilon}{\varepsilon+is}
\right)^l \:.\spc \spc
\end{eqnarray*}
Applying the standard formula
\[ 1 + a + \cdots + a^n \;=\; \frac{1-a^{n+1}}{1-a} \]
gives the result.
\QED

In the regularization tails, we want to have a power
behavior~$\sim |s|^{-\alpha}$ or a log-power behavior~$\sim \log(s)\, |s|^{-\alpha}$, where~$\alpha$ can be any positive real number.
The method is to work with noninteger powers
of~$\omega$ in the Fourier integral:
\begin{Lemma} \label{lemmareg2}
For all~$\omega \in \R$ and~$\alpha, \rho>0$,
\begin{eqnarray*}
\int_0^\infty e^{-i \omega s -\rho \omega} \: \omega^{\alpha-1}\:
d\omega &=& \Gamma(\alpha) \: \exp \Big(\!-\alpha
\log(\rho + i s) \Big) \\
\int_0^\infty e^{-i \omega s -\rho \omega} \: \omega^{\alpha-1}\:
\left(\log \omega - \frac{\Gamma'(\alpha)}{\Gamma(\alpha)}
\right) \: d\omega &=& -\log \left( \rho+is \right)
\Gamma(\alpha) \: \exp \Big(-\alpha
\log(\rho + i s) \Big) .
\end{eqnarray*}
\end{Lemma}
{\Proof} It is most convenient to present the integrals
as~$\lambda$-derivatives of a generating functional~$F(\lambda)$,
\[ \int_0^\infty e^{-i \omega s -\rho \omega} \:
\omega^{\alpha-1}\: \log^p \omega\: d\omega \;=\;
\frac{d^p}{d\lambda^p} F(\lambda) \Big|_{\lambda=0} \:, \]
which is computed as follows,
\begin{eqnarray*}
F(\lambda) &=& \int_0^\infty e^{-i \omega s - \rho \omega}
\: \omega^{\alpha-1}\: e^{\lambda \log \omega} d\omega
\;=\; \int_0^\infty e^{-i \omega s - \rho \omega}
\: \omega^{\lambda+\alpha-1}\: d\omega \\
&=& (\rho + is)^{-\lambda-\alpha} \int_0^\infty
u^{\lambda+\alpha-1}\:e^{-u}\:du \;=\;
\Gamma(\lambda+\alpha) \: (\rho + is)^{-\lambda-\alpha} \:.
\end{eqnarray*}
In the last line we introduced the new integration variable
$u=(\rho +is) \omega$ and rotated the contour around the origin
in the complex plane. \QED
Here the parameter~$\rho$ gives the length scale at which the
regularization tail comes into play. Namely, in the
regime~$|s| \gg \rho$, we can simplify
the above formulas using the expansions
\begin{eqnarray}
\log (\rho + is) &=& \left( \log|s|\;
+ \frac{i \pi}{2} \:\epsilon(s) \right)
\left(1 + {\mathcal{O}}(\rho/s) \right) \\
\exp \Big(\!-i \alpha \, \log(\rho + i s) \Big) &=&
|s|^{-\alpha}\: \exp \left(-\frac{i \pi}{2}\: \alpha\:
\epsilon(s) \right) \left(1 + {\mathcal{O}}(\rho/s) \right) \label{id2}
\end{eqnarray}
(where~$\epsilon$ is again the step function), to obtain the
desired power and log-power behavior in~$s$. However,
the Fourier integrals of Lemma~\ref{lemmareg2} have two disadvantages.
First, in~(\ref{id2}) we have an error term linear in~$\rho/s$;
it would be preferable that the error term is of
higher order~$(\rho/s)^n$. Second, the Fourier integrals
in Lemma~\ref{lemmareg2} are bounded in the limit~$s \rightarrow 0$
(by~$\rho^{-\alpha}$ and~$\rho^{-\alpha} \log \rho$, respectively),
but it would be useful that the Fourier integrals even decay
for small~$|s|$. The following more general regularization functions
remedy these disadvantages by a polynomial smoothing and
additional overall~$\omega$-derivatives.
\begin{Def} For given integers~$p, q \geq 0$ and
positive parameters~$\rho, \tilde{\rho}, \alpha$
with~$\rho < \tilde{\rho}$, we introduce the following functions:
\begin{eqnarray}
\hRf^{p,q}(\rho, \alpha, \omega) &=&
\frac{1}{\Gamma(\alpha+p)}
\left(\frac{d}{d\omega} \right)^{p} \left[ e^{-\rho \omega}
\sum_{l=0}^q \frac{(\rho \omega)^l}{l!}
\omega^{\alpha+p-1} \right] \label{Rfdef} \\
\hRflog^{p,q}(\rho, \alpha, \omega) &=&
-\frac{1}{\Gamma(\alpha+p)}
\left(\frac{d}{d\omega} \right)^p \!\left[ e^{-\rho \omega}
\sum_{l=0}^q \frac{(\rho \omega)^l}{l!} \omega^{\alpha+p-1}
\left(\log \omega - \frac{\Gamma'(\alpha)}{\Gamma(\alpha)} \right)
\right] \qquad \label{Rflogdef} \\
\hRf^{p,q}(\rho, \tilde{\rho}, \alpha, \omega)
&=& \hRf^{p,q}(\rho, \alpha, \omega) - \hRf^{p,q}(\tilde{\rho}, \alpha, \omega) \label{Rf2} \\
\hRflog^{p,q}(\rho, \tilde{\rho}, \alpha, \omega)
&=& \hRflog^{p,q}(\rho, \alpha, \omega)
- \hRflog^{p,q}(\tilde{\rho}, \alpha, \omega) \:. \label{Rf2log} 
\end{eqnarray}
\end{Def}

\begin{figure}[tb]
\begin{center}
\begin{picture}(0,0)%
\includegraphics{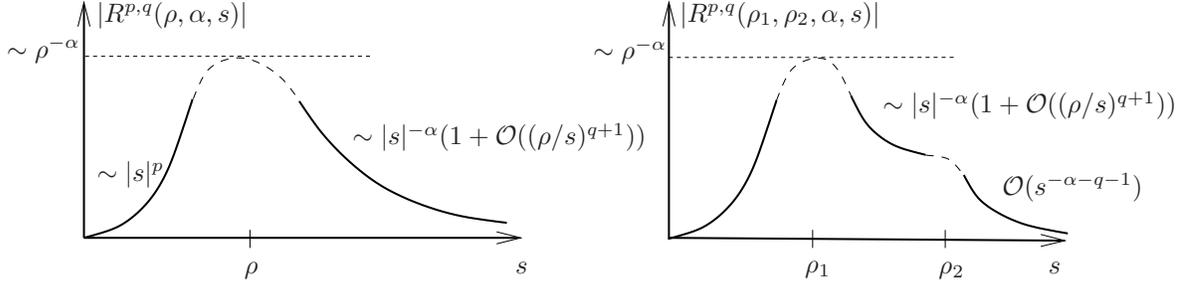}%
\end{picture}%
\setlength{\unitlength}{1657sp}%
\begingroup\makeatletter\ifx\SetFigFont\undefined%
\gdef\SetFigFont#1#2#3#4#5{%
  \reset@font\fontsize{#1}{#2pt}%
  \fontfamily{#3}\fontseries{#4}\fontshape{#5}%
  \selectfont}%
\fi\endgroup%
\begin{picture}(15916,4246)(31,-7690)
\put(3599,-7575){\makebox(0,0)[lb]{\smash{{\SetFigFont{10}{12.0}{\familydefault}{\mddefault}{\updefault}$\rho$}}}}
\put(15627,-7577){\makebox(0,0)[lb]{\smash{{\SetFigFont{10}{12.0}{\familydefault}{\mddefault}{\updefault}$s$}}}}
\put(13157,-5161){\makebox(0,0)[lb]{\smash{{\SetFigFont{10}{12.0}{\familydefault}{\mddefault}{\updefault}$\sim |s|^{-\alpha} (1+{\mathcal{O}}((\rho/s)^{q+1}))$}}}}
\put(11989,-7560){\makebox(0,0)[lb]{\smash{{\SetFigFont{10}{12.0}{\familydefault}{\mddefault}{\updefault}$\rho_1$}}}}
\put(14941,-6395){\makebox(0,0)[lb]{\smash{{\SetFigFont{10}{12.0}{\familydefault}{\mddefault}{\updefault}${\mathcal{O}}(s^{-\alpha-q-1})$}}}}
\put(8820,-4359){\makebox(0,0)[lb]{\smash{{\SetFigFont{10}{12.0}{\familydefault}{\mddefault}{\updefault}$\sim\rho^{-\alpha}$}}}}
\put(13988,-7576){\makebox(0,0)[lb]{\smash{{\SetFigFont{10}{12.0}{\familydefault}{\mddefault}{\updefault}$\rho_2$}}}}
\put(5221,-5641){\makebox(0,0)[lb]{\smash{{\SetFigFont{10}{12.0}{\familydefault}{\mddefault}{\updefault}$\sim |s|^{-\alpha} (1+{\mathcal{O}}((\rho/s)^{q+1}))$}}}}
\put(7651,-7576){\makebox(0,0)[lb]{\smash{{\SetFigFont{10}{12.0}{\familydefault}{\mddefault}{\updefault}$s$}}}}
\put(2521,-3796){\makebox(0,0)[b]{\smash{{\SetFigFont{10}{12.0}{\familydefault}{\mddefault}{\updefault}$|R^{p,q}(\rho,\alpha,s)|$}}}}
\put( 46,-4336){\makebox(0,0)[lb]{\smash{{\SetFigFont{10}{12.0}{\familydefault}{\mddefault}{\updefault}$\sim\rho^{-\alpha}$}}}}
\put(11611,-3796){\makebox(0,0)[b]{\smash{{\SetFigFont{10}{12.0}{\familydefault}{\mddefault}{\updefault}$|R^{p,q}(\rho_1,\rho_2,\alpha,s)|$}}}}
\put(1396,-6181){\makebox(0,0)[lb]{\smash{{\SetFigFont{10}{12.0}{\familydefault}{\mddefault}{\updefault}$\sim |s|^p$}}}}
\end{picture}%
\caption{Asymptotics of the functions~$\Rf^{p,q}$.}
\label{fig4}
\end{center}
\end{figure}

In the next proposition we collect some properties of the Fourier
transforms of the functions~$\hRf^{p,q}(\rho, \alpha, \omega)$
and~$\hRflog^{p,q}(\rho, \alpha, \omega)$.
\begin{Prp} {\bf{(regularization tails)}} \label{prpreg2}
The two Fourier integrals
\begin{eqnarray*}
\Rf^{p,q}(\rho, \alpha, s) &:=&
\int_0^\infty e^{-i \omega s}\, \hRf^{p,q}(\rho, \alpha, \omega)\:d\omega \\
\Rflog^{p,q}(\rho, \alpha, s) &:=&
\int_0^\infty e^{-i \omega s}\, \hRflog^{p,q}(\rho, \alpha, \omega)\:d\omega
\end{eqnarray*}
in the region~$|s| \gg \rho$ have the asymptotic expansions
\begin{eqnarray}
\Rf^{p,q}(\rho, \alpha, s) &=&
|s|^{-\alpha}\: \exp \left(-\frac{i \pi}{2}\: \alpha\:
\epsilon(s) \right) \left(1 + {\mathcal{O}}\!\left(
(\rho/s)^{q+1} \right) \right) \\
\Rf^{p,q}(\rho, \alpha, s) &=& \left( \log|s|\;
+ \frac{i \pi}{2} \:\epsilon(s) \right)
|s|^{-\alpha}\: \exp \left(-\frac{i \pi}{2}\: \alpha\:
\epsilon(s) \right) \left(1 + {\mathcal{O}}\!\left(
(\rho/s)^{q+1} \right) \right) .\qquad
\end{eqnarray}
Moreover, there are constants~$c=c(p,q)$ such that
\beq
|\Rf^{p,q}(\rho, \alpha, s)|
\;\leq\; c(p,q)\:\rho^{-\alpha-p}\: |s|^p \:,\qquad
|\Rflog^{p,q}(\rho, \alpha, s)|
\;\leq\; c(p,q)\:|\log \rho|\: \rho^{-\alpha-p}\: |s|^p
\:.
\eeq
The functions~$\hRf^{p,q}$ and~$\hRflog^{p,q}$ satisfy
for suitable constants~$C=C(p,q)$ the bounds
\beq \label{momes}
|\hRf^{p,q}(\rho, \alpha, \omega)|
\;\leq\; C(p,q)\: \rho^{-\alpha+1}\:,\qquad
|\hRflog^{p,q}(\rho, \alpha, \omega)| \;\leq\;
C(p,q)\: |\log \rho| \:\rho^{-\alpha+1} \:.
\eeq
\end{Prp}
{\Proof} The estimate~(\ref{momes}) follows from a simple
scaling argument.
Integrating by parts, one sees that the differential
operator~$(\partial_\omega)^p$ corresponds in position space to a
multiplication by~$(is)^p$. Using this fact, we can compute
the Fourier transform of every summand in~(\ref{Rfdef})
and~(\ref{Rflogdef}) using Lemma~\ref{lemmareg2}.
This gives the result.
\QED
Thus in the region~$|s| \ll \rho$, the regularization functions~$R^{p,q}(\rho, \alpha, s)$
can be made arbitrarily small by choosing~$p$ sufficiently large. In the
region~$|s| \gg \rho$, on the other hand, the regularization tail
has the desired power behavior~$\sim s^{-\alpha}$, up to an error
term which can be made arbitrarily small by increasing~$q$.
This asymptotics is shown on the left of Figure~\ref{fig4}.
The regularization tails~$\Rf^{p,q}(\rho, \tilde{\rho}, \alpha, s)$
and~$\Rflog^{p,q}(\rho, \tilde{\rho}, \alpha, s)$, which involve
two length scales~$\rho$ and~$\tilde{\rho}$, are illustrated on the
right of Figure~\ref{fig4}. The second
parameter~$\tilde{\rho}$ gives an upper scale for the power
behavior~$\sim |s|^{-\alpha}$ of the tail. If~$|s| \gg \tilde{\rho}$,
the regularization function can be made arbitrarily small
by choosing~$q$ large.

Combining the results of Propositions~\ref{prpreg1} and~\ref{prpreg2}, we
are led to regularization functions of the form
\[ e^{\varepsilon \omega} \left(\sum_{l=0}^n \frac{(-\rho \omega)^l}{l!}
\right) + \kappa_1 \,\hRf^{p_1,q_1}(\rho_1, \alpha_1, -\omega)
+ \kappa_2 \,\hRflog^{p_2,q_2}(\rho_2, \alpha_2, -\omega)
+ \cdots , \]
where we are still free to adjust the parameters
of the regularization tails as functions of~$\varepsilon$,\
and `$\cdots$' may be a sum of additional regularization tails.
It is important to keep in mind that the functions describing the
regularization tails should be small in momentum space and should
tend to zero as~$\varepsilon \searrow 0$. In view
of~(\ref{momes}), this can be ensured by arranging that
\beq \label{kappacond}
\kappa_i\, \rho_i^{-\alpha_i+1} \log \rho_i \ll 1\:.
\eeq

\section{The Outer Strip} \label{sec4}
\setcounter{equation}{0}
We now begin the study of~${\mathcal{M}}[A^\varepsilon_{xy}]$ (see~(\ref{0}, \ref{chain}))
for the fermionic projector~$P^\varepsilon$ with spherically symmetric regularization
(see Section~\ref{sec3}). In order to work out the underlying mechanisms, we begin by discussing
the effect of simple contributions to the regularized fermionic
projector and proceed by taking into account additional, more
complicated contributions. This will lead us to the general
construction of Proposition~\ref{prpos}.

Recall that away from the light cone and without regularization, ${\mathcal{M}}[A_{xy}]$
is given by~(\ref{3}, \ref{fdef}). Naively, one might expect that this formula should
to good approximation also be valid for~${\mathcal{M}}[A^\varepsilon_{xy}]$, except
in the strip~$||t|-r| \lesssim \varepsilon$ near the light cone, where the regularization
should be important. In order to explain why this naive picture is not correct, we
consider for simplicity {\em{one Dirac sea}}
and disregard the regularization of the
functions~$K$, $f$ and~$h$ in~(\ref{Pansatz}) (i.e. we omit the index~$\beta$ and
choose~$K$, $f$ and~$h$ as in~(\ref{mex}, \ref{mex2})). For the function~$g$
we choose a polynomially smoothed
exponential regularization plus a regularization tail,
\beq \label{greg}
g(\omega) \;=\; e^{\varepsilon \omega} \left( \frac{1}{16 \pi}
\sum_{l=0}^n \frac{(-\varepsilon \omega)^l}{l!} + \frac{\delta}{\Gamma(\gamma)}\,
|\omega|^{\gamma-1} \right) \Theta(-\omega)
\eeq
with parameters~$\gamma>1$, $n \in \N$ and~$\delta>0$. The regularization tail should
be small compared to the first term, meaning that
\beq \label{dcond} \delta \;\ll\; \varepsilon^{\gamma-1}\:. \eeq
Considering the leading terms of the mass expansion of Lemma~\ref{lemma2} and computing the remaining one-di\-men\-sio\-nal Fourier transform with the help
of Propositions~\ref{prpreg1} and~\ref{prpreg2},
we obtain to leading order in~$s/r$ the following regularization effect,
\begin{eqnarray}
P^\varepsilon - P &\asymp& -\frac{i}{r}\: (\gamma^0 - \gamma^r)
\int_{-\infty}^0 \omega \left(g - \frac{1}{16 \pi} \right) e^{i \omega s}\:d\omega
\:+\: \frac{1}{r^2}\: \gamma^r
\int_{-\infty}^0 \left(g - \frac{1}{16 \pi} \right) e^{i \omega s} \:d\omega \nonumber \\
&=& \left( -\frac{\gamma^0 - \gamma^r}{r} \: \frac{\partial}{\partial s}
+ \frac{\gamma^r}{r^2} \right) \left[ \frac{i}{s} \left(\frac{\varepsilon}{\varepsilon + is} \right)^{n+1}
+ \delta\, e^{-\gamma \, \log(\varepsilon + is)} \right] \label{aux}
\end{eqnarray}
(the symbol ``$\asymp$'' means that we pick a particular contribution to~$P^\varepsilon - P$).
If~$s=\varepsilon$, the condition~(\ref{dcond}) ensures that the contribution by the polynomially smoothed exponential regularization is much larger than
the regularization tail,
\[ \frac{i}{s} \left(\frac{\varepsilon}{\varepsilon + is} \right)^{n+1} \Big|_{s=\varepsilon}
\;\sim\; \frac{1}{\varepsilon} \;\gg\; \delta\, \varepsilon^{-\gamma} \;\sim\;
\delta\, e^{-\gamma \, \log(\varepsilon + i s)} \Big|_{s=\varepsilon} . \]
(Note that differentiating with respect to~$s$ changes the scaling only by an overall factor~$\varepsilon^{-1}$, and thus the last inequality can also be applied to the derivative term in~(\ref{aux}).)
But, by choosing~$n$ large, we can arrange that the contribution by the
polynomially smoothed exponential regularization decays faster in~$s$ than
the regularization tail, and thus
the regularization tail dominates if~$s \gg \varepsilon$. Furthermore,
using the identity~(\ref{id2}) we obtain the following asymptotic formula
for the regularization effect,
\begin{eqnarray}
P^\varepsilon - P &\asymp&
\left( -\frac{\gamma^0 - \gamma^r}{r} \: \frac{\partial}{\partial s}
+ \frac{\gamma^r}{r^2} \right)
\delta\, |s|^{-\gamma} \Big( \cos(\pi \gamma/2) - i \epsilon(s)\,
\sin(\pi \gamma/2) \Big) \nonumber \\
&=& \delta\, \frac{\gamma^0 - \gamma^r}{r}\:\gamma\: |s|^{-\gamma-1}
\Big( \epsilon(s)\, \cos(\pi \gamma/2) - i \sin(\pi \gamma/2) \Big) \label{regeff1} \\
&&+\delta\, \frac{\gamma^r}{r^2}\, |s|^{-\gamma} \Big( \cos(\pi \gamma/2) - i \epsilon(s)\,
\sin(\pi \gamma/2) \Big) . \label{regeff2}
\end{eqnarray}
Let us carefully consider in which range this asymptotic formula is valid.
Since we took into account only the leading contribution in~$s/r$,
we clearly need to assume that~$|s| \ll r$. Furthermore, working with
the regularization tails is justified only if~$s \gg \varepsilon$.
Finally, the mass expansion requires that
$m^2 \gg |t^2-r^2| = |s^2 + 2 s r|$ and thus~$s \ll m^{-1}, m^{-2}/r$.
We conclude that the above asymptotic formula is valid for~$s$ in the range
\beq \label{srange}
\varepsilon \;\ll\; |s| \;\ll\; \min(m^{-1},r, m^{-2}/r) \:.
\eeq

Comparing with the leading contribution to the unregularized fermionic projector,
\beq
P \;\asymp\; -\frac{i}{16 \pi^3}\: (\gamma^0 - \gamma^r)\: \frac{1}{rs^2}
\:-\:\frac{i}{16 \pi^3}\:\gamma^r\: \frac{1}{r^2 s} \:, \label{unreg}
\eeq
one sees that the regularization terms~(\ref{regeff1}, \ref{regeff2}) are very small.
However, as a major difference to the purely imaginary expression~(\ref{unreg}), they
also have real components. As a consequence, the regularization terms behave
differently in composite expressions. More precisely, a direct calculation yields for the
traceless part of~$A^\varepsilon_{xy}$ to leading order in~$s/r$ the formula
\begin{eqnarray}
A^\varepsilon_{xy} - \frac{1}{4}\, \Tr(A^\varepsilon_{xy}) &\asymp&
\frac{\delta}{8 \pi^3}\: \frac{i \gamma^0 \gamma^r}{r^3} \:(1-\gamma)\: |s|^{-\gamma-2}\:
\cos(\pi \gamma/2) \nonumber \\
&&-\frac{m^3}{256 \pi^5}\: \frac{\gamma^0-\gamma^r}{r s^2}\:\Theta(s)
\:-\:\frac{m^3}{256 \pi^5}\: \frac{\gamma^r}{r^2 s}\: \Theta(s) \:, \label{traceless1}
\end{eqnarray}
valid in the range~(\ref{srange}).
We see that the regularization terms give rise to a {\em{bilinear contribution}}
to~$A^\varepsilon_{xy}$.
This component involves no powers of~$m$, and has thus been
``amplified'' compared to the vector component, which is~$\sim m^3$ (see also
the discussion on page~\pageref{discussion}).
We remark that there are also contributions by the regularization tail to the
vector components of the form~$\sim \delta m \gamma^0$
and~$\sim \delta m \gamma^r$; for clarity we postpone their discussion
(see~(\ref{dvc})).

To compute the corresponding~${\mathcal{M}}[A^\varepsilon_{xy}]$, we introduce for the
decomposition of~$A^\varepsilon_{xy}$ into its scalar, vector and bilinear components the notation
\beq \label{Agen}
A^\varepsilon_{xy} \;=\; A^s\: \1 \:+\: A^0\, \gamma^0 \:-\: A^r\, \gamma^r + A^b\, i \gamma^0 \gamma^r\:.
\eeq
Then the roots of the characteristic polynomial of~$A^\varepsilon_{xy}$ are given by
\beq \label{discr}
\lambda_\pm \;=\; A^s \pm \sqrt{(A^0)^2 - (A^r)^2 - (A^b)^2}\:.
\eeq
Since~$A^s$ is real, the~$\lambda_\pm$ form a complex conjugate pair if the discriminant
is negative, and in this case the argument given after~(\ref{F}) yields
that~${\mathcal{M}}[A^\varepsilon_{xy}]$ vanishes.
Using~(\ref{regeff1}, \ref{regeff2}), we conclude that
\beq \label{Min}
{\mathcal{M}}[A^\varepsilon_{xy}] \;=\; 0 \qquad
{\mbox{if }} s \;\leq\; s_1 := \left( \frac{32 \pi^2\, \delta}{\sqrt{2}\,m^3}\: |\cos (\pi \gamma/2)|
\right)^{\frac{2}{2 \gamma+1}} r^{-\frac{3}{2 \gamma+1}} \:.
\eeq
We shall never choose~$\gamma$ equal to an odd integer, so that the
factor~$\cos(\pi \gamma/2)$ will always be nonzero.
If conversely~$s>s_1$, the discriminant is positive, and thus the~$\lambda_\pm$ are real.
Let us verify that they have the same sign. One possible method would be a direct calculation
similar as in the proof of Lemma~\ref{lemma11}. It is more elegant to proceed as follows.
We can clearly assume that~$\lambda_+ \neq \lambda_-$. From~(\ref{Agen}) one sees
that~$A^\varepsilon_{xy}$ commutes with the matrix
\[ \kappa \;:=\; \frac{1}{2} \left( \1 + \rho \gamma^0 \gamma^r \right) \]
(where~$\rho$ is the pseudoscalar matrix). It is easily verified that~$\kappa$ projects
on a two-dimensional subspace, and that~$A^\varepsilon_{xy}$ restricted to this subspace
is not a multiple of the identity matrix. Hence the characteristic polynomial of the
matrix~$A^\varepsilon_{xy}|_{{\mbox{\scriptsize{Im}}}\, \kappa} : {\mbox{Im}}\, \kappa
\rightarrow {\mbox{Im}}\, \kappa$ has precisely the simple roots~$\lambda_+$ and~$\lambda_-$.
Thus
\begin{eqnarray*}
\lambda_+ \lambda_- &=& \det A^\varepsilon_{xy}|_{{\mbox{\scriptsize{Im}}}\, \kappa} \;=\;
\det \Big( P^\varepsilon(x,y)|_{P^\varepsilon(y,x)\,{\mbox{\scriptsize{Im}}}\, \kappa}\;
P^\varepsilon(y,x)|_{{\mbox{\scriptsize{Im}}}\, \kappa} \Big) \\
&=& \det \Big( (P^\varepsilon(y,x)|_{{\mbox{\scriptsize{Im}}}\, \kappa})^*\;
P^\varepsilon(y,x)|_{{\mbox{\scriptsize{Im}}}\, \kappa} \Big) \;=\;
\Big| \det (P^\varepsilon(y,x)|_{{\mbox{\scriptsize{Im}}}\, \kappa}) \Big|^2 \;\geq\; 0\:,
\end{eqnarray*}
and we conclude that~$\lambda_+$ and~$\lambda_-$ indeed have the same sign. Therefore, ${\mathcal{M}}$ can be computed exactly as explained before~(\ref{0}),
\begin{eqnarray}
\lefteqn{ {\mathcal{M}}[A^\varepsilon_{xy}] \;=\; 2 A^\varepsilon_{xy} - \frac{1}{2}\, \Tr (A^\varepsilon_{xy})
\qquad {\mbox{if $s>s_1$}} } \label{Moutm} \\
&\asymp&
\frac{\delta}{4 \pi^3}\: \frac{i \gamma^0 \gamma^r}{r^3} \:(1-\gamma)\: |s|^{-\gamma-2}\:
\cos(\pi \gamma/2)
\:-\: \frac{m^3}{128 \pi^5}
\left( \frac{\gamma^0-\gamma^r}{r s^2} + \frac{\gamma^r}{r^2 s} \right) \Theta(s) \:. \label{Mout}
\end{eqnarray}

\begin{figure}[tb]
\begin{center}
\begin{picture}(0,0)%
\includegraphics{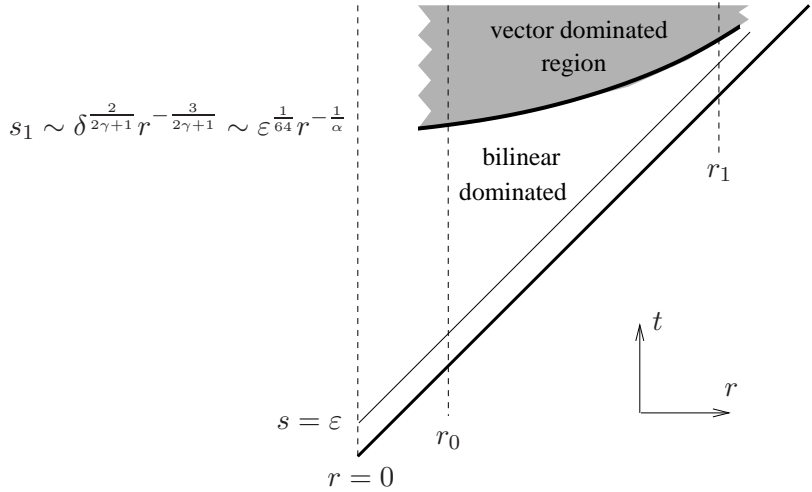}%
\end{picture}%
\setlength{\unitlength}{1657sp}%
\begingroup\makeatletter\ifx\SetFigFont\undefined%
\gdef\SetFigFont#1#2#3#4#5{%
  \reset@font\fontsize{#1}{#2pt}%
  \fontfamily{#3}\fontseries{#4}\fontshape{#5}%
  \selectfont}%
\fi\endgroup%
\begin{picture}(12029,7399)(-3884,-7393)
\put(901,-7261){\makebox(0,0)[lb]{\smash{{\SetFigFont{11}{13.2}{\familydefault}{\mddefault}{\updefault}$r=0$}}}}
\put(5761,-4921){\makebox(0,0)[lb]{\smash{{\SetFigFont{11}{13.2}{\familydefault}{\mddefault}{\updefault}$t$}}}}
\put(6841,-5911){\makebox(0,0)[lb]{\smash{{\SetFigFont{11}{13.2}{\familydefault}{\mddefault}{\updefault}$r$}}}}
\put(136,-6406){\makebox(0,0)[lb]{\smash{{\SetFigFont{11}{13.2}{\familydefault}{\mddefault}{\updefault}$s=\varepsilon$}}}}
\put(2521,-6631){\makebox(0,0)[lb]{\smash{{\SetFigFont{11}{13.2}{\familydefault}{\mddefault}{\updefault}$r_0$}}}}
\put(6571,-2626){\makebox(0,0)[lb]{\smash{{\SetFigFont{11}{13.2}{\familydefault}{\mddefault}{\updefault}$r_1$}}}}
\put(-3869,-1996){\makebox(0,0)[lb]{\smash{{\SetFigFont{11}{13.2}{\familydefault}{\mddefault}{\updefault}$s_1 \sim \delta^{\frac{2}{2\gamma+1}} r^{-\frac{3}{2\gamma+1}} \sim \varepsilon^{\frac{1}{64}}r^{-\frac{1}{\alpha}}$}}}}
\end{picture}%
\caption{The outer strip.}
\label{fig2}
\end{center}
\end{figure}

Let us briefly discuss our results so far. If~$s \leq s_1$, the bilinear component
of~$A^\varepsilon_{xy}$ makes the discriminant negative. In this so-called
{\em{bilinear dominated regime}}, ${\mathcal{M}}[A^\varepsilon_{xy}]$ vanishes
identically. The region~$s>s_1$, on the other hand, is {\em{vector dominated}},
and~${\mathcal{M}}[A^\varepsilon_{xy}]$ is nontrivial. The common boundary
of these two regions is the surface~$s=s_1$. This surface is not Lorentz invariant;
instead it has a {\em{power law scaling}}, $s_1 \sim r^{-\frac{3}{2 \gamma+1}}$.
Clearly, the above considerations are only valid if the
conditions~(\ref{srange}) and~(\ref{dcond}) are satisfied.
This implies in particular that~$r$ must lie in an interval
$(r_0, r_1)$ with
\beq \label{r01def}
r_0 \;\gg\; \delta^{\frac{1}{\gamma+2}} \,,\qquad  r_1\;\ll\;
\varepsilon^{-\frac{2\gamma+1}{3}}\, \delta^{\frac{2}{3}} \:.
\eeq
The region
\[ r_0 < r < r_1 \:, \spc s_1 < s \ll r \]
(and similarly also the region $r_0 < r < r_1$,
$s_1 < -t-r \ll r$ in the past) is referred to as the {\em{outer strip}}.
The outer strip and our scalings are illustrated in Figure~\ref{fig2}
(with the parameter~$\alpha$ and the power of~$\varepsilon$ as they will
used in Proposition~\ref{prpos} below).

The moments of~${\mathcal{M}}[A^\varepsilon_{xy}]$ can be introduced similar
to~(\ref{Itdef}--\ref{hmdef}). However, we are not allowed to take the
limit~$s_0 \searrow 0$, because this limit must be performed {\em{after}}
taking the limit~$\varepsilon \searrow 0$. Furthermore, we need here to take into
account the bilinear component, without counterterm. We thus define
\begin{eqnarray*}
I_{\varepsilon}(s_0, r) &=& \int_{-s_0}^{s_0}
{\mathcal{M}}[A^\varepsilon_{xy}] \:ds
\:+\: (\gamma^0 - \gamma^r) \left(\frac{\m_3}{4r s_0} - \frac{\m_5}{2}\: \log s_0 \right)
\:-\: \gamma^r \: \frac{\m_3}{4r^2}\, \log s_0 \\
J_{\varepsilon}(s_0,r) &=& \int_{-s_0}^{s_0}
s\, {\mathcal{M}}[A^\varepsilon_{xy}] \:ds \:-\:
(\gamma^0 - \gamma^r)\, \frac{\m_3}{4r}\, \log s_0 \\
K_{n,\varepsilon}(s_0,r) &=& \int_{-s_0}^{s_0}
s^n\, {\mathcal{M}}[A^\varepsilon_{xy}] \:ds\, , \spc n \geq 2\:,
\end{eqnarray*}
and denote the vector and bilinear components as in~(\ref{Agen}) by
a superscript. Using~(\ref{Min}, \ref{Mout}), we can immediately
compute the contribution of the outer strip to the moments. For example,
\begin{eqnarray}
J^0_\varepsilon &=& \int_{s_1}^{s_0} s\, {\mathcal{M}}^0[A^\varepsilon_{xy}]\, ds
\:+\: \frac{m^3}{128 \pi^5}\, \frac{1}{r}\, \log(s_0) \nonumber \\
&\asymp&-\int_{s_1}^{s_0} \frac{m^3}{128 \pi^5}\: \frac{1}{r s}\, ds
\:+\: \frac{m^3}{128 \pi^5}\, \frac{1}{r}\, \log(s_0) \;=\;
\frac{m^3}{128 \pi^5}\, \frac{1}{r}\, \log(s_1) \label{Jt} \\
J^b_\varepsilon &=& \int_{s_1}^{s_0} s\, {\mathcal{M}}^b[A^\varepsilon_{xy}]\, ds
\;\asymp\; \int_{s_1}^{s_0} \frac{\delta}{4 \pi^3}\: \frac{1}{r^3} \:(1-\gamma)\:
s^{-\gamma-1}\: \cos(\pi \gamma/2) \nonumber \\
&=& \frac{\delta}{4 \pi^3}\,\frac{\gamma-1}{\gamma}\,\cos(\pi \gamma/2)
\left(s_0^{-\gamma} - s_1^{-\gamma} \right) . \label{Jb}
\end{eqnarray}
Removing the regularization corresponds to the simultaneous
limits~$\varepsilon, \delta \searrow 0$, where~$\delta=\delta(\varepsilon)$
is to be chosen in agreement with~(\ref{dcond}). We see from~(\ref{Jt})
and~(\ref{Min}) that $J^0_\varepsilon$ will diverge in this limit,
\beq \label{Jtsing}
J^0_\varepsilon \;\sim\; \log s_1 \;\sim\; \log \delta \:.
\eeq
A contribution to a moment which diverges when the regularization
is removed is called a {\em{singular contribution}}. The appearance
of singular contributions shows that the constructions so far are not yet sufficient; we must find a way to compensate for these divergences.
However, before doing so, we discuss the structure of the singular
contributions in more detail.
The moment~$I^0_\varepsilon$ is also singular and diverges even like a negative
power of~$\delta$,
\[ I^0_\varepsilon \;\sim\; s_1^{-1} \;\sim\;
\delta^{-\frac{2}{2\gamma+1}}\:. \]
The bilinear component is more regular
than the corresponding vector component, but~$I^b$ is nevertheless singular,
\[ |J^b_\varepsilon| \;\lesssim\; \delta\, s_1^{-\gamma} \;\sim\;
\delta^{\frac{1}{2\gamma+1}} \:,\qquad
I^b_\varepsilon \;\sim\; \delta\, s_1^{-1-\gamma} \;\sim\;
\delta^{-\frac{1}{2\gamma+1}} \:. \]
More generally, the singularities of the moments satisfy the following
scaling rules. First, we observe that the vector component
in~(\ref{Mout}) is, to leading order in~$s/r$,
proportional to the nilpotent matrix~$\gamma^0-\gamma^r$.
This observation can also be expressed by saying that the {\em{vector
component of}}~${\mathcal{M}}[A^\varepsilon_{xy}]$ {\em{is null on the
light cone}}. Since the singular contributions to the moments
are determined by the behavior of~${\mathcal{M}}[A^\varepsilon_{xy}]$ on the light cone,
we come to the following simple conclusion.
\begin{description}
\item[(R1)] The leading singular contributions to the radial and time components of a moment
coincide.
\end{description}
The fact that the vector component of~${\mathcal{M}}[A^\varepsilon_{xy}]$ is null on the
light cone also yields a relation between the vector and bilinear components of the
moments. Namely, from~(\ref{Mout}) one sees that
\[ ({\mathcal{M}}^0)^2 - ({\mathcal{M}}^r)^2 \;\sim\; \frac{s}{r}\, ({\mathcal{M}}^0)^2\:. \]
Since the discriminant in~(\ref{discr}) vanishes at~$s_1$,
\[ \frac{s_1}{r}\, ({\mathcal{M}}^0)^2 |_{s=s_1}\;\sim\;
({\mathcal{M}}^0)^2 - ({\mathcal{M}}^r)^2 |_{s=s_1} \;=\; ({\mathcal{M}}^b)^2 |_{s=s_1}\:, \]
and thus
\[ {\mathcal{M}}^b(s_1) \;\sim\; \sqrt{\frac{s_1}{r}}\, {\mathcal{M}}^0(s_1)\:. \]
Since~${\mathcal{M}}$ has a power law behavior in~$s$, integrating over s
from~$s_1$ to~$s_0$ changes the scaling simply by an additional factor~$s_1$.
This explains the following rule.
\begin{description}
\item[(R2)] The leading singular contribution to the bilinear component of a moment is smaller than the corresponding
vector component by a scaling factor~$(s_1/r)^{\frac{1}{2}}$.
\end{description}

Let us now think about how to compensate for the singular contributions to the momenta.
One method is to take into account the regularization of the scalar component
of~$P^\varepsilon$. To this end, we choose the function~$h$
in~(\ref{Pansatz}) in analogy to~(\ref{greg}, \ref{dcond}) as
\beq \label{hreg}
h(\omega) \;=\; e^{\varepsilon \omega} \left( \frac{m}{16 \pi}
\sum_{l=0}^n \frac{(-\varepsilon \omega)^l}{l!} + \frac{\kappa\, m}{\Gamma(\alpha)}\,
|\omega|^{\alpha-1} \right) \Theta(-\omega)
\eeq
with parameters~$\alpha>1$ and
\beq \label{kcond} \kappa \;\ll\; \varepsilon^{\alpha-1}\:. \eeq
After choosing~$n$ sufficiently large, we can again restrict attention to the
regularization tail. The leading contribution in~$s/r$ of the regularization
tail to the fermionic projector is
\beq \label{stail}
P^\varepsilon \;\asymp\;
-\kappa \:\frac{im}{r}\: |s|^{-\alpha}
\Big( \cos(\pi \alpha/2) - i \epsilon(s)\, \sin(\pi \alpha/2) \Big)\:.
\eeq
This gives rise to the following contribution to the traceless part
of~$A^\varepsilon_{xy}$,
\beq \label{traceless2}
A^\varepsilon_{xy} - \frac{1}{4}\, \Tr(A^\varepsilon_{xy}) \;\asymp\;
\frac{\kappa m}{8 \pi^3}\:
\cos(\pi \alpha/2) \left( \frac{\gamma^0-\gamma^r}{r^2}\:|s|^{-\alpha-2}
+ \frac{\gamma^r}{r^3}\:\epsilon(s)\, |s|^{-\alpha-1} \right) ,
\eeq
again valid in the range~(\ref{srange}).
Note that we get no bilinear contribution. The vector contribution is of
lower order in~$m$ than that in~(\ref{traceless1}), and thus the
regularization tail has again been ``amplified'' (we shall always choose~$\alpha$
such that $\cos(\pi \alpha/2) \neq 0$). The vector component is again
null on the light cone.

We next compute the matrix~${\mathcal{M}}[A^\varepsilon_{xy}]$ corresponding
to~(\ref{traceless1})+(\ref{traceless2}). A short calculation shows that,
to leading order in~$s/r$,
\beq \label{Atcond}
(A^0)^2 - (A^r)^2 \;=\; \frac{2s}{r}\: (A^0)^2 \:,
\eeq
and thus we can say that the discriminant is positive if and only if
\beq \label{cvd0}
s>0 \quad {\mbox{and}} \quad \sqrt{\frac{2s}{r}}\: |A^0| \;\geq\; |A^b|\:.
\eeq
In particular, for negative~$s$ the discriminant is negative and~${\mathcal{M}}[A^\varepsilon_{xy}]$
vanishes. Hence in what follows we may restrict attention to the region~$s>0$.
Putting in the detailed formulas and abbreviating the appearing
constants by~$c_1, c_2, c_3$, we thus obtain
\beq \label{cvd}
{\mbox{$A^\varepsilon_{xy}$ vector dominated}} \;\; \Longleftrightarrow \;\;
\left| -c_1\, r^{-\frac{3}{2}}\, s^{-\frac{3}{2}} + c_2 \, \kappa\, r^{-\frac{5}{2}}\,
s^{-\alpha-\frac{3}{2}} \right| \;\geq\; |c_3 \,\delta|\, r^{-3}\, s^{-\gamma-2}\:.
\eeq
If condition~(\ref{cvd}) is satisfied, we find similar to~(\ref{Moutm}) that
\beq \label{Moutn}
{\mathcal{M}}[A^\varepsilon_{xy}] \;=\; 2 A^\varepsilon_{xy} - \frac{1}{2}\, \Tr (A^\varepsilon_{xy})
\spc {\mbox{if $A^\varepsilon_{xy}$ vector dominated}}\:,
\eeq
whereas~${\mathcal{M}}[A^\varepsilon_{xy}]=0$ otherwise.

The situation described by~(\ref{cvd}, \ref{Moutn}) is complicated for
the following reason.
Since we want the new term~$\sim \kappa$ to compensate for the singularity of
the vector component, we want to choose~$\kappa$ positive.
Then~$A^0$ changes sign, having a zero at
\beq \label{s2def2}
s_2 \;=\; \left( \frac{c_2\, \kappa}{c_1\, r} \right)^{\frac{1}{\alpha}}\:.
\eeq
Since the bilinear component of~$A$ does {\em{not}} vanish at~$s_2$,
it will dominate the vector component in a neighborhood of~$s_2$.
This gives rise to a so-called {\em{intermediate bilinear dominated strip}}
around~$s=s_2$. The size~$\Delta s$ of this strip is of the
order~$s_1$. Thus it is impossible to treat the
intermediate bilinear dominated strip with a Taylor expansion in~$\Delta s/s_1$,
making the analytic details rather difficult.
In order to avoid this difficulty, it is convenient to take into account
the regularization tail of the next term in the mass expansion of the vector
component of~$P^\varepsilon$ by choosing~$\alpha(\omega)$ in such a way that
\[ \alpha(\omega)\: g(\omega) \;=\;
 e^{\varepsilon \omega} \left( \frac{1}{16 \pi}
\left( |\omega| - \sqrt{\omega^2 - k^2} \right)
\sum_{l=0}^n \frac{(-\varepsilon \omega)^l}{l!}
+ \frac{m^2\, \delta_1}{\Gamma(\gamma)}\,
|\omega|^{\gamma-\alpha-1} \Theta(-\omega) \right) . \]
The corresponding contribution to the fermionic projector is
\[ P^\varepsilon \;\asymp\;
-i \delta_1 m^2 \: (\gamma^0-\gamma^r)\:(\gamma-\alpha)
\: |s|^{-\gamma+\alpha-1} \Big( \epsilon(s)\, \cos(\pi(\gamma-\alpha)/2) -i
\sin(\pi (\gamma-\alpha)/2) \Big) , \]
and this gives rise to an additional bilinear contribution
to~$A^\varepsilon_{xy}$,
\beq
A^\varepsilon_{xy} - \frac{1}{4}\, \Tr(A^\varepsilon_{xy}) \;\asymp\;
\frac{\delta_1}{8 \pi^3}\: \frac{i \gamma^0 \gamma^r}{r^2} \:(\gamma-\alpha)\:
|s|^{-\gamma+\alpha-2}\: \epsilon(s)\, \sin(\pi (\gamma-\alpha)/2) \:.
\label{traceless3}
\eeq
By choosing~$\delta_1$ appropriately, we can arrange that
the bilinear component of~(\ref{traceless1})+(\ref{traceless3})
also has a zero at~$s_2$, (\ref{s2def2}).
Thus the vector component dominates also in a neighborhood of~$s_2$,
and the bilinear dominated intermediate strip disappears.
Clearly, this method works only if~$\gamma > \alpha+1$ and
if the resulting~$\delta_1$ is sufficiently small, but we shall see
below that these conditions will automatically be satisfied.
Using this method, the bilinear and vector dominated regions
look again as in Figure~\ref{fig2}.
The left boundary of the vector dominated region is again denoted by~$s_1$.

The following {\em{radial scaling argument}} \label{radscal}
gives us a relation between the
parameters~$\alpha$ and~$\gamma$. We want to have cancellations in the vector component
when computing a certain moment. Even without specifying in which moment the cancellation should
appear, the fact that the cancellation should occur for all values of~$r$
allows us to conclude that the graphs in Figure~\ref{fig2} should be independent
of~$r$ except for scalings of the coordinates. In particular, the
quotient~$s_1/s_2$ should be independent of the radius,
\[ \frac{s_1}{s_2} \;\sim\; r^0\:. \]
As a consequence, using~(\ref{s2def2}),
\begin{eqnarray}
\lefteqn{ -c_1\, r^{-\frac{3}{2}}\, s_1^{-\frac{3}{2}} + c_2 \, \kappa\, r^{-\frac{5}{2}}\,
s_1^{-\alpha-\frac{3}{2}} \;=\; -c_1\, r^{-\frac{3}{2}}\, s_2^{-\frac{3}{2}} \left[
\left(\frac{s_1}{s_2}\right)^{-\frac{3}{2}} -  \left(\frac{s_1}{s_2}\right)^{-\alpha-\frac{3}{2}}
\right] } \label{sz} \\
&\sim& r^{-\frac{3}{2}} \left( r^{-\frac{1}{\alpha}} \right)^{-\frac{3}{2}}
\;=\; r^{\frac{-3\alpha+3}{2 \alpha}} \spc\spc\spc\spc\spc\spc\spc\qquad \label{sa} \\
\lefteqn{ c_3\, \delta\, r^{-3}\, s_1^{-\gamma-2} \;=\;
c_3\, \delta\, r^{-3}\, s_2^{-\gamma-2} \left(\frac{s_1}{s_2}\right)^{-\gamma-2}
\;\sim\; r^{-3} \left( r^{-\frac{1}{\alpha}} \right)^{-\gamma-2} \;=\;
r^{\frac{-3\alpha+\gamma+2}{\alpha}}\:. } \label{sb}
\end{eqnarray}
Since at~$s_1$, (\ref{cvd}) holds with equality, we conclude that the scalings
in~(\ref{sa}) and~(\ref{sb}) coincide, and thus
\beq \label{garel} 2 \gamma+1 \;=\; 3 \alpha\:. \eeq

We next discuss the vector component~$\sim \delta$
to~${\mathcal{M}}[A^\varepsilon_{xy}]$, which was disregarded
in~(\ref{traceless1}). The contribution~(\ref{regeff1}, \ref{regeff2})
to the fermionic projector gives rise to a corresponding contribution
to the vector component of the closed chain,
\beq \label{dvc}
A^\varepsilon_{xy} \;\asymp\; -\delta\, \frac{m \gamma}{8 \pi^3}\:
(\gamma^0-\gamma^r)\: \frac{|s|^{-2-\gamma}}{r^2}\:
\cos(\pi \gamma/2) \:-\:
\delta\: \frac{m}{8 \pi^3}\, \gamma^r\:\epsilon(s)\:
\frac{|s|^{-1-\gamma}}{r^3}\: \cos(\pi \gamma/2)\:.
\eeq
Using~(\ref{Min}, \ref{dcond}), one sees that at~$s_1$,
(\ref{dvc}) is smaller than the vector component in~(\ref{traceless1})
by a scaling factor $\sqrt{s_1 r}$.
Hence, it gives rise to a contribution to the vector component
of the moments, which is smaller
than the leading contribution by a scaling factor~$\sqrt{s_1 r}$.
Comparing with~(\ref{traceless2}), one sees
that the contribution~$\sim(\gamma^0 - \gamma^r)$ of~(\ref{dvc})
can be compensated by inserting an additional
scalar regularization tail into~(\ref{hreg}) of the form
\[ \frac{\delta \gamma\, m}{\Gamma(\gamma)}\,
|\omega|^{\gamma-1} \:. \]
Note that this does {\em{not}} compensate for the contribution~$\sim \gamma^r$
of~(\ref{dvc}). However, this last term is smaller by an additional scaling
factor~$s/r$, and as a consequence it will not contribute
to any of the moments.
Another complication is that~(\ref{dvc}) also contributes to the discriminant.
This gives rise to corrections to~$s_1$, which in turn
yield corrections to the moments which are again smaller by
a scaling factor~$\sqrt{s_1 r}$ than the leading contributions.
Fortunately, all these corrections can be compensated by
taking into account additional regularization tails of the scalar
component of~$P^\varepsilon$. The last complication is that these
additional regularization tails as well as~(\ref{dvc}) also
yield corrections
to the zeros of the vector and bilinear components of~$A_{xy}^\varepsilon$.
In particular, these two zeros no longer coincide, giving
rise to an intermediate bilinear dominated strip near~$s_2$.
However, the size~$\Delta s$ of this strip now scales
like~$s_1^{3/2} \sqrt{r}$ and can thus be treated by
a simple expansion in powers of~$\Delta s/s_1$ (for details see the
proof of Proposition~\ref{prpos} below).

Before coming to the general construction,
we point out a structural difference between the vector components
in~(\ref{traceless1}, \ref{traceless2}) and~(\ref{dvc}):
whereas in~(\ref{traceless1}, \ref{traceless2}) the vector component
is a scalar multiple of the matrix
\beq \label{Lorentzform}
\cdots \left( (\gamma^0-\gamma^r) + \frac{s}{r}\, \gamma^r \right) ,
\eeq
in~(\ref{dvc}) the radial and time components have a different relative form.
This can be understood as follows. As shown before~(\ref{2ctr}), a Lorentz
invariant contribution is always of the form~(\ref{Lorentzform}). This
immediately explains why~(\ref{traceless1}) satisfies~(\ref{Lorentzform}).
For the term~(\ref{traceless2}), which is not Lorentz invariant, we
observe that~(\ref{traceless2}) comes about as the product of the scalar
regularization tail~(\ref{stail}) with the unregularized vector component
of the fermionic projector. The unregularized vector component is clearly
of the form~(\ref{Lorentzform}), and this structure is preserved when
multiplying by the scalar regularization tail.
In the contribution~(\ref{dvc}), however, the vector regularization
tail~(\ref{regeff1}, \ref{regeff2}) comes into play, which
violates~(\ref{Lorentzform}).
It is actually very helpful that~(\ref{dvc}) is negligible, so that
the vector component of~${\mathcal{M}}[A^\varepsilon_{xy}]$
satisfies~(\ref{Lorentzform}). Namely, integrating over~$s$, we can
restate~(R1) in the following stronger form:
\begin{description}
\item[(R1')] The leading singular contributions to the radial and time components
of the moments satisfy the following relations,
\[ I_\varepsilon^r \;=\; I_\varepsilon^0 - \frac{1}{r}\, J_\varepsilon^0 \:,\qquad
J_\varepsilon^r \;=\; J_\varepsilon^0 - \frac{1}{r}\, K_{2,\varepsilon}^0 \:,\qquad
K_{n,\varepsilon}^r \;=\; K_{n,\varepsilon}^0 - \frac{1}{r}\, K_{n+1,\varepsilon}^0\:. \]
\end{description}
In view of~(\ref{2c}), it will be sufficient to consider the time component of
the moments; the corresponding relations for the radial component will
follow automatically.

We can now state the main result of this section.
Anticipating that the construction of Section~\ref{sec5} will allow us to
modify~${\mathcal{M}}[A^\varepsilon_{xy}]$ on an ``intermediate layer'' $s
\approx s_I$ with~$s_I \ll s_1$, giving a significant contribution to~$I_\varepsilon$, but leaving all other moments unchanged, we do not need to consider~$I_\varepsilon$ here.
Thus we want to use the scalar regularization tail~(\ref{stail}, \ref{traceless2})
to compensate for the singular contribution to~$J_\varepsilon^0$.
According to~(\ref{Jtsing}), we must compensate for a logarithmic divergence.
The logarithm does not cause major problems; the only difference
compared to our above discussion is that we must also
use the {\em{log-power regularization
tails}}~$\sim |\omega|^{\alpha-1}\, \log |\omega|$ as considered in
Lemma~\ref{lemmareg2}.
Another modification is that we model the regularization tails more
generally with the functions~$\hRf^{p,q}$ and~$\hRflog^{p,q}$
introduced in Definition~\ref{Rfdef}.
We choose the length scale~$\rho$ of the regularization tails
equal to~$\varepsilon^{\frac{3}{64}} \gg \varepsilon$.
Choosing~$\rho$ so large makes it is easier
to satisfy the condition~(\ref{kappacond}).
Furthermore, we will choose the parameter~$q$ so large that the
error terms of order~$(\rho/s)^q$ will vanish as~$\varepsilon \searrow 0$.
The parameter~$p$, on the other hand, remains undetermined
(it will be specified later, see Proposition~\ref{prpil}).

\begin{Prp} \label{prpos}
For any~$c_0 \in \R$ and~$\alpha>1$ in the range
\beq \label{alpharange}
1 < \alpha < \frac{3}{2}\:,
\eeq
there is a family of fermionic projectors~$(P^\varepsilon)_{\varepsilon>0}$
with spherically symmetric regularization~(\ref{Pansatz})
having the following properties.
The region $r_0 < r < r_1$, $s_1 < s \ll r$ with
\beq \label{s0fdef}
r_0 \;\gg\; \varepsilon^{\frac{\alpha}{64(\alpha+1)}}
\:,\qquad r_1 \;\ll\;
\min \left( \varepsilon^{-\frac{\alpha}{32}},
\varepsilon^{-\frac{\alpha}{64(\alpha-1)}} \right) ,\qquad
s_1(r) \;=\; \varepsilon^{\frac{1}{64}} \: r^{-\frac{1}{\alpha}}
\eeq
is an outer strip in the sense that
\[ {\mathcal{M}}[A^\varepsilon_{xy}] \;\equiv\; 0 \spc
{\mbox{for~$\varepsilon \ll s \leq s_1$}} \:, \]
whereas the region~$s_1 \leq s \ll r$ is vector dominated,
except for an intermediate bilinear dominated strip of the
size~$\Delta s \sim s_1^{3/2} \sqrt{r}$.
As~$\varepsilon \searrow 0$, the contribution of the outer strip to the moments
behaves like
\begin{eqnarray}
\lefteqn{ \hspace*{-5cm} \lim_{s_0 \searrow 0} \lim_{\varepsilon \searrow 0}
\left\{ I^0_\varepsilon(s_0, r) - r^{-1+\frac{1}{\alpha}}\:
\varepsilon^{-\frac{1}{64}}
\left(u_1 + u_2 \log r + u_3 \log \epsilon \right)
\right. } \nonumber \\
&&\hspace*{-3cm}
\:-\:  r^{-\frac{1}{2}+\frac{1}{2\alpha}}\:
\varepsilon^{-\frac{1}{128}} \left( v_1 + v_2 \log r + v_3 \log \epsilon \right) \nonumber \\
&&\hspace*{-3cm}
\:-\:  r^{-2} \left( w_1 + w_2 \log r + w_3 \log \epsilon \right)
\Big\} \;=\; 0 \label{Itreg} \\
\lefteqn{ \hspace*{-5cm} \lim_{s_0 \searrow 0} \lim_{\varepsilon \searrow 0}
\left\{ I^b_\varepsilon(s_0, r)
-r^{-\frac{3}{2}+\frac{1}{2\alpha}}\: \varepsilon^{-\frac{1}{128}}
\left( b_1 + b_2 \log r + b_3 \log \epsilon \right) \right\} \;=\; 0 }
\label{Ibreg} \\
\lim_{s_0 \searrow 0} \lim_{\varepsilon \searrow 0} J_\varepsilon^0(s_0, r)
&=& \frac{\m_3}{4}\: \frac{\log 2r}{r} \:+\: \frac{\m_3 -2 c_0}{8r} \label{Jtreg} \\
\lim_{s_0 \searrow 0} \lim_{\varepsilon \searrow 0} J_\varepsilon^b(s_0, r)
&=& 0 \label{Jberg} \\
\lim_{s_0 \searrow 0} \lim_{\varepsilon \searrow 0} K_{n,\varepsilon}(s_0, r) &=& 0 \label{Kreg}
\end{eqnarray}
with suitable real parameters~$v_i$, $w_i$ and~$b_i$.
The radial component of the moments is given by the rule~(R1').
\end{Prp}
{\Proof} We choose a spherically symmetric regularization~(\ref{Pansatz})
with~$K_\beta$ and~$\alpha_\beta$ according to~(\ref{mex},\ref{alphadef})
and~$f_\beta \equiv 0$.
The regularization functions~$g$, $h$ all vanish for positive~$\omega$, whereas for negative~$\omega$ they should satisfy the conditions
\begin{eqnarray}
\lefteqn{ \sum_{\beta=1}^3 h_\beta(\omega)
\;=\; e^{\varepsilon \omega}\:\frac{M_1}{16 \pi}\:
\sum_{l=0}^n \frac{(-\varepsilon \omega)^l}{l!} } \label{TZ} \\
&&+ \kappa\, M_1 \left[ \hRf^{p_1, q_1}\!\left(\varepsilon^{\frac{3}{64}}, \alpha,
-\omega \right)
+ l_h \,\hRflog^{p_1, q_1}\!\left(\varepsilon^{\frac{3}{64}}, \alpha, -\omega \right) \right] \label{TA} \\
&&+ \nu_1 M_1 \left[ \hRf^{p_1, q_1}\!\left(\varepsilon^{\frac{3}{64}},
\frac{3\alpha-1}{2}, -\omega \right)
+ l_{\nu1} \,\hRflog^{p_1, q_1}\!\left(\varepsilon^{\frac{3}{64}},
\frac{3\alpha-1}{2}, -\omega \right) \right] \label{TB} \\
\lefteqn{ \sum_{\beta=1}^3 g_\beta(\omega) \;=\; e^{\varepsilon \omega}\:
\frac{M_0}{16 \pi} \sum_{l=0}^n \frac{(-\varepsilon \omega)^l}{l!} }
\label{TC} \\
&&+\delta\,M_3 \left[ \hRf^{p_1, q_1}\!\left(\varepsilon^{\frac{3}{64}},
\frac{3\alpha-1}{2}, -\omega \right)
+ l_g \,\hRflog^{p_1, q_1}\!\left(\varepsilon^{\frac{3}{64}},
\frac{3\alpha-1}{2}, -\omega \right) \right] \label{TD} \\
\lefteqn{ \sum_{\beta=1}^3  \alpha_\beta(\omega)\: h_\beta(\omega) \;=\;
e^{\varepsilon \omega} \sum_{\beta=1}^3
\frac{\rho_\beta}{16 \pi} \left( |\omega| - \sqrt{|\omega|^2 - K_\beta(\omega)^2} \right) \sum_{l=0}^n \frac{(-\varepsilon \omega)^l}{l!} } \label{TF} \\
&&+\nu_2\,M_3 \left[ \hRf^{p_1, q_1}\!\left(\varepsilon^{\frac{3}{64}},
\frac{\alpha-1}{2}, -\omega \right)
+ l_{\nu2} \,\hRflog^{p_1, q_1}\!\left(\varepsilon^{\frac{3}{64}},
\frac{\alpha-1}{2}, -\omega \right) \right] \label{TG} \\
&&+\kappa_1\,M_3 \left[ \hRf^{p_1, q_1}\!\left(\varepsilon^{\frac{3}{64}},
\alpha-1, -\omega \right)
+ l_{h1} \,\hRflog^{p_1, q_1}\!\left(\varepsilon^{\frac{3}{64}},
\alpha-1, -\omega \right) \right] \label{TH} \\
\lefteqn{ \sum_{\beta=1}^3  \alpha_\beta(\omega)\: g_\beta(\omega)
\;=\; e^{\varepsilon \omega} \sum_{\beta=1}^3
\frac{\rho_\beta}{16 \pi} \left( |\omega| - \sqrt{|\omega|^2 - K_\beta(\omega)^2} \right) \sum_{l=0}^n \frac{(-\varepsilon \omega)^l}{l!} } \label{TI} \\
&&+ \delta_1\,M_2\: \hRf^{p_1, q_1}\!\left(\varepsilon^{\frac{3}{64}},
\frac{\alpha-1}{2}, -\omega \right) ,  \label{TJ}
\end{eqnarray}
where we introduced the constants~$M_n = \sum_{\beta=1}^3 \rho_\beta \,m_\beta^n$,
and~$p_1$, $q_1$ are two integers to be determined later.
These are four linear equations for the nine unknown functions
$h_\beta$, $g_\beta$ and~$\alpha_\beta$.
A short consideration shows that the above conditions can all
be satisfied. Note that our ansatz involves the 11 free
real parameters $\kappa$, $\delta$, $\kappa_1$, $\delta_1$,
$\nu_1$, $\nu_2$,
$l_h$, $l_g$, $l_{h1}$, $l_{\nu1}$ and~$l_{\nu2}$.
These parameters are determined by a lengthy, but straightforward
calculation. We here give the individual calculation steps.

Let us first consider
only the leading contributions to~$A^\varepsilon_{xy}$.
The scalar regularization tail~(\ref{TA}) leads to a vector component
of~$A_{xy}^\varepsilon$ of the form
\begin{eqnarray}
A^0 &=& \frac{\kappa\, C_1}{r^2 s^{2+\alpha}} \left\{ 1 + C_2\, l_h
\log(s)\:+\: C_3\, \kappa^{-1} r s^{\alpha} \right\} \label{Atasy} \\
A^r &=& \left(1 - \frac{s}{r} \right) A^0
\end{eqnarray}
(this is very similar to~(\ref{traceless1}, \ref{traceless2}),
with the only difference that now also logarithms of~$s$ appear).
The vector regularization tails~(\ref{TD}, \ref{TJ}) give rise to
a bilinear component of~$A_{xy}^\varepsilon$ of the form
\beq \label{Abasy}
A^b \;=\; \frac{\delta\, C_4}{r^3 s^{2+\frac{3\alpha-1}{2}}}
\left\{ 1 + C_5\, l_g \log(s)\:+\: C_6\, \delta_1\, \delta^{-1}\,
r s^{\alpha} \right\}
\eeq
(this is analogous to the bilinear component in~(\ref{traceless1})
and~(\ref{traceless3})). We choose~$l_g$ and~$\delta_1$ such that
the curly brackets in~(\ref{Atasy}) and~(\ref{Abasy}) coincide.
This ensures that the zeros of the
vector and bilinear components coincide, thereby avoiding
an intermediate bilinear dominated strip of size~$\Delta s \sim s_1$
(see the discussion after~(\ref{s2def2})).
The parameter~$\delta$ is determined by the condition
\[ \sqrt{\frac{2 s_1}{r}} \:A^0(s_1) \;=\; A^b(s_1)\:, \]
whereas the condition
\[ \int_{s_1}^{s_0} s\, A^0(s)\: ds \;=\;
\frac{\m_3}{4}\: \frac{\log 2r}{r} \:+\: \frac{\m_3 -2 c_0}{8r}
\:+\: \rho(s_0) \:+\: {\mathcal{O}}(s_1) \]
fixes~$\kappa$ and~$l_h$ (here~$\rho$ is any function which
takes care of the boundary values at~$s_0$).

Next we need to take into account different kinds of correction terms.
First, by choosing~$q_1$ sufficiently large, we can make the
error terms of the regularization tails as small as we like.
Hence we do not need to consider these error terms here.
Furthermore, the vector regularization tail~(\ref{TD}, \ref{TJ}) also
leads to a contribution to the vector component of~$A^\varepsilon_{xy}$,
which we denote by a subscript~$g$,
\begin{eqnarray}
A_g^0 &=& \frac{\delta\, C_7}{r^2 s^{2+\frac{3\alpha-1}{2}}}
\left\{ 1 + C_4\, l_g \log(s)\:+\: C_5\, \delta_1\, \delta^{-1}\,
r s^{\alpha} \right\} \label{Agt} \\
A_g^r &=& A^0 \:+\: \frac{\delta\, C_8}{r^3 s^{1+\frac{3\alpha-1}{2}}}
\left\{ 1 + C_4\, l_g \log(s)\:+\: C_5\, \delta_1\, \delta^{-1}\,
r s^{\alpha} \right\} , \label{Agr}
\end{eqnarray}
which at~$s_1$ is smaller than~(\ref{Atasy}) by a scaling factor~$\sqrt{s_1 r}$
(see the discussion after~(\ref{dvc})).
The contribution~$A_g$ gives rise to a correction~$\Delta s_1$
to the left boundary~$s_1$ of the vector dominated region, which scales
like~$\Delta s_1 \sim s_1^{3/2} r^{1/2}$.
Both~$A_g^{t/r}$ and the correction~$\Delta s_1$ give rise to
additional contributions to the vector component of the moments,
which are smaller by a scaling factor~$\sqrt{s_1 r}$ than the
leading contributions.
In order to compensate for these additional contributions,
we use the scalar regularization tails~(\ref{TB}, \ref{TG}).
They give rise to an additional contribution to the vector
component of~$A_{xy}^\varepsilon$, which we denote by a
subscript~$\nu$. This contribution also yields
a correction to~$s_1$ and to the moments.
We choose the constants~$\nu_1$, $\nu_2$ and~$l_{\mu1}$, $l_{\mu2}$
such that all the corrections to~$s_1$ and to the
moment~$J^0$ cancel each other,
\[ \Delta s_1 \;=\; 0 \:,\spc
\int_{s_1}^\infty s\, (A^0_g - A^0_\nu)\,ds \;=\; 0\:. \]
For clarity, we point out that it is impossible to
compensate for both~(\ref{Agt}) and~(\ref{Agr}) completely by scalar regularization
tails, because~$A_\nu$ is of the form~(\ref{Lorentzform}),
whereas~$A_g$ is not. As a consequence, (\ref{Agt}, \ref{Agr})
give rise to an intermediate bilinear dominated strip near
the zero~$s_2$ of~$A^0$, whose size~$\Delta s$ is of the
order~$s_1^{3/2} r^{1/2}$. Since the leading contributions of
both~$A^{t/r}$ and~$A^b$ vanish at~$s_2$, the contribution of this
intermediate bilinear dominated strip to the moments will be even
of the order~$s_1 r$ smaller than the leading contribution
(see below).

So far, we only discussed a few selected contributions of the
regularization tails to the fermionic projector, and there are indeed many
other contributions which all give rise to additional correction terms.
Generally speaking, the corrections to~$A^\varepsilon_{xy}$ can be
classified as follows. The {\em{mass expansion}} gives rise
to powers of~$s r$, whereas the so-called {\em{regularization expansion}}
gives powers of~$s/r$.
From power counting one sees that the contributions of higher order
in~$s$ to the bilinear component of the moments vanish.
For the vector component of the moments, we need to take into account
only the first orders in the mass and regularization expansion.
The first order mass expansion terms can be compensated by
the scalar regularization tail~(\ref{TH}). The first order
regularization expansion terms, however, are at most logarithmically
divergent and give rise precisely to the terms involving~$w_i$
in~(\ref{Itreg}).

Now all the free parameters have been determined. They scale
in~$\varepsilon$ as follows,
\begin{eqnarray*}
\kappa &\sim& \varepsilon^{\frac{\alpha}{64}} \left(
1 + C \log \varepsilon \right) \;\:,\spc
\kappa_1 \;\sim\; \varepsilon^{\frac{\alpha}{64}} \left(
1 + C \log \varepsilon \right) \\
\delta &\sim& \varepsilon^{\frac{3 \alpha}{128}} \left(
1 + C \log \varepsilon \right) \:,\spc
\delta_1 \;\sim\; \frac{\delta}{\kappa} \\
\nu_1 &\sim& \varepsilon^{\frac{3 \alpha}{128}} \left(
1 + C \log \varepsilon \right) \:,\spc
\nu_2 \;\sim\; \varepsilon^{\frac{\alpha}{128}} \left(
1 + C \log \varepsilon \right) ,
\end{eqnarray*}
where~$C$ stands for a different constant each time.
Evaluating the condition~(\ref{kappacond}) for all regularization
functions gives rise to the condition~$\alpha<2$.
Likewise, the conditions~(\ref{srange}) yield the values
for~$r_0$ and~$r_1$.

Now that all the free parameters have been determined, a straightforward
calculation shows that the resulting regularized fermionic
projectors~$P^\varepsilon$ have the desired properties.
\hspace*{1cm}
\QED
We point out that the conditions~(\ref{TA}--\ref{TJ}) could
not be satisfied if we had only one generation.
Furthermore, it is noteworthy that $s_1 \sim \varepsilon^\frac{1}{64}
\gg \varepsilon$, and thus the outer region lies
away from the strip~$s \sim \varepsilon$ where~$P(x,y)$ is affected
considerably by the regularization.
We remark for clarity that the precise
power~$s_1 \sim \varepsilon^\frac{1}{64}$
introduced in~(\ref{s0fdef}) was only a matter of convenience.
We could just as well have realized any other power
law~$s_1 \sim \varepsilon^\nu$ with~$0<\nu<1$.

It is a natural question whether, by taking into account additional
regularization tails, one can compensate for some of the singular
contributions in~(\ref{Itreg}) or~(\ref{Ibreg}).
Our attempts in this direction were
not successful. This is clearly no definite answer, but it gives
nevertheless an indication that it should indeed not be possible
to compensate for the terms in~(\ref{Itreg}, \ref{Ibreg}), for the following general reason. Compensating the vector component~(\ref{Itreg})
seems impossible because, using a radial
scaling argument, one can compensate only for the
singular vector component of~$J^0_\varepsilon$ or of~$I^0_\varepsilon$, but
not for both moments at a time.
Compensating the bilinear contribution~(\ref{Ibreg}), on the
other hand, would make it necessary to consider a regularization tail of
the functions~$f_\beta$.
Again using a radial scaling argument, one finds that the corresponding
bilinear contribution does not fall off in~$s$ fast enough, so that the
outer strip is no longer vector dominated.

\section{The Intermediate Layers} \label{sec5}
\setcounter{equation}{0}
In Proposition~\ref{prpos}, the region~$\varepsilon \ll s \ll s_1$ was
bilinear dominated, and thus~${\mathcal{M}}[A^\varepsilon_{xy}]$ was
trivial. In this section, we want to introduce additional
structures in this region, making it possible to compensate for
the singular contributions to the moments in~(\ref{Itreg}, \ref{Ibreg}).
To explain the basic idea, we again restrict attention to one Dirac sea
and consider the two regularization tails
\beq \label{vttails}
g(\omega) \;\asymp\; e^{\varepsilon \omega}\,
\frac{\delta}{\Gamma(\gamma)}\: |\omega|^{\gamma-1}\; \Theta(-\omega) \:,\qquad
f(\omega) \;\asymp\; e^{\varepsilon \omega}\,
\frac{\nu}{\Gamma(\beta)}\: |\omega|^{\beta-1}\; \Theta(-\omega)
\eeq
(where the parameters~$\delta$, $\gamma$ and~$\beta$ are different
from those in the previous section). These regularization tails
give a large contribution to the bilinear component of~$A^\varepsilon_{xy}$,
\beq \label{Abinn}
A^b \;\asymp\;
-\frac{\delta}{8 \pi^3}\: (\gamma-1)\:\cos(\pi \gamma/2)
\: r^{-3}\: |s|^{-\gamma-2} \:-\:
\frac{\nu}{8 \pi^3}\: \sin(\pi \beta/2)
\: r^{-2}\: |s|^{-\beta-2}\:\epsilon(s) \:.
\eeq
By making this contribution sufficiently large, we can arrange that
the bilinear contribution dominates, so that~${\mathcal{M}}[A^\varepsilon_{xy}]$
vanishes. In order to avoid for~${\mathcal{M}}[A^\varepsilon_{xy}]$ to be
trivial everywhere, we let~$\delta$ and~$\nu$ have opposite sign.
Then the bilinear component has a positive zero at
\beq \label{s2rel}
s_2 \;=\; \left(-\frac{\sin(\pi \beta/2)}{(\gamma-1)\,
\cos(\pi \gamma/2)}\:
\frac{\nu\, r}{\delta} \right)^{\frac{1}{\beta-\gamma}} \:.
\eeq
Then a small strip in a neighborhood of~$s_2$ will be vector
dominated, and thus~${\mathcal{M}}[A^\varepsilon_{xy}]$ will be
nonzero inside this strip.
We choose the parameters~$\nu, \delta, \beta, \gamma$ in such
a way that~$\varepsilon \ll s_2 \ll s_1$.
Also, the contribution~(\ref{Abinn}) at~$s \approx s_2$ should be
much larger than the contributions to~$A^\varepsilon_{xy}$ considered
in the previous section. Conversely, the contribution~(\ref{Abinn})
should decay so rapidly in~$s$ that it is negligible inside the
outer region. In this way, the outer strip and the region~$s \approx s_2$
can be analyzed independent of each other.
We refer to the region~$s \approx s_2$ as the
{\em{intermediate layer}}.

In order to model the vector component of~$A^\varepsilon_{xy}$
in the intermediate layer, the simplest method is to work
similar to~(\ref{hreg}) with a scalar regularization tail. Unfortunately,
this leads to the following problem.
When working with a scalar regularization tail, the vector component
of~$A^\varepsilon_{xy}$ satisfies the condition~(\ref{Atcond}), and the
intermediate layer is thus determined by the inequalities~(\ref{cvd0}). Setting
\beq \label{abdef}
a \;=\; A^0(s_2) \spc {\mbox{and}} \spc b \;=\; (A^b)'(s_2)\:,
\eeq
the width~$\Delta s_2$ of the intermediate layer is given in linear approximation by
\beq \label{dsdef}
\Delta s_2 \;=\; 2 \,\sqrt{\frac{2s_2}{r}} \:\frac{a}{b}\:.
\eeq
For this approximation to be justified, we need to assume that~$\Delta s_2 \ll s_2$, and thus
\beq \label{consis}
\frac{a}{b} \;\ll\; \sqrt{s_2\,r}\:.
\eeq
The leading contribution of the intermediate layer to the moments is computed to be
\begin{eqnarray*}
I^0 &\sim& a\, \Delta s_2 \;\sim\; \frac{a^2}{b}\, s_2^{\frac{1}{2}}\,
r^{-\frac{1}{2}} \\
I^b &\sim& (A^b)''\, (\Delta s_2)^3 \;\sim\; \frac{a^3}{b^2}\, s_2^{\frac{1}{2}}
\, r^{-\frac{3}{2}}\:,
\end{eqnarray*}
where in the last step we used the natural scaling~$(A^b)''(s_2) \sim b/s_2$.
Using~(\ref{consis}), we conclude that
\beq \label{qu1}
\frac{I^0}{I^b} \;\sim\; \frac{b}{a}\, r \;\gg\;
\sqrt{\frac{r}{s_2}} \:.
\eeq
On the other hand, taking the quotient of the leading singular
contributions in Proposition~\ref{prpos}, which we need to compensate,
we see from~(\ref{Itreg}, \ref{Ibreg}, \ref{s0fdef}) that
\beq \label{qu2}
\frac{I^0}{I^b} \;\sim\; \varepsilon^{-\frac{1}{64}}\:
r^{\frac{1}{2} + \frac{1}{2 \alpha}} \;=\; \sqrt{\frac{r}{s_1}}\:.
\eeq
The scalings~(\ref{qu1}) and~(\ref{qu2}) contradict our
assumption~$s_2 \ll s_1$. We conclude that
it is impossible to compensate for the leading singular contributions
in~(\ref{Itreg}) and~(\ref{Ibreg}).

One idea for avoiding the above contradiction is to work
with a vector component of~$A^\varepsilon_{xy}$
which violates~(\ref{Atcond}). To explain the method, we consider
the two regularization tails
\[ \alpha(\omega)\, g(\omega) \;\asymp\; e^{\varepsilon \omega}\,
\frac{\delta_1 m^2}{\Gamma(\sigma)}\: |\omega|^{\sigma-1}\; \Theta(-\omega) \:,\qquad
f(\omega) \;\asymp\; e^{\varepsilon \omega}\,
\frac{\tilde{\nu}}{\Gamma(\tilde{\beta})}\: |\omega|^{\tilde{\beta}-1}\; \Theta(-\omega)\:. \]
They yield contributions to the bilinear component
of~$A^\varepsilon_{xy}$ of very similar form,
\[ A^b \;\asymp\;
\frac{\delta_1 m^2}{8 \pi^3}\:\sigma\, \sin(\pi \sigma/2)
\: r^{-2}\: |s|^{-\sigma-2}\:\epsilon(s)
\:-\: \frac{\tilde{\nu}}{8 \pi^3}\: \sin(\pi \tilde{\beta}/2)
\: r^{-2}\: |s|^{-\tilde{\beta}-2}\:\epsilon(s) \:, \]
and by choosing~$\tilde{\beta}=\sigma$ and~$\tilde{\nu} =
m^2 \sigma\, \delta_1$, we can arrange for these contributions
to cancel each other. However, then the regularization
tails still contribute to the vector component of~$A^\varepsilon_{xy}$,
\beq \label{notl}
A^0 \;\asymp\; \frac{\delta_1\, m^3}{8 \pi^3}\:\sigma \, \sin(\pi \sigma/2)
\:r^{-1}\: |s|^{-\sigma-2}\: \epsilon(s) \:,\spc
A^r \;\asymp\; \left\{ 1 - \frac{s}{2 r} \right\} A^0 \:.
\eeq
The important point is that, due to the extra factor~$1/2$ inside the
curly brackets, this vector component is not of the form~(\ref{Lorentzform}),
and thus~(\ref{Atcond}) is violated.
By combining~(\ref{notl}) with a vector contribution to~$A^\varepsilon_{xy}$
which comes from a scalar regularization tail and thus
satisfies~(\ref{Lorentzform}), we get the freedom to adjust the
time and radial component of~$A^\varepsilon_{xy}$ independently, without
influence on the bilinear component of~$A^\varepsilon_{xy}$.
This additional freedom can actually be used to modify the
scaling in~(\ref{qu1}), thus resolving the above contradiction.
Nevertheless, the method does not allow us to compensate for
all the moments in~(\ref{Itreg}) and~(\ref{Ibreg}), as the following
argument shows. Suppose that~$A^0$, $A^r$ and~$A^b$ can be chosen
independently. Setting
\[ a \;=\; A^0(s_2) \:, \spc b \;=\; (A^b)'(s_2)
\spc {\mbox{and}} \spc \nu \;=\; (A^0 - A^r)(s_2)\:, \]
we can correct the scaling in~(\ref{qu1}) by choosing
\[ \nu \;\gg\; a \: \frac{s_2}{r}\:, \]
because then
\begin{eqnarray}
\Delta s_2 &\sim& \frac{\sqrt{a \nu}}{b} \:,\spc I^0 \;\sim\; b\, \Delta s_2
\label{eq1} \\
I^b &\sim& \frac{b}{s_2}\: (\Delta s_2)^3
\;\stackrel{(\ref{eq1})}{=}\; \frac{\Delta s_2}{s_2}\:
\sqrt{I^0}\: \sqrt{\nu\, \Delta s_2}\:. \label{eq2}
\end{eqnarray}
By choosing~$a$, $b$ and~$\nu$ appropriately, we can indeed
compensate for the leading singularities in~(\ref{Itreg}) and~(\ref{Ibreg}).
However, if this is done, ~(\ref{eq2}) implies that
\[ \varepsilon^{-\frac{1}{64}} \;\sim\;
\frac{\Delta s_2}{s_2}\:
\varepsilon^{-\frac{1}{64}}\: \sqrt{\nu\, \Delta s_2}\:, \]
where for simplicity we omitted the scaling in~$r$.
Clearly, $\Delta s_2$ should be much smaller than~$s_2$, because
otherwise we do not have a ``strip.'' We conclude that
\[ \nu\, \Delta s_2 \;\gg\; 1\:. \]
Note that, by definition of~$\nu$,
\[ I^r \;=\; I^0 - \nu \Delta s_2 \:. \]
Hence $I^r$ has a nonzero contribution which violates
the rule~(R1'). Unfortunately, it seems impossible to compensate for this
additional contribution to~$I^r$. This leads us to conclude that
working with a vector component of~$A^\varepsilon_{xy}$
which violates~(\ref{Atcond}) does not resolve our problem.

\begin{figure}[tb]
\begin{center}
\begin{picture}(0,0)%
\includegraphics{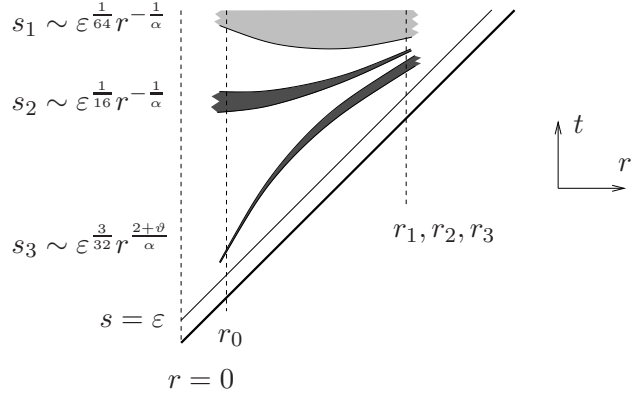}%
\end{picture}%
\setlength{\unitlength}{1243sp}%
\begingroup\makeatletter\ifx\SetFigFont\undefined%
\gdef\SetFigFont#1#2#3#4#5{%
  \reset@font\fontsize{#1}{#2pt}%
  \fontfamily{#3}\fontseries{#4}\fontshape{#5}%
  \selectfont}%
\fi\endgroup%
\begin{picture}(12279,7752)(-2039,-7881)
\put(2091,-6799){\makebox(0,0)[lb]{\smash{{\SetFigFont{11}{13.2}{\familydefault}{\mddefault}{\updefault}$r_0$}}}}
\put(10076,-3394){\makebox(0,0)[lb]{\smash{{\SetFigFont{11}{13.2}{\familydefault}{\mddefault}{\updefault}$r$}}}}
\put(5584,-4711){\makebox(0,0)[lb]{\smash{{\SetFigFont{11}{13.2}{\familydefault}{\mddefault}{\updefault}$r_1, r_2, r_3$}}}}
\put(9176,-2622){\makebox(0,0)[lb]{\smash{{\SetFigFont{11}{13.2}{\familydefault}{\mddefault}{\updefault}$t$}}}}
\put(1084,-7722){\makebox(0,0)[lb]{\smash{{\SetFigFont{11}{13.2}{\familydefault}{\mddefault}{\updefault}$r=0$}}}}
\put(-269,-6496){\makebox(0,0)[lb]{\smash{{\SetFigFont{11}{13.2}{\familydefault}{\mddefault}{\updefault}$s=\varepsilon$}}}}
\put(-2024,-601){\makebox(0,0)[lb]{\smash{{\SetFigFont{11}{13.2}{\familydefault}{\mddefault}{\updefault}$s_1\sim\varepsilon^{\frac{1}{64}}r^{-\frac{1}{\alpha}}$}}}}
\put(-2024,-2176){\makebox(0,0)[lb]{\smash{{\SetFigFont{11}{13.2}{\familydefault}{\mddefault}{\updefault}$s_2\sim\varepsilon^{\frac{1}{16}}r^{-\frac{1}{\alpha}}$}}}}
\put(-2024,-5011){\makebox(0,0)[lb]{\smash{{\SetFigFont{11}{13.2}{\familydefault}{\mddefault}{\updefault}$s_3\sim\varepsilon^{\frac{3}{32}}r^{\frac{2+\vartheta}{\alpha}}$}}}}
\end{picture}%
\caption{The intermediate layers.}
\label{fig3}
\end{center}
\end{figure}

The above consideration explains why it seems difficult
to compensate for the leading singular contributions in~(\ref{Itreg})
and~(\ref{Ibreg}) using one intermediate layer. Our way out is to
work with {\em{two}} intermediate layers, see Figure~\ref{fig3}.

\begin{Prp} \label{prpil}
We choose parameters~$\vartheta$ and~$\tau$ in the range
\beq \label{txrange}
\alpha \;<\; \vartheta \;<\; 2\:,\qquad 3 \;<\; \tau \;<\; \frac{15}{4} \:.
\eeq
Then for any~$c_0, c_1 \in \R$, there is a family of fermionic
projectors~$(P^\varepsilon)_{\varepsilon>0}$ which satisfies
the conditions of Proposition~\ref{prpos} and moreover has the
following properties. Choosing the parameters
\[ r_0 \gg \varepsilon^{\frac{\alpha}{64\,(2+\vartheta)}} \:,\qquad r_2 \ll \varepsilon^{-\frac{\alpha}{64}}\:,\quad
r_3 \ll \varepsilon^{-\frac{\alpha}{64\,(2+ \vartheta)}} \]
and setting
\begin{eqnarray*}
s_2 &=& r^{-\frac{1}{\alpha}}\: \varepsilon^{\frac{1}{16}} \:,\qquad\:
\Delta s_2 \;=\; r^{-\frac{2-\alpha+\vartheta}{2 \alpha}}\:
\varepsilon^{\frac{5}{64}} \\
s_3 &=& r^{\frac{2+\vartheta}{\alpha}}\: \varepsilon^{\frac{3}{32}} \:,\qquad
\Delta s_3 \;=\; r^{\frac{1}{2} + \frac{(2+\vartheta)(5+2 \tau)}{2 \alpha}}\:
\varepsilon^\frac{1}{8} \:,
\end{eqnarray*}
the two layers
\begin{eqnarray}
r_0 < r < r_2,\qquad |s-s_2| &<& \frac{\Delta s_2}{2} \label{s2def} \\
r_0 < r < r_3,\qquad |s-s_3| &<& \frac{\Delta s_3}{2}  \label{s3def}
\end{eqnarray}
are intermediate layers in the sense that they are vector dominated,
whereas the regions outside these layers and in the
range~$\varepsilon \ll s < s_1$ are bilinear dominated.
The contributions of the intermediate layers to the moments satisfy the
rule~(R1'). Furthermore, they combine
with~(\ref{Itreg}) and~(\ref{Ibreg}) in such a way that
all singular contributions cancel. The total contributions of the
outer strip and the intermediate layers to the moments~$I^0_\varepsilon$
and~$I^b_\varepsilon$ behave like
\begin{eqnarray*}
\lim_{s_0 \searrow 0} \lim_{\varepsilon \searrow 0}
I^0_\varepsilon(s_0, t) &=& \frac{\m_3}{8}\: \frac{1}{r^2} \:+\: \frac{\m_5}{2}\: \log 2r \:+\: \frac{\m_3+c_1}{2} \\
\lim_{s_0 \searrow 0} \lim_{\varepsilon \searrow 0}
I^b_\varepsilon(s_0, t) &=& 0 \:.
\end{eqnarray*}
The contribution of the intermediate layers to the moments~$J^b_\varepsilon$
and~$J^0_\varepsilon$ vanishes in the limit~$\varepsilon \searrow 0$.
\end{Prp}
{\Proof} We choose the regularization functions as in the
proof of Proposition~\ref{prpos}, but add regularization tails as
follows. In order to construct the intermediate layer at~$s_2$, we
consider the tails
\begin{eqnarray}
\sum_{\beta=1}^3 h_\beta(\omega) \!\!\!&\asymp&\!\!\!
\kappa_2
\left[ \hRf^{p_2, q_2}\!\left(\varepsilon^{\frac{9}{128}}, 
\varepsilon^{\frac{3}{64}}, \vartheta, -\omega \right)
+ l_{h2} \,\hRflog^{p_2, q_2}\!\left(\varepsilon^{\frac{9}{128}}, 
\varepsilon^{\frac{3}{64}}, \vartheta, -\omega \right) \right] \label{Th2} \\
\sum_{\beta=1}^3 g_\beta(\omega) \!\!\!&\asymp&\!\!\!
\delta_2 \left[ \hRf^{p_2, q_2}\!\left(\varepsilon^{\frac{5}{64}}, \frac{3 \vartheta-1}{2}, -\omega \right) + l_{g2} \,\hRflog^{p_2, q_2}\!\left(\varepsilon^{\frac{5}{64}}, \frac{3 \vartheta-1}{2}, -\omega \right) \right] \label{Tg2} \\
\sum_{\beta=1}^3 f_\beta(\omega) \!\!\!&\asymp&\!\!\!
\nu_2 \left[ \hRf^{p_2, q_2}\!\left(\varepsilon^{\frac{5}{64}}, \frac{3 \vartheta-1}{2}-\alpha, -\omega \right) \!+\! l_{f2} \,\hRflog^{p_2, q_2}\!\left(\varepsilon^{\frac{5}{64}}, \frac{3
\vartheta-1}{2}-\alpha, -\omega \right) \!\right] \qquad\; \label{Tf2}
\end{eqnarray}
with integers~$p_2$, $q_2$ and real parameters~$\kappa_2$, $\delta_2$,
$\nu_2$ and $l_{f2}$, $l_{g2}$, $l_{h2}$.
Note that in~(\ref{Th2}) we use tails involving two length
scales~(\ref{Rf2}, \ref{Rf2log}). For clarity, we postpone the
discussion of the length scales in the regulation tails
to Sections~\ref{sec6a} and~\ref{sec6b}. Here it suffices to keep in
mind that the tail is ``active'' in the considered region of~$s$,
meaning that $\varepsilon^{\frac{5}{64}} \ll |s| \ll \varepsilon^{\frac{3}{64}}$.

Similar to~(\ref{Abinn}), the tails~(\ref{Tg2}) and~(\ref{Tf2})
yield a bilinear contribution to~$A^\varepsilon_{xy}$. We choose~$\delta_2$
and~$l_{g2}$ such that this bilinear contribution vanishes
at~$s_2$, (\ref{s2def}). We set~$b=(A^b)'(s_2)$.
The scalar tail~(\ref{Th2}) gives rise to
a vector contribution to~$A^\varepsilon_{xy}$ which
satisfies~(R2'). Setting~$a=(A^0)'(s_2)$, the width~$\Delta s_2$
of the layer is in linear approximation given by~(\ref{dsdef}).
We choose~$\nu_2$ and~$l_{f2}$ such that the resulting value
for~$\Delta s_2$ agrees with~(\ref{s2def}).
Using Taylor expansions near~$s=s_2$, the contribution of the intermediate
layer to the bilinear moment is computed to be
\[ I^b_\varepsilon \;\asymp\;
-\frac{1}{12}\, (A^b)''(s_2)\: (\Delta s_2)^3 +
\frac{1}{2} \left( A^0\; \sqrt{\frac{2s}{r}} \right)'\!\!(s_2)\:
\: (\Delta s_2)^2 \:. \]
We choose~$\kappa_2$ and~$l_{h2}$ such that this contribution
compensates for the singular contribution in~(\ref{Ibreg}).
In agreement with our scalings~(\ref{qu1}), the intermediate layer at~$s_2$
gives rise to a singular contribution to the vector component of the form
\beq \label{Iterror}
I^0 \;=\; \varepsilon^{-\frac{7}{64}}\, r^{\frac{2-3\alpha+\vartheta}
{2 \alpha}}
\left( c_1 + c_2\, \log \varepsilon + c_3 \, \log r \right) \:.
\eeq
The just-determined parameters have the following scaling in~$\varepsilon$,
\begin{eqnarray*}
\kappa_2 &\sim& \varepsilon^{-\frac{1}{128} + \frac{\vartheta}{16}}
\left( 1 + C \log \varepsilon \right) \;\:,\spc
\delta_2 \;\sim\; \varepsilon^{-\frac{3}{128}
+\frac{3\vartheta}{32}} \left( 1 + C \log \varepsilon \right) \\
\nu_2 &\sim& \varepsilon^{-\frac{3}{128}
- \frac{\alpha}{16}+\frac{3\vartheta}{32}} \left(
1 + C \log \varepsilon \right) \:.
\end{eqnarray*}
A short calculation shows that, using the upper bound for~$\vartheta$
in~(\ref{txrange}), the condition~(\ref{kappacond}) is satisfied
for all regularization tails. Furthermore, choosing~$p_1>2$,
the regularization tails considered in Proposition~\ref{prpos} are very
small near~$s=s_2$ and give rise to corrections which all vanish as~$\varepsilon \searrow 0$. Likewise, using the lower bound for~$\vartheta$
in~(\ref{txrange}), the tails~(\ref{Th2}--\ref{Tf2}) are so small in the
region~$s>s_1$ that the corresponding error terms in the outer strip
vanish in the limit~$\varepsilon \searrow 0$. Finally, by
choosing~$q_2$ sufficiently large, we can arrange that the error terms
of the tails give no contribution to the moments in the layer at~$s_2$.

The next step is to construct the intermediate layer near~$s_3$ in such a way that its
contribution to~$A^0$ compensates for~(\ref{Iterror}), whereas its contribution
to~$A^b$ vanishes as~$\varepsilon \rightarrow 0$.
To this end, we introduce the additional regularization tails
\begin{eqnarray}
\sum_{\beta=1}^3 h_\beta(\omega) &\asymp&
\kappa_3  \left[ \hRf^{p_3, q_3}\!\left(\varepsilon^\frac{7}{64}, \varepsilon^\frac{11}{128}, \tau, -\omega \right)
+ l_{h3} \,\hRflog^{p_3, q_3}\!\left(\varepsilon^\frac{7}{64}, \varepsilon^\frac{11}{128}, \tau, -\omega \right) \right] \label{Th3} \\
\sum_{\beta=1}^3 g_\beta(\omega) &\asymp&
\delta_3 \left[ \hRf^{p_3, q_3}\!\left(\varepsilon^\frac{7}{64}, 2 \tau+1, -\omega \right) + l_{g3} \,\hRflog^{p_3, q_3}\!\left(\varepsilon^{\frac{7}{64}}, 2 \tau+1, -\omega \right) \right] \label{Tg3} \\
\sum_{\beta=1}^3 f_\beta(\omega) &\asymp&
\nu_3 \left[ \hRf^{p_3, q_3}\!\left(\varepsilon^\frac{7}{64}, 1+2 \tau
+ \frac{\alpha}{2+\vartheta}, -\omega \right) \right. \nonumber \\
&&\left.\qquad + l_{f3} \,\hRflog^{p_3, q_3}\!\left(\varepsilon^{\frac{7}{64}}, 1+2 \tau + \frac{\alpha}{2+\vartheta}, -\omega \right) \right] \!.\;\;
\qquad \label{Tf3}
\end{eqnarray}
With a similar calculation as in the layer near~$s_2$ above, we
determine the parameters~$\delta_3$, $l_{g3}$ and~$\nu_3$, $l_{f3}$ such
as to comply with~(\ref{s3def}). We choose~$\kappa_3$ and~$l_3$
such that the contribution of the layer near~$s_3$
to~$I^0$ compensates for~(\ref{Iterror}).
The just-determined parameters have the following scaling
in~$\varepsilon$,
\begin{eqnarray*}
\kappa_3 &\sim& \varepsilon^{\frac{1}{128} + \frac{3 \tau}{32}}
\left( 1 + C \log \varepsilon \right) \;\:,\spc
\delta_3 \;\sim\; \varepsilon^{\frac{15}{128}
+\frac{3 \tau}{16}} \left( 1 + C \log \varepsilon \right) \\
\nu_3 &\sim& \varepsilon^{\frac{15}{128}
 + \frac{\alpha}{64+32 \vartheta}+\frac{3\tau}{16}} \left(
1 + C \log \varepsilon \right) \:,
\end{eqnarray*}
where~$C$ again stands for a different constant each time.
A short calculation shows that the upper bound for~$\tau$
in~(\ref{txrange}) ensure that the condition~(\ref{kappacond}) is satisfied
for all regularization tails. Furthermore, choosing~$p_1$ and~$p_2$
sufficiently large,
the contributions of the tails~(\ref{Th2}--\ref{Tf2}) as well as the tails
in Proposition~\ref{prpos} to the moments in the
layer near~$s_3$ all vanish in the limit~$\varepsilon \searrow 0$.
Likewise, using the lower bound for~$\tau$ in~(\ref{txrange}),
the tails~(\ref{Th3}--\ref{Tf3}) do not contribute to the moments in
the layer near~$s_2$ nor in the outer strip.
Moreover, it is straightforward to check that the resulting contribution
to~$I^b$ vanishes as~$\varepsilon \searrow 0$. Finally,
by choosing~$p_2$ and~$q_3$ sufficiently large, we can arrange that
the tails~(\ref{Th2}--\ref{Tf2}) as well as the error terms of
the tails~(\ref{Th3}--\ref{Tf3}) do not contribute to the moments
in the layer at~$s_2$.

Next we arrange with suitable scalar regularization tails that~$I^0$ has the
desired value. To this end, we consider the scalar regularization tails
\begin{eqnarray}
\lefteqn{ \sum_{\beta=1}^3 h_\beta(\omega) \;\asymp\;
\kappa_3 \left[ \hRf^{0,0}\!\left(\varepsilon^\frac{1}{8}, \tau-\frac{\alpha-\vartheta}{4+2 \vartheta},
-\omega \right)
+ l_{h3} \,\hRflog^{0,0}\!\left(\varepsilon^\frac{1}{8}, \tau-\frac{\alpha-\vartheta}{4+2 \vartheta}, -\omega \right) \right] } \nonumber \\
&&+ \kappa_4 \left[ \hRf^{0,0}\!\left(\varepsilon^\frac{1}{8}, \tau-\frac{2\alpha-\vartheta-1}{4+2 \vartheta},
-\omega \right)
+ l_{h4} \,\hRflog^{0,0}\!\left(\varepsilon^\frac{1}{8},  \tau-\frac{2\alpha-\vartheta-1}{4+2 \vartheta}, -\omega \right) \right] \nonumber \\
&&+ \kappa_5 \left[ \hRf^{0,0}\!\left(\varepsilon^\frac{1}{8}, \tau+\frac{2+\alpha+\vartheta}{4+2 \vartheta},
-\omega \right)
+ l_{h5} \,\hRflog^{0,0}\!\left(\varepsilon^\frac{1}{8}, \tau + \frac{2+\alpha+\vartheta}{4+2 \vartheta}, -\omega \right) \right] \label{TailB}
\end{eqnarray}
and choose the parameters~$\kappa_i, l_{hi}$ such that the corresponding contribution in the intermediate layer
near~$s_3$ compensates for all the terms in~(\ref{Itreg}) as well as the first summand on the right of~(\ref{2a}).
The second and third summands in~(\ref{2a}) can be constructed similarly using the regularization tail
\begin{eqnarray}
\sum_{\beta=1}^3 h_\beta(\omega)
&\asymp&
\kappa_6 \left[ \hRf^{0,0}\!\left(\varepsilon^\frac{1}{8}, \tau+\frac{1}{2} - \frac{3 \alpha}{4+2 \vartheta},
-\omega \right) \right. \nonumber \\
&&\left. \qquad+\: l_{h6} \,\hRflog^{0,0}\!\left(\varepsilon^\frac{1}{8}, \tau+\frac{1}{2}
-\frac{3 \alpha}{4+2 \vartheta}, -\omega \right) \right] .\quad \label{TailC}
\end{eqnarray}

A short calculation shows that the intermediate layers do not
contribute to the higher moments~$J$ or~$K_n$. Thus it remains
to consider the error terms. More precisely, we need to compensate for the
vector component of the moment when we expand in powers of~$\Delta s_2/s_2$
and~$\Delta s_3/s_3$. Moreover, the vector regularization tails give
contributions to the vector component of~$A^\varepsilon_{xy}$, which
in analogy to the term~(\ref{dvc}) in the outer strip, are by a
factor~$\sqrt{sr}$ smaller than the leading terms.
Further error terms arise in an expansion
in powers of~$s/r$. All the resulting contributions to the moments
can be compensated in a straightforward way by additional
regularization tails.

In order to specify the range of~$r$ for which the
above arguments hold, one evaluates the conditions~(\ref{srange})
for all appearing regularization tails.

Finally, we need to verify that the region~$-s_1<s \ll -\varepsilon$ is everywhere bilinear dominated.
Since we used the scalar tails to arrange a nonzero vector component of~$A$,
this vector component is of the form~(\ref{Lorentzform}). As a consequence,
$(A^0)^2 - (A^r)^2<0$, implying that the region is indeed bilinear dominated.
\QED

\section{The Innermost Layer} \label{sec6z}
\setcounter{equation}{0}
With the constructions of the previous Sections~\ref{sec4}
and~\ref{sec5}, we arranged that~$\M[A^\varepsilon_{xy}]$ converges on the
light cone~$t=r>0$ in the distributional sense to~$\tilde{\mathcal{M}}(\xi)$.
However, so far we have not considered the momentum cone conditions
(see Definitions~\ref{defsmcc} and~\ref{defvmcc}).
In this section, we shall satisfy these additional conditions by introducing
another layer, the so-called {\em{innermost layer}},
which lies even closer to the light cone than the intermediate layers.
Sometimes we also refer to the intermediate layers together with the
innermost layer as the {\em{inner layers}}.
We will also have a closer look at the momentum cone
conditions, and we will slightly modify them in order to take into account
the regularization of~$P$ and the fact that~$P$ is supported on hyperbolas instead of a cone (compare Figures~\ref{fig1} and~\ref{figcon2}).

In preparation, we consider the matrix structure of the contributions to
the regularized product~${\mathcal{M}}[A_{xy}]\, P^\varepsilon(x,y)$.
We can clearly restrict attention to a vector~$\xi$
for which~${\mathcal{M}}[A_{xy}]$ is nonzero. From general properties
of the characteristic polynomial (see \cite[Lemma~5.2.1]{PFP})
we know that
\beq \label{Qsymm}
Q^\varepsilon(x,y) \;=\;
\frac{1}{2}\: \M[A^\varepsilon_{xy}]\, P^\varepsilon(x,y)
\;=\; \frac{1}{2}\: P^\varepsilon(x,y)\: \M[A^\varepsilon_{yx}]\:.
\eeq
Furthermore, from~(\ref{0}) we see that~$\M[A^\varepsilon_{xy}]$ equals
twice the trace-free part of~$A_{xy}$,
\[ \M[A^\varepsilon_{xy}] \;=\;
2 \left( A^0_{xy}\, \gamma^0 \:-\: A^r_{xy}\, \gamma^r
\:+\: A^b_{xy}\, i \gamma^0 \gamma^r \right) . \]
Let us study the symmetry under the transformation~$\xi \rightarrow -\xi$.
The bilinear component of~$A_{xy}$ arises because vector
contributions to~$P(x,y)$ and~$P(y,x)$ anticommute,
\[ A^b_{xy}\, i \gamma^0 \gamma^r \;=\;
\frac{1}{2} \left[ P(x,y),\, P(y,x) \right] \]
and thus, using that exchanging~$x$ and~$y$ also flips the sign
of~$\gamma^r$,
\beq \label{bsymm}
A^b_{xy} \;=\; A^b_{yx}\:.
\eeq
The vector component of~$A_{xy}$, on the other hand, arises if the
vector component of~$P(x,y)$ is multiplied by the scalar component
of~$P(y,x)$, or vice versa. In both cases, the corresponding contributions
to~$P(x,y)$ and~$P(y,x)$ commute, and thus
\beq \label{vsymm}
A^0_{xy} \;=\; A^0_{yx} \qquad {\mbox{and}} \qquad
A^r_{xy} \;=\; -A^r_{yx}\:.
\eeq
Combining the symmetry relations~(\ref{Qsymm}) and~(\ref{bsymm}, \ref{vsymm}),
we can compute~$Q^\varepsilon(x,y)$,
\begin{eqnarray}
Q^\varepsilon(x,y) &=& \frac{1}{2} \left[ A^b_{xy}\:i \gamma^0 \gamma^r,\,
P(x,y) \right] \:+\: \frac{1}{2} \Big\{ A^0_{xy} \,\gamma^0
- A^r_{xy} \,\gamma^r,\, P(x,y) \Big\} \nonumber \\
&=& i A^b \left( P^r \gamma^0 - P^0 \gamma^r \right)
\:+\: (A^0 \gamma^0 - A^r \gamma^r)\, P^s
\:+\: A^0 P^0 - A^r P^r\:, \label{Qm}
\end{eqnarray}
where in the last line for convenience we omitted the subscript
arguments~$x$ and~$y$. It is remarkable that~$Q^\varepsilon$ has
no bilinear component.

Clearly, the difficulty in evaluating the momentum cone conditions
is to handle the singularity on the light cone. Let us first bring
the leading singularity into a more convenient form.
We restrict attention to the future light cone~$t \approx r$.
For the scalar condition, we again choose the variables~$r$
and~$s=t-r$ and expand in powers of~$s/r$,
\beq \label{sls}
\int_0^\infty \frac{\M[A^\varepsilon_{xy}]}{t^2-r^2}\:dt \;=\;
\int_0^\infty \frac{\M[A^\varepsilon_{xy}]}{(2r+s)\, s}\: dt \;=\;
\int_0^\infty \frac{\M[A^\varepsilon_{xy}]}{2 r s}\:dt
\:+\: {\mbox{(distributional)}}\:.
\eeq
The vector condition is more difficult to handle because of the
radial derivative. Since the derivative~$\partial_t+\partial_r$
tangential to the light cone is less singular than a transversal
derivative, it is useful to rewrite the integral in~(\ref{vmcc}) as
\begin{eqnarray}
\lefteqn{ \int_0^{\infty} \frac{\partial}{\partial r} \Big( r\,
\M[A^\varepsilon_{xy}]\, i \xi\slsh \Big) \: \frac{t\, dt}{t^2-r^2} } 
\nonumber \\
&=& -\int_0^{\infty} \frac{\partial}{\partial t} \Big( r\,
\M[A^\varepsilon_{xy}]\, i \xi\slsh \Big) \: \frac{t\, dt}{t^2-r^2}
\:+\: \int_0^{\infty} \left(\partial_t + \partial_r \right) \Big( r\,
\M[A^\varepsilon_{xy}]\, i \xi\slsh \Big) \: \frac{t\, dt}{t^2-r^2}
\nonumber \\
&=& -\int_0^{\infty} (\M[A^\varepsilon_{xy}]\, i \xi\slsh) \:
\frac{r\, (t^2+r^2)}{(t^2-r^2)^2}\:dt
\:+\: \int_0^{\infty} \left(\partial_t + \partial_r \right) \Big( r\,
\M[A^\varepsilon_{xy}]\, i \xi\slsh \Big) \: \frac{t\, dt}{t^2-r^2} \:,
\label{tangder}
\end{eqnarray}
where in the last line we integrated by parts.
In the first integral we can again expand in powers of~$s/r$ to obtain
\beq \label{vls}
\int_0^{\infty} \frac{\partial}{\partial r} \Big( r\,
\M[A^\varepsilon_{xy}]\, i \xi\slsh \Big) \: \frac{t\, dt}{t^2-r^2}
\;=\; -\frac{r}{2} \int_0^{\infty}
\frac{\M[A^\varepsilon_{xy}]\, i \xi\slsh}{s^2} \:dt \:+\:
{\mbox{(less singular terms)}}\:.
\eeq

Let us evaluate the momentum cone conditions for the leading singular
terms~(\ref{sls}) and~(\ref{vls}) (all the correction terms
will be treated in the proof of Proposition~\ref{prpinnermost} below).
Since $Q^\varepsilon$ has no bilinear component~(\ref{Qm}), 
we may simply disregard the bilinear contributions to~(\ref{smcc})
and~(\ref{vmcc}). Thus it suffices to consider the scalar and vector components.
As $\M$ has no scalar component, in~(\ref{smcc}) we get only a
vector contribution. Combining the symmetry property~(\ref{vsymm})
with~(\ref{sls}), we obtain for the leading singularities the condition
\beq \label{cv1}
\int_0^\infty \frac{\M^0}{s}\, dt \;\asymp\;
\left( \frac{c_0}{r} + c_2\, r + \cdots \right) \left(1 + o(\varepsilon^0)
\right)
\eeq
(here~$o(\varepsilon^0)$ stands for terms which vanish as~$\varepsilon
\searrow 0$).
Considering again regularizations with the general property~(R2),
the corresponding condition for the radial component is automatically
satisfied. Similarly, we can compute the scalar contribution
to~(\ref{vmcc}). Again restricting attention to regularizations where the
leading vector contribution satisfies~(\ref{Lorentzform}),
combining~(\ref{vsymm}) with~(\ref{vls}) gives
\beq \label{cv2}
\int_0^\infty \frac{\M^0}{s}\, dt \;\asymp\;
\left( c_1\, r + c_3\, r^3 + \cdots \right) \left(1 + o(\varepsilon^0)
\right) ,
\eeq
and this condition is consistent with and stronger than~(\ref{cv1}).
Finally, the bilinear component of~$\M$ gives rise to a vector
contribution to~(\ref{vls}). Using the symmetry property~(\ref{bsymm}),
we get the condition
\beq \label{cb}
\int_0^\infty \frac{\M^b}{s^2}\, dt \;\asymp\;
\left( \frac{c_0}{r} + c_2\, r + \cdots \right) \left(1 + o(\varepsilon^0)
\right) .
\eeq

Starting from the conditions~(\ref{cv2}, \ref{cb}) we can now explain
our method for satisfying the momentum cone conditions.
Unfortunately, we cannot arrange that the contributions of the
outer strip and the intermediate layers are of the
form~(\ref{cv2}, \ref{cb}), in particular because those contributions
involve logarithms of~$r$.
However, using the same method as for the intermediate layers, we can
construct an additional layer at~$s \approx s_4$ with
$0<s_4 \ll s_3$. This {\em{innermost layer}} should have the
following properties.
The contribution of the innermost layer to all the moments~$I$, $J$,
and~$K^{(n)}$ should vanish as~$\varepsilon \searrow 0$, so that the
results of Propositions~\ref{prpos} and~\ref{prpil} remain valid.
Moreover, its contribution to the integrals in~(\ref{cv2}, \ref{cb})
should be of required form, and should be much larger than the
corresponding contributions of the outer strip and the intermediate
layers. Then the conditions~(\ref{cv2}, \ref{cb}) are satisfied
for the leading singular contributions, but the less singular
contributions (in particular the contribution by the outer strip
and the intermediate layers) will violate the momentum cone conditions.
Then our strategy is to perturb the innermost layer by additional
regularization tails so as to compensate for all the contributions
to~$\M[A^\varepsilon_{xy}]$ which violate the momentum cone conditions.

Before entering the details of the construction, we need to explain
how in principle to compute the less singular contributions to the
momentum cone conditions. First of all, there are the higher orders
in the~$s/r$ expansion, which were left out in~(\ref{sls}) and~(\ref{vls});
these terms are straightforward to compute.
To explain the method for handling the tangential derivatives
in~(\ref{tangder}), we consider as an example the contribution by the
intermediate layer at~$s_2$. Approximating it by a $\delta$-layer and
for simplicity leaving out the $\log$-terms, we obtain
\[ \M[A^\varepsilon_{xy}] \;\asymp\;
\left( c_1\, \varepsilon^{-\frac{7}{128}}\, r^{-\frac{1+3\alpha+\vartheta}{2 \alpha}}\: \frac{\xi\slsh}{r}
\:-\: b_1\, \varepsilon^{-\frac{1}{128}}\, r^{-\frac{3}{2}+\frac{1}{2\alpha}}
\,i \gamma^0 \gamma^r \right) \delta \!\left(s - \varepsilon^{\frac{1}{16}}\,
r^{\frac{1}{\alpha}} \right) \]
and consequently
\beq \label{lastf}
r\, \M[A^\varepsilon_{xy}] \xi\slsh \;\asymp\;
\left( 2 c_1\, \varepsilon^{-\frac{7}{128}}\, r^{-\frac{1+\alpha+\vartheta}{2 \alpha}}\: s\:+\: i b_1\, \varepsilon^{-\frac{1}{128}}\,
r^{-\frac{1}{2}+\frac{1}{2\alpha}} \,\xi\slsh \right)
\delta \!\left(s - \varepsilon^{\frac{1}{16}}\,
r^{\frac{1}{\alpha}} \right) ,
\eeq
where we omitted the higher orders in~$s/r$. The tangential derivative
$\partial_t+\partial_r$ is expressed
in the coordinates~$(s,r)$ simply by~$\partial_r$.
Hence by differentiating~(\ref{lastf}) with respect to~$r$ and
integrating with respect to~$t$, we can compute the tangential derivative
term in~(\ref{tangder}). The terms of higher order in~$s/r$ as well
as the correction terms arising from the finite size~$\Delta s_2$ of
the intermediate strip are straightforward to compute.

Finally, there are many correction terms which take into account all
the simplifications made in the derivation of the momentum cone conditions.
In order to treat these corrections systematically, we first note that
to derive momentum cone conditions without any simplifications, instead
of~(\ref{ci4}) we would have to consider the convolution integral
\[ B^\varepsilon \;:=\;
\int \frac{d^4p}{(2 \pi)^4}\: \hM^\varepsilon(p)\: \hat{H}^\varepsilon(q-p)
\:,\]
where $\hat{H}^\varepsilon$ is the high-energy part of the regularized
fermionic projector,
\[ \hat{H}^\varepsilon(k) \;=\; \hat{P}^\varepsilon(k)\:
\Theta(-k^0-\Omega)\:. \]
The Fourier transform of~$\hat{H}^\varepsilon$ can be determined in
detail using the formulas of Lemma~\ref{lemma1}, and this
makes it possible to compute~$B^\varepsilon$ in position space
in analogy to~(\ref{ci5}) by
\[ B^\varepsilon \;=\; \int e^{i \Omega t}
\M[A^\varepsilon_{xy}]\, H^\varepsilon(\xi)\: d^4 \xi\:. \]
Let us be more specific, for simplicity only for the scalar component
of~$P^\varepsilon$ and for one Dirac sea.
Then~$H^\varepsilon$ is obtained from the formula of Lemma~\ref{lemma1}
simply by changing the integration range,
\[ H^\varepsilon(\xi) \;=\; \frac{i}{r} \int_{-\infty}^{-\Omega}
d\omega\: h(\omega) \:e^{i \omega t}
\left( e^{-i K(\omega) \:r} - e^{i K(\omega) \:r} \right) \:. \]
Setting~$h(\omega)=1/(16 \pi^3)$ and~$K=-\omega$, we
recover our earlier formula for $H_{\mbox{\tiny{scal}}}$
(see~(\ref{hscalf})). Similar to Lemma~\ref{lemma2}, we can now perform
the mass expansion,
\[ H^\varepsilon(\xi) \;=\; \frac{i}{r}
\sum_{k=0}^\infty \int_{-\infty}^{-\Omega}
d\omega\: h(\omega) \:e^{i \omega t}
\left( e^{i \omega r} \:\frac{(i \alpha r)^k}{k!}
- e^{-i \omega r}  \:\frac{(-i \alpha r)^k}{k!} 
\right) . \]
When computing the effect of the regularization terms,
changing the integration range to the set~$\Xi$ keeps the
integral unchanged up to a contribution which tends to zero as~$\varepsilon
\searrow 0$. Therefore, the effect of the regularization terms is
exactly as computed for $P^\varepsilon$ in Section~\ref{sec3}, up to
small corrections which turn out to be negligible.
Computing the unregularized contributions of higher order in the mass,
we are led to integrals of the form
\[ \int_{-\infty}^{-\Omega} \frac{e^{i \omega (t \pm r)}}{\omega^n} \, d\omega \:. \]
By iteratively integrating by parts, one can reduce to the case~$n=1$, which
can be expressed in terms of the incomplete $\Gamma$ function
\[ \int_{-\infty}^{-\Omega} \frac{e^{i \omega x}}{\omega} \, d\omega
\;=\; -\int_{\Omega x}^\infty \frac{e^{-i u}}{u} \, du \;=\;
-\Gamma(0, i \Omega x)\:,  \]
and this can be asymptotically expanded in a straightforward way.
Expanding~$H^\varepsilon$ in this way, we obtain correction terms
to~$H^\varepsilon_{\mbox{\tiny{scal}}}$ and similarly
to~$H^\varepsilon_{\mbox{\tiny{vect}}}$. Computing the time integral
of the resulting expressions, we get corrections to~(\ref{smcc})
and~(\ref{vmcc}), which we refer to as the {\em{mass}} and
{\em{regularization corrections}} to the momentum cone conditions.

We are now ready to state the main result of this section.
\begin{Prp} \label{prpinnermost}
We choose a parameter~$\sigma$ in the range
\[ 8 \;<\; \sigma \;<\; 9\: . \]
For any~$c_0, c_1 \in \R$, there is a family of fermionic
projectors~$(P^\varepsilon)_{\varepsilon>0}$ which satisfies
the conditions of Propositions~\ref{prpos} and~\ref{prpil}
and moreover has the following properties. For any~$r$ in the
range~$r_0 < r < r_4$ with
\[ r_0 \gg \varepsilon^{\frac{3}{104}\:(3+2 \sigma)} \:,\qquad
r_4 \ll \varepsilon^{-\frac{1}{832}\:(3+2\sigma)} \:, \]
the layer
\beq \label{s4def}
|s-s_4| \;<\; \frac{\Delta s_4}{2} \qquad {\mbox{with}} \qquad
s_4 \;=\; r^{-\frac{13}{3+2 \sigma}}\: \varepsilon^{\frac{31}{64}}
\:,\qquad\;
\Delta s_4 \;=\; r^{-\frac{24 + 3 \sigma}{3+2\sigma}}\:\varepsilon
\eeq
is an inner layer in the sense that it is vector dominated,
whereas the regions outside this layer and in the
range~$\varepsilon \ll s < s_3$ are bilinear dominated.
The family~$(P^\varepsilon)_{\varepsilon>0}$ satisfies the momentum cone conditions,
including the mass and regularization corrections.
\end{Prp}
{\Proof} We begin with the regularization functions as in the
proof of Propositions~\ref{prpos} and~\ref{prpil} and keep
adding regularization tails. We first make the ansatz
\begin{eqnarray}
\sum_{\beta=1}^3 h_\beta(\omega) &\asymp&
\kappa_7 \:\hRf^{p_4, q_4}\!\left(\varepsilon^{\frac{31}{64}}, 
\varepsilon^{\frac{7}{64}}, \sigma,
-\omega \right) \label{h4tail} \\
\sum_{\beta=1}^3 g_\beta(\omega) &\asymp&
\delta_7 \:\hRf^{p_4, q_4}\!\left(\varepsilon^\frac{1}{2}, \frac{6+17 \sigma}{13},
-\omega \right) \\
\sum_{\beta=1}^3 f_\beta(\omega) &\asymp&
\nu_7 \:\hRf^{p_4, q_4}\!\left(\varepsilon^\frac{1}{2}, \frac{3+15 \sigma}{13}, -\omega \right) \label{h4tailz}
\end{eqnarray}
and choose~$\nu_7$ such that the corresponding bilinear component~(\ref{Abinn})
vanishes at~$s_4$. Then we choose~$\delta_7$ and~$\kappa_7$ so as to
satisfy the momentum cone conditions~(\ref{cv2}) and~(\ref{cb}).
More precisely, abbreviating the integrals in~(\ref{cv2}, \ref{cb}) by
\[ X \;=\; \int_0^\infty \frac{\M^0}{s}\:ds\:,\qquad
Y \;=\; \int_0^\infty \frac{\M^b}{s^2}\:ds\:, \]
we arrange that the innermost layer gives the following contributions,
\beq \label{XYscal}
X \;\sim\; r^3\: \varepsilon^{-\frac{59}{128}}\:,\qquad
Y \;\sim\; r\: \varepsilon^{-\frac{3}{16}}\:.
\eeq
The just determined parameters scale in~$\varepsilon$ as follows,
\[ \kappa_7 \;\sim\; \varepsilon^{\frac{62 \sigma-1}{128}}\:,\qquad
\delta_7 \;\sim\; \varepsilon^{\frac{527 \sigma-48}{832}}\:,\qquad
\nu_7 \;\sim\; \varepsilon^{\frac{465 \sigma-141}{832}}\:. \]

Next we need to compensate for the singular contributions of the outer
strip and the intermediate layers to~$X$ and~$Y$. We explain the method
only for the most singular contributions; all the other terms can be
compensated similarly. The leading contributions to~$X$ and~$Y$ comes
from the intermediate strip at~$s_3$,
\beq \label{XYerror}
X \;\asymp\; r^{-\frac{2+3\alpha+\vartheta}{2 \alpha}}\:
\varepsilon^{-\frac{19}{128}}\:,\qquad
Y \;\asymp\; r^{\frac{2-3\alpha+\vartheta+4 \tau + 2 \vartheta \tau}
{2 \alpha}}\:
\varepsilon^{-\frac{21}{128}}\:.
\eeq
In order to compensate for these contributions, we proceed as follows.
Introducing similar to~(\ref{abdef}) the notation
\[ a \;=\; A^0(s_4) \spc {\mbox{and}} \spc b \;=\; (A^b)'(s_4)\:, \]
the contributions of the innermost layer to~$X$ and~$Y$ can be
expressed by
\[ X(a,b) \;\asymp\; \frac{a^2}{b \sqrt{s_4 r}}\:,\spc
Y(a,b) \;\asymp\; \frac{a^3}{b^2 \,(s_4 r)^\frac{3}{2}}\:. \]
By perturbing~$a$ and~$b$ one can easily perturb these contributions.
It is convenient, however, to keep the zero~$s_4$ of the bilinear
component fixed, because perturbing~$s_4$ would have a rather complicated
influence on~$a$ and~$b$. We thus introduce the following tails,
\begin{eqnarray}
\sum_{\beta=1}^3 h_\beta(\omega) &\asymp&
\kappa_8
\left[ \hRf^{p_4, q_4}\!\left(\varepsilon^\frac{1}{2},
\varepsilon^{\frac{7}{64}}, \vartheta_h, -\omega \right)
+ l_{h8} \,\hRflog^{p_4, q_4}\!\left(\varepsilon^\frac{1}{2},
\varepsilon^{\frac{7}{64}}, \vartheta_h, -\omega \right) \right]
\label{TailA} \\
\sum_{\beta=1}^3 g_\beta(\omega) &\asymp&
\delta_8 \left[ \hRf^{p_4, q_4}\!\left(\varepsilon^\frac{1}{2}, \vartheta_g, -\omega \right) + l_{g8} \,\hRflog^{p_4, q_4}\!\left(\varepsilon^\frac{1}{2}, \vartheta_g, -\omega \right) \right] \\
\sum_{\beta=1}^3 f_\beta(\omega) &\asymp&
\nu_8 \left[ \hRf^{p_4, q_4}\!\left(\varepsilon^\frac{1}{2}, \vartheta_f, -\omega \right) + l_{f8} \,\hRflog^{p_4, q_4}\!\left(\varepsilon^\frac{1}{2}, \vartheta_f, -\omega \right) \right] . \label{TailZ}
\end{eqnarray}
We determine~$\nu_8$, $\vartheta_f$ and~$l_{f8}$ such that the
bilinear contribution of these tails vanishes at~$s_4$.
To compensate for the contribution to~$X$ in~(\ref{XYerror}), we then choose~$\delta_8$,
$\vartheta_g$ and~$l_{g8}$ such that the perturbation of~$Y$
vanishes. We finally determine~$\kappa_8$, $\vartheta_h$ and~$l_{h8}$
such that the perturbation of~$X$ just cancels the term on the left
of~(\ref{XYerror}). Likewise, to compensate for~$Y$ in~(\ref{XYerror}),
we can use the free parameters~$\kappa_8$, $\vartheta_h$ and~$l_{h8}$.

Next we need to consider all the error terms. By choosing the
parameters~$p_3$ (in Proposition~\ref{prpil}) and~$q_4$
sufficiently large, we can arrange that the tails
considered in Proposition~\ref{prpil} as well as the error
terms of the tails considered in the proof of the present Proposition
do not contribute to the moments in the layer at~$s_3$.
The error terms resulting from expansions in~$\Delta s_4/s_4$
and in~$s/r$ are all compensated in a straightforward way by
suitable regularization tails.

It remains to satisfy the mass and regularization corrections to the
momentum cone conditions. The regularization corrections tend to
zero as~$\varepsilon \searrow 0$. The mass corrections can be
compensated in a straightforward manner. The only difficulty is
that the tangential derivatives like in~(\ref{tangder}) destroy the
general form~(\ref{Lorentzform}) of the vector component. But the
resulting radial error term can be compensated in a straightforward
way by a contribution~(\ref{notl}) which violates~(\ref{Lorentzform}).
\QED

\section{The Regularization Tails near the Origin} \label{sec6a}
\setcounter{equation}{0}
In the calculations of Sections~\ref{sec4}--\ref{sec6z} we took into
account only the leading order in~$s/r$, and therefore these calculations
do not apply near~$r=0$. In this section we shall analyze the effect
of the regularization tails near the origin. More precisely, we will
compute~$\M[A^\varepsilon_{xy}]$ asymptotically near~$x=y$, and we will
match these asymptotics to the outer strips and inner layers of the
preceding sections.
Our starting point is Lemma~\ref{lemma1}. Expanding in powers of~$r$,
all the negative powers of~$r$ cancel, and we obtain to leading order
\begin{eqnarray*}
P^\varepsilon(x,y) &=& \sum_{\beta=1}^3
\int_{\Xi_\beta} 2 K_\beta(\omega) \, h_\beta(\omega)
\: e^{i \omega t} \, d\omega \:+\: {\mathcal{O}}(r) \\
&&+\gamma^0 \sum_{\beta=1}^3 \int_{\Xi_\beta}
2 K_\beta(\omega)\, (f_\beta(\omega)+\omega \:g_\beta(\omega))\:
e^{i \omega t}\: d\omega \:+\: \gamma^0 {\mathcal{O}}(r)\\
&&+ \gamma^r \sum_{\beta=1}^3 \int_{\Xi_\beta}
\frac{2ir}{3}\, K_\beta(\omega)^3\, g_\beta(\omega)\:
e^{i \omega t}\: d\omega \:+\: \gamma^r {\mathcal{O}}(r^2)\:.
\end{eqnarray*}
Note that the leading radial contribution is linear in~$r$. This
fact, which can already be understood from the smoothness
of~$P^\varepsilon$ (note that $r \gamma^r = \vec{\xi} \vec{\gamma}$ is
smooth, whereas $\gamma^r = \vec{\xi} \vec{\gamma}/|\vec{\xi}|$ is not),
has an important consequence: Since at~$r=0$ the radial component
of~$P^\varepsilon$ vanishes, the bilinear
component of~$A^\varepsilon_{xy}$ is also zero.
The vector component of~$A^\varepsilon_{xy}$, on the other hand,
is proportional to~$\gamma^0$.
Combining these two observations, we conclude that the {\em{line~$r=0$ is
vector dominated}}.

Let us analyze in more detail how the outer strip and the inner layers behave
asymptotically near the line~$r=0$. The {\em{outer strip}} was constructed using
the regularization tail~(\ref{greg}), which led to a bilinear component
of~$A^\varepsilon_{xy}$, (\ref{traceless1}). The boundary~$s_1$ of the
outer strip separated the corresponding bilinear and vector dominated
regimes~(\ref{Min}) (all the other regularization tails in Section~\ref{sec4}
affected the details of the outer strip, but are irrelevant here).
Taking into account the regularization tail~(\ref{greg}), the
traceless part of the closed chain near the origin has the form
\begin{eqnarray}
A^\varepsilon_{xy} - \frac{1}{4}\, \Tr(A^\varepsilon_{xy}) &\asymp&
-\frac{m^3}{64 \pi^5}\: \frac{\gamma^0}{|t|^3} \left(1+{\mathcal{O}}(r) \right) \:+\: \gamma^r \:{\mathcal{O}}(r) \nonumber \\
&&\:+\: \frac{\delta r}{3 \pi^3}\:\epsilon(t)
\: \frac{i \gamma^0 \gamma^r}{|t|^{6+\gamma}}\;
\gamma(1-\gamma^2)\, \cos(\pi \gamma/2) \left(1+{\mathcal{O}}(r) \right) .
\label{traceless4}
\end{eqnarray}
The corresponding boundary between the vector and bilinear dominated regimes
is given by
\beq \label{ror}
r(t) \;=\; \frac{3m^3\, \sec(\pi \gamma/2)}{64 \pi^2 \,\gamma\,(\gamma^2-1)}\:
\frac{|t|^{3+\gamma}}{\delta}\:\left(1+{\mathcal{O}}(t) \right)\:.
\eeq
We thus obtain a vector dominated cone centered at the origin, which has zero opening angle. We refer to this cone as the {\em{vector dominated cusp}}
at the origin. Our expansion near~$r=0$ can be matched to the expansion in powers of~$s/r$ as performed in Section~\ref{sec4}. Namely, the
expansion~(\ref{ror}) is valid if
\[ r \;\lesssim\; t \qquad {\mbox{and thus}} \qquad
r,t \;\lesssim\; \delta^{\frac{1}{2+\gamma}}\;. \]
The formula~(\ref{Min}), on the other hand, holds in the range
\[ s \;\lesssim\; r \qquad {\mbox{and thus}} \qquad
r \;\gtrsim\; \delta^{\frac{1}{2+\gamma}} \:. \]

We next discuss the effect of an additional scalar regularization
tail of the form~(\ref{hreg}). Its contribution to the trace-free
part of the closed chain is computed to be
\beq \label{traceless5}
A^\varepsilon_{xy} - \frac{1}{4} \, \Tr(A^\varepsilon_{xy}) \;\asymp\;
\frac{\kappa m}{\pi^3}\:
\alpha \cos(\pi \alpha/2) \:\gamma^0\: |t|^{-\alpha-4}
\:+\: {\mathcal{O}}(r) \:.
\eeq
Assuming that~$\alpha<\gamma+2$ and that~$|t| \ll \kappa^{\frac{1}{1+\alpha}}$, from~(\ref{traceless4}, \ref{traceless5}) we again find
a vector dominated cusp which scales like
\[ r(t) \;\sim\; \frac{\kappa}{\delta}\: |t|^{2+\gamma-\alpha}\:. \]
Interestingly, we can arrange that the vector dominated cusp gives
singular contributions to~$\M[A^\varepsilon_{xy}]$ and/or to the
momentum cone conditions, without affecting the outer strip and the inner
layers. Here we merely explain the basic method, which will be very
useful in order to {\em{compensate for singular contributions at the origin}}
(see Section~\ref{sec6} for details).
By choosing the vector and scalar tails~(\ref{greg}, \ref{hreg}) appropriately,
we want to arrange a contribution to ${\mathcal{M}}[A^\varepsilon_{xy}]$
near the origin~$x \approx y$, without getting a contribution on the light
cone away from the origin (in Section~\ref{sec6}, (\ref{hreg}) will be an additional
scalar tail, and the following argument will make sure that this additional tail
has no effect on the outer strip and the inner layers which are already present).
Hence a neighborhood of the light cone should be bilinear dominated,
and thus the bilinear component in~(\ref{traceless1}) should dominate the vector
component~(\ref{traceless2}). This can be achieved most easily
by choosing
\[ \alpha \;=\; \gamma+ \frac{1}{2}
\spc {\mbox{and}} \spc \kappa \;\ll\; \delta\:. \]
Then the boundary of the vector dominated cusp has the scaling
\[ r(t) \;\sim\; \frac{\kappa}{\delta}\: |t|^{\frac{3}{2}} \:,\spc
t(r) \;\sim\; \left( \frac{\delta r}{\kappa} \right)^{\frac{2}{3}}\: , \]
and thus
\begin{eqnarray*}
r^2 \int_0^\infty \M[A^\varepsilon_{xy}]\: dt
&\sim& r^2 \int_{t(r)}^\infty A^0\: d\tau
\;\sim\; r^2\; \kappa\: t(r)^{-\alpha-3} \;\sim\;
\kappa^{3+\frac{2 \alpha}{3}}\, \delta^{-2-\frac{2\alpha}{3}}\:
r^{-\frac{2\alpha}{3}} \\
r^2 \int_0^\infty \frac{\M[A^\varepsilon_{xy}]}{t^2-r^2}\: dt
&\sim& r^2 \int_{t(r)}^\infty \frac{A^0}{\tau^2}\: d\tau
\;\sim\; r^2\; \kappa\: t(r)^{-\alpha-5} \;\sim\;
\kappa^{\frac{13+2 \alpha}{3}}\, \delta^{-\frac{10+2\alpha}{3}}\:
r^{-\frac{4+2\alpha}{3}}\:.
\end{eqnarray*}
The point is that, after choosing~$\alpha$ sufficiently large, we obtain
non-integrable poles at~$r=0$. 
By arranging that the scalar tail is present up to some small
radius~$\tilde{r}>0$, we can thus make the contribution of the
vector dominated cusp to the above expressions arbitrarily large
in the distributional sense.

\begin{figure}[tb]
\begin{center}
\begin{picture}(0,0)%
\includegraphics{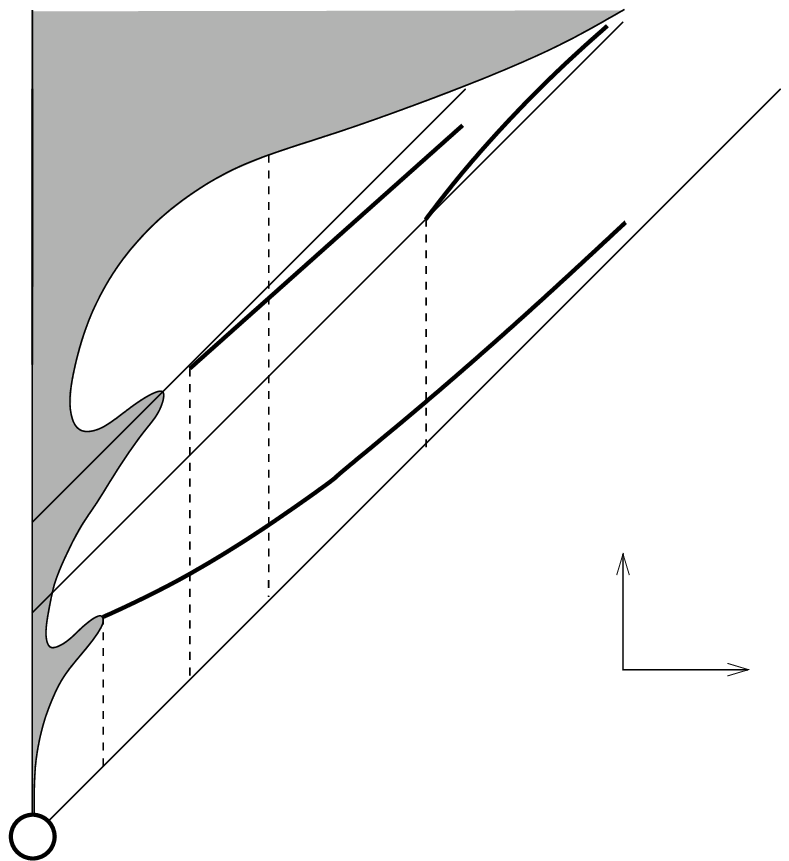}%
\end{picture}%
\setlength{\unitlength}{1657sp}%
\begingroup\makeatletter\ifx\SetFigFont\undefined%
\gdef\SetFigFont#1#2#3#4#5{%
  \reset@font\fontsize{#1}{#2pt}%
  \fontfamily{#3}\fontseries{#4}\fontshape{#5}%
  \selectfont}%
\fi\endgroup%
\begin{picture}(10072,10154)(-1049,-8435)
\put(7484,1380){\makebox(0,0)[lb]{\smash{{\SetFigFont{11}{13.2}{\familydefault}{\mddefault}{\updefault}$s_1$}}}}
\put(7471,-5191){\makebox(0,0)[lb]{\smash{{\SetFigFont{11}{13.2}{\familydefault}{\mddefault}{\updefault}$t$}}}}
\put(8371,-5866){\makebox(0,0)[lb]{\smash{{\SetFigFont{11}{13.2}{\familydefault}{\mddefault}{\updefault}$r$}}}}
\put(1261,-7711){\makebox(0,0)[lb]{\smash{{\SetFigFont{11}{13.2}{\familydefault}{\mddefault}{\updefault}$\varepsilon^{\frac{31(3+2\sigma)}{128(8+\sigma)}}$}}}}
\put(2251,-6721){\makebox(0,0)[lb]{\smash{{\SetFigFont{11}{13.2}{\familydefault}{\mddefault}{\updefault}$\varepsilon^{\frac{\alpha}{64}}$}}}}
\put(3241,-5866){\makebox(0,0)[lb]{\smash{{\SetFigFont{11}{13.2}{\familydefault}{\mddefault}{\updefault}$\varepsilon^{\frac{\alpha}{64(1+\alpha)}}$}}}}
\put(5446,119){\makebox(0,0)[lb]{\smash{{\SetFigFont{11}{13.2}{\familydefault}{\mddefault}{\updefault}$s_2$}}}}
\put(6886,524){\makebox(0,0)[lb]{\smash{{\SetFigFont{11}{13.2}{\familydefault}{\mddefault}{\updefault}$s_3$}}}}
\put(7111,-916){\makebox(0,0)[lb]{\smash{{\SetFigFont{11}{13.2}{\familydefault}{\mddefault}{\updefault}$s_4$}}}}
\put(-1034,-466){\makebox(0,0)[lb]{\smash{{\SetFigFont{11}{13.2}{\familydefault}{\mddefault}{\updefault}$\varepsilon^{\frac{\alpha}{64(1+\alpha)}}$}}}}
\put(-629,-4606){\makebox(0,0)[lb]{\smash{{\SetFigFont{11}{13.2}{\familydefault}{\mddefault}{\updefault}$\varepsilon^{\frac{3}{64}}$}}}}
\put(-629,-5641){\makebox(0,0)[lb]{\smash{{\SetFigFont{11}{13.2}{\familydefault}{\mddefault}{\updefault}$\varepsilon^{\frac{7}{64}}$}}}}
\put(4861,-4246){\makebox(0,0)[lb]{\smash{{\SetFigFont{11}{13.2}{\familydefault}{\mddefault}{\updefault}$\varepsilon^{\frac{\alpha}{64(2+\vartheta)}}$}}}}
\end{picture}%
\caption{Behavior of $\M[A^\varepsilon_{xy}]$ near the origin.}
\label{figorigin}
\end{center}
\end{figure}

We next analyze the {\em{inner layers}} near the origin.
We thus consider the tails~(\ref{vttails}).
Near the origin, the bilinear component of the closed chain becomes
\[ A^b \;=\; -\frac{\delta}{3 \pi^3}\: \gamma\,(\gamma^2-1)\: \cos(\pi \gamma/2)\:
r\: |t|^{-6-\gamma} \:\epsilon(t)
+ \frac{\nu}{\pi^3}\:\beta\, \sin(\pi \beta/2)\:r\: |t|^{-5-\beta}\: \epsilon(t)
\:+\: {\mathcal{O}}(r^2) \,. \]
The zero of the bilinear component is at fixed time
\[ t \;=\; \left(
\frac{3 \beta\: \sin(\pi \beta/2)}{\gamma\,(\gamma^2-1)\: \cos(\pi \gamma/2)}\:
 \frac{\nu}{\delta} \right)^{\frac{1}{\beta-\gamma-1}} \:.  \]
This asymptotic form is again consistent with the expansion in powers of~$s/r$,
(\ref{s2rel}).
Taking into account the scalar tail~(\ref{hreg}), we obtain again
a vector dominated region near~$r=0$.
The relevant scalings are shown in Figure~\ref{figorigin}, where we
also indicate the possible singular contributions at the origin
by a white circle.

\section{The Regularization Tails near Infinity} \label{sec6b}
\setcounter{equation}{0}
The structure of the outer strip and of the inner layers persists
only up to maximal radii~$r_1, \ldots, r_4$
(see Propositions~\ref{prpos}, \ref{prpil}
and~\ref{prpinnermost}), and therefore we must analyze the behavior
of~$\M[A^\varepsilon_{xy}]$ for even larger radii. The basic difficulty
for large~$\xi$ can already be understood by considering the
{\em{outer layer}} of Proposition~\ref{prpos}, whose boundary is given by
\beq \label{osb}
s_1 \;=\; \varepsilon^{\frac{1}{64}}\, r^{-\frac{1}{\alpha}}\:.
\eeq
In Section~\ref{sec4} we made use of the mass expansion, and
thus~(\ref{osb}) holds only if~$s_1 \,r \ll m^{-2}$. This leads to the
condition
\[ \varepsilon^{\frac{1}{64}}\: r^{1-\frac{1}{\alpha}} \;\ll\; m^{-2}
\qquad {\mbox{or equivalently}} \qquad
r \;\ll\; \varepsilon^{-\frac{\alpha}{64\,(\alpha+1)}}\:. \]
Thus if~$r \gg r_1$, the bilinear dominated regime enters the region
$r s \gg m^{-2}$ where the fermionic projector has an oscillatory
behavior. These oscillations of~$P(x,y)$ also lead to an oscillatory
behavior of the closed chain, making it difficult to describe the
bilinear dominated region explicitly.

In order to bypass this difficulty, we shall now modify the fermionic
projector by changing the direction of the vector field~$k\slsh$
in~(\ref{Pansatz}). In agreement with the more general notion
introduced in~\cite[\S~4.4]{PFP}, we refer to this mechanism
as introducing a {\em{shear of the surface states}}.
The basic idea already becomes clear if we consider the unregularized
fermionic projector and ``shorten'' the Dirac matrix $\gamma^0$
by the transformation
\beq \label{g0short}
\gamma^0 \;\longrightarrow\; (1-\theta)\, \gamma^0
\eeq
with a positive parameter~$\theta \ll 1$. Clearly, this transformation
does not change the singular set~$t=\pm r$ of the
distribution~$P(x,y)$. But it does change the square of the
matrix~$\xi\slsh$,
\[ (\xi\slsh)^2 \;\longrightarrow\; (1-\theta)^2 \,t^2 - r^2\:. \]
As a consequence, computing~$\M[A_{xy}]$ naively similar
to~(\ref{3}), we find that
\[ \M[A_{xy}]=0 \quad {\mbox{if}} \quad
|t| < \frac{r}{1-\theta}\:. \]
In particular, $\M[A_{xy}]$ now vanishes identically in a neighborhood
of the light cone. For our purpose, it is very helpful that the
boundary of the region where~$\M[A_{xy}]$ vanishes is easy to describe:
it consists simply of the two cones~$t=\pm r/(1-\theta)$.

The transformation~(\ref{g0short}) can also be expressed by
choosing the functions~$f_\beta$ in the spherically symmetric
regularization~(\ref{Pansatz}) as
\beq \label{fb}
f_\beta(\omega) \;=\; - \frac{\delta}{16 \pi^3}\: \omega\:.
\eeq
This transformation is too simple for our application, because
we want~(\ref{g0short}) to be active only in
the region~$s \gg \epsilon^{\frac{3}{8}}$, so that it affects the
outer strip, but not the inner layers. To this end, we choose
\[ r_\infty \;=\; \varepsilon^{-\frac{\alpha}{64}} \:, \qquad
\theta \;=\; \frac{s_1(r_\infty)}{r_\infty}
\;=\; \varepsilon^{\frac{1}{32}+\frac{\alpha}{64}} \]
and introduce a contribution to the functions~$f_\beta$ of the form
\beq \label{finf}
f_\beta \;\asymp\;  -\frac{\theta}{16 \pi^3} \:
\hRf^{p_\infty, q_\infty} \!\left(\varepsilon^{\frac{3}{64}}, 2, \omega
\right) ,
\eeq
where~$p_\infty, q_\infty$ are integer parameters.
This has the following effect. If $s \ll \varepsilon^\frac{3}{64}$,
the Fourier transform~$\hRf^{p_\infty, q_\infty}(s)$
of~$\Rf^{p_\infty, q_\infty}(s)$ decays
like~$(s/\varepsilon^\frac{3}{64})^{p_\infty}$
(see Lemma~\ref{prpreg2} and Figure~\ref{fig4}). Thus by
by choosing~$p_\infty$ sufficiently large we can arrange
that~(\ref{finf}) has no effect on the inner layers.
However, if~$s \ll \varepsilon^{\frac{3}{64}}$, only the behavior
of~$\Rf^{p_\infty, q_\infty}$ for~$\omega \ll \varepsilon^{-\frac{3}{64}}$
is relevant. Considering the asymptotics of~(\ref{Rfdef})
for small~$|\omega|$, we get agreement with~(\ref{fb}), with an
error term which can be made arbitrarily small by increasing~$q_\infty$.
Hence the regularization functions~(\ref{finf}) influence the outer
strip exactly as~(\ref{fb}), but have no effect on the inner layers.

\begin{figure}[tb]
\begin{center}
\begin{picture}(0,0)%
\includegraphics{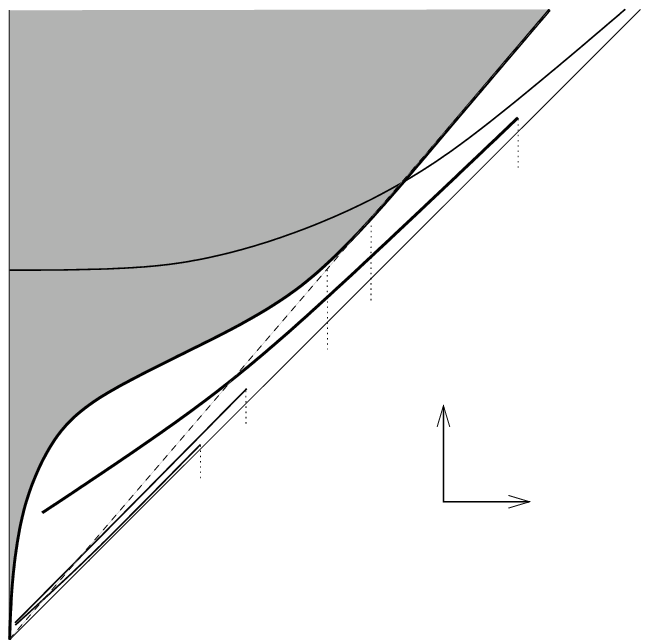}%
\end{picture}%
\setlength{\unitlength}{1657sp}%
\begingroup\makeatletter\ifx\SetFigFont\undefined%
\gdef\SetFigFont#1#2#3#4#5{%
  \reset@font\fontsize{#1}{#2pt}%
  \fontfamily{#3}\fontseries{#4}\fontshape{#5}%
  \selectfont}%
\fi\endgroup%
\begin{picture}(7822,7866)(1191,-8191)
\put(3824,-6787){\makebox(0,0)[lb]{\smash{{\SetFigFont{11}{13.2}{\familydefault}{\mddefault}{\updefault}$r_4$}}}}
\put(4315,-6174){\makebox(0,0)[lb]{\smash{{\SetFigFont{11}{13.2}{\familydefault}{\mddefault}{\updefault}$r_3$}}}}
\put(5179,-5264){\makebox(0,0)[lb]{\smash{{\SetFigFont{11}{13.2}{\familydefault}{\mddefault}{\updefault}$r_\infty$}}}}
\put(1206,-5824){\makebox(0,0)[lb]{\smash{{\SetFigFont{11}{13.2}{\familydefault}{\mddefault}{\updefault}$s_3$}}}}
\put(7451,-3286){\makebox(0,0)[lb]{\smash{{\SetFigFont{11}{13.2}{\familydefault}{\mddefault}{\updefault}$r_2$}}}}
\put(5745,-4728){\makebox(0,0)[lb]{\smash{{\SetFigFont{11}{13.2}{\familydefault}{\mddefault}{\updefault}$r_1$}}}}
\put(2449,-6797){\makebox(0,0)[lb]{\smash{{\SetFigFont{11}{13.2}{\familydefault}{\mddefault}{\updefault}$s_2$}}}}
\put(7949,-652){\makebox(0,0)[lb]{\smash{{\SetFigFont{11}{13.2}{\familydefault}{\mddefault}{\updefault}$s=\Theta r$}}}}
\put(6099,-1486){\makebox(0,0)[lb]{\smash{{\SetFigFont{11}{13.2}{\familydefault}{\mddefault}{\updefault}$s=\Theta r$}}}}
\put(7001,-5761){\makebox(0,0)[lb]{\smash{{\SetFigFont{11}{13.2}{\familydefault}{\mddefault}{\updefault}$t$}}}}
\put(7563,-6211){\makebox(0,0)[lb]{\smash{{\SetFigFont{11}{13.2}{\familydefault}{\mddefault}{\updefault}$r$}}}}
\put(2232,-3586){\makebox(0,0)[lb]{\smash{{\SetFigFont{11}{13.2}{\familydefault}{\mddefault}{\updefault}$\xi^2=m^{-2}$}}}}
\put(2791,-5011){\makebox(0,0)[lb]{\smash{{\SetFigFont{11}{13.2}{\familydefault}{\mddefault}{\updefault}$s_1$}}}}
\end{picture}%
\caption{Behavior of $\M[A^\varepsilon_{xy}]$ near infinity.}
\label{figinf}
\end{center}
\end{figure}

In order to quantify the influence of~(\ref{finf}) on the outer strip,
we now compute the boundary~$s_1$ of the bilinear and vector dominated
regions in the asymptotic regime
\[ \varepsilon^{-\frac{\alpha}{64}} =
r_\infty \;\ll\; r \;\ll\; (m^2\, \theta)^{-\frac{1}{2}} =
\frac{1}{m}\: \varepsilon^{-\frac{1}{64} - \frac{\alpha}{128}}\:, \]
where~$\theta$ plays in important role, but nevertheless the mass
expansion can be used. In this regime~(\ref{traceless1}) is modified to
\begin{eqnarray*}
A^\varepsilon_{xy} - \frac{1}{4}\, \Tr(A^\varepsilon_{xy}) &\asymp&
\frac{\delta}{8 \pi^3}\:(1-\theta)\: \frac{i \gamma^0 \gamma^r}{r^3} \:(1-\gamma)\: |s|^{-\gamma-2}\:
\cos(\pi \gamma/2) \nonumber \\
&&-\frac{m^3}{256 \pi^5}\: \frac{1}{r^2 s^2}\:\Theta(s)
\left((1-\theta)\, t \gamma^0 - r \gamma^r \right) \:.
\end{eqnarray*}
Writing $s=\theta r + \Delta s$, the correction~$\Delta s$ is to
leading order determined from the equation
\[ \delta\: \frac{(\theta r)^{-\gamma-2}}{r^3} \;\sim\;
\frac{1}{r^2\, (\theta r)^2}\: \sqrt{\Delta s\: r}\:, \]
and thus
\[ s_1 \;=\; \theta r + \delta^2\, r^{-2\gamma-3} \, \theta^{-2\gamma}
\:+\: \cdots \:. \]
This shows that the function~$s_1(r) - \theta r$ tends to zero fast
and becomes negligible before~$s_1$ enters the oscillatory region of the
fermionic projector, see Figure~\ref{figinf}.

We next consider the behavior of the {\em{inner layers}} for large~$r$.
According to the above construction, they are not affected
by the regularization tail~(\ref{finf}).
The layers at~$s_2$ and~$s_4$ have the property that~$s$
decreases for increasing~$r$ (see~(\ref{s2def}) and~(\ref{s4def})).
According to~(\ref{Th2}, \ref{h4tail}), we choose the length scale
of the scalar regularization tail such that this  tail dies off
at a smaller radius than the vector tails. As a consequence,
the corresponding inner layers
``fade away smoothly'' on the scales~$r_2$ and~$r_4$, respectively.
For the layer at~$s_3$, the situation is somewhat different
because~$s_3$ increases with~$r$. Due to the second scale in~(\ref{Th3}),
the corresponding scalar regularization tail dies off if~$s \gtrsim
\varepsilon^{\frac{11}{128}}$ (see the right of Figure~\ref{fig4}).
We thus conclude that the inner layer at~$s_3$
also fades away smoothly on the scale~$r_3$.

We finally remark that, by changing the scales of the scalar
regularization tails~(\ref{Th2}, \ref{Th3}, \ref{h4tail}),
we could still modify the radii~$r_2$, $r_3$ and~$r_4$. But we will not need
this freedom here.

\section{The Continuum Limits of~$\M$ and~$\M \!\cdot\! P$} \label{sec6}
\setcounter{equation}{0}
We are now ready to prove the main theorem. \\[.5em]
{\em{Proof of Theorem~\ref{thm1}. }}
We consider a family of regularizations~$(P_\varepsilon)_{\varepsilon>0}$,
to which Propositions~\ref{prpos}, \ref{prpil} and~\ref{prpinnermost}
apply, and in addition the shear of the surface states~(\ref{finf}) is
present. Then clearly~(1) holds. To prove~(2), for a given
test function~$f \in {\mathcal{S}}(\R^4)$ we must show that
\beq \label{lim0}
\lim_{\varepsilon \searrow 0} \int \left(\M[A^\varepsilon_{xy}] - \tM(\xi)
\right) f(\xi)\: d^4 \xi \;=\; 0 \:.
\eeq
In order to first analyze the situation in a compact set away from the
origin, we consider for given parameters~$\delta, \gamma \geq 0$ the
integral
\[ \int \left(\M[A^\varepsilon_{xy}] - \tM(\xi)
\right) f(\xi) \left( \eta_{\varepsilon^{-\gamma}}(\xi) - \eta_\delta(\xi)
\right) d^4 \xi \:. \]
Obviously, in any compact set away from the light cone,
the function~$\M[A^\varepsilon_{xy}]$ tends to~$\tM(\xi)$ 
as~$\varepsilon \searrow 0$. Moreover, comparing Lemma~\ref{lemma21}
with Propositions~\ref{prpos} and~\ref{prpil}, one sees
that~$\M[A^\varepsilon_{xy}]-\tM(\xi)$ tends to zero on the light cone
in the distributional sense. This means that for every~$\delta>0$
and after choosing~$\gamma=0$,
\beq \label{lim1}
\lim_{\varepsilon \searrow 0} \int \left(\M[A^\varepsilon_{xy}] - \tM(\xi)
\right) f(\xi) \Big( \eta_{\varepsilon^{-\gamma}}(\xi) - \eta_\delta(\xi)
\Big) d^4 \xi \;=\; 0 \:.
\eeq
Since all the scales in our regularization are powers of~$\varepsilon$,
(\ref{lim1}) remains valid if~$\gamma$ is chosen positive, but
sufficiently small.

We next consider the integral
\beq \label{int2}
\int \left(\M[A^\varepsilon_{xy}] - \tM(\xi)
\right) f(\xi) \Big( 1 - \eta_{\varepsilon^{-\gamma}}(\xi)
\Big) d^4 \xi .
\eeq
The contribution by~$\tilde{M}$ clearly tends to zero as~$\varepsilon
\searrow 0$. To control the contribution by~$\M[A^\varepsilon_{xy}]$,
we first note that~$\M[A^\varepsilon_{xy}]$ is
obviously bounded by a negative power of~$\varepsilon$, i.e.\
there is $n \in \N$ such that
\[ \| \M[A^\varepsilon_{xy}] \| \;\leq\; C\, \varepsilon^{-n}
\spc {\mbox{for all~$\xi$}}. \]
This negative power of~$\varepsilon$ is compensated in~(\ref{int2})
by the rapid decay of~$f$. We conclude that
\beq \label{lim2}
\lim_{\varepsilon \searrow 0} \int \left(\M[A^\varepsilon_{xy}] - \tM(\xi)
\right) f(\xi) \Big( 1 - \eta_{\varepsilon^{-\gamma}}(\xi)
\Big) d^4 \xi \;=\; 0\:.
\eeq

To finish the proof of~(2) it remains to show that
\beq \label{lim3}
\lim_{\delta \searrow 0}\;
\lim_{\varepsilon \searrow 0} \int \left(\M[A^\varepsilon_{xy}] - \tM(\xi)
\right) f(\xi)\: \eta_\delta(\xi)\: d^4 \xi \;=\; 0 \:,
\eeq
because~(\ref{lim0}) then follows immediately by combining~(\ref{lim1}, \ref{lim2}, \ref{lim3}).
In view of Lemma~\ref{lemma21}, we only need to consider
the vector dominated cusps, see~(\ref{ror}) and Figure~\ref{figorigin}.
The leading contributions to the integral in~(\ref{lim3}) have the
following scaling,
\beq \label{116}
\int \left(\M[A^\varepsilon_{xy}] - \tM(\xi)
\right) f(\xi)\: \eta_\delta(\xi)\: d^4 \xi
\;\sim\; A^t\, t \,r^3\, f(0) + A^b\, t \,r^4\: \nabla f(0)\:.
\eeq
A short calculation shows that all these contributions tend to zero
as~$\varepsilon \searrow 0$. This concludes the proof of~(2).

Next we want to show that, for any~$k$ not on the mass cone,
$\hM^\varepsilon(k)$ converges even pointwise,
\beq \label{Mconvp}
\lim_{\varepsilon \searrow 0} \hM^\varepsilon(k) \;=\; \tM(k)
\qquad {\mbox{if $k^2 \neq 0$}}.
\eeq
To this end, we consider the Fourier integral
\[ \hM^\varepsilon(k) \;=\; \int \M[A^\varepsilon_{xy}]\,
e^{-i k \xi}\: d^4 \xi\:. \]
Choosing polar coordinates $(t,r,\vartheta, \varphi)$ and carrying
out the angular variables as in the proof of Lemma~\ref{lemma1}, we
obtain an expression involving two-dimensional Fourier transforms of the
following form:
\[ \int_{-\infty}^\infty dt \int_0^\infty r^p dr\: \M^\varepsilon(t,r)\:
e^{-i \omega t \pm i k r}\:,\qquad p \in \N_0\:. \]
For~$t$ and~$r$ in a compact set, we can argue exactly as
in~(\ref{lim1}) and~(\ref{lim3}). Hence it remains to consider the
integral near infinity,
\[ \int_0^\infty r^p  \:dr\: e^{-i (\omega \mp k) r}\:
\eta_{\varepsilon^{-\gamma}}(r)
\int_{-\infty}^\infty ds\: e^{-i \omega s}  \M^\varepsilon(t,r) \:, \]
where we again chose the light-cone coordinate~$s=t-r$.
Since~$\omega \mp k \neq 0$, we can use the identity
\[ e^{-i (\omega \mp k) r} \;=\; \frac{i}{\omega \mp k}\:
\frac{d}{dr} e^{-i (\omega \mp k) r} \]
and integrate by parts in the variable~$r$. This gives a scaling
factor~$\varepsilon^{\gamma}$. We iterate this procedure until
the expression scales like a positive power of~$\varepsilon$.
Then we can take the limit~$\varepsilon \searrow 0$ to obtain~(\ref{Mconvp}).

For the proof of~(3) we need to analyze similar to~(\ref{ci})
the following convolution integral,
\beq \label{convolv3}
\int \frac{d^4p}{(2 \pi)^4}\; \hM^\varepsilon(p)\: P^\varepsilon(q-p)\:.
\eeq
If~$P^\varepsilon$ were replaced by a family of distributions with compact support, the convolution integral as well as the limit~$\varepsilon \searrow 0$ would be well defined as the convolution of distributions. Using furthermore
the pointwise convergence~(\ref{Mconvp}), one sees that it
is indeed sufficient to verify the momentum cone conditions.
In Proposition~\ref{prpinnermost}, these conditions were satisfied
on the light cone, in an annulus~$\delta <r < \varepsilon^{-\gamma}$.
For large~$r > \varepsilon^{-\gamma}$ we can in~(\ref{Bsr}, \ref{Bvr}) iteratively substitute the identity
\[ e^{-i \Omega r} \;=\; \frac{i}{\Omega}\: \frac{d}{dr} e^{-i \Omega r} \]
and integrate by parts to show that the corresponding contribution
to the convolution integral~(\ref{convolv3}) tends to zero
as~$\varepsilon \searrow 0$.
Hence it remains to consider the momentum cone conditions
near the origin~$\xi=0$. Using the symmetries~(\ref{bsymm}, \ref{vsymm}),
the leading contributions have the following form,
\begin{eqnarray}
B^\varepsilon_{\mbox{\tiny{\rm{scal}}}} &\sim& 
\frac{A^t}{t^2}\; t\, r^3 \:+\: \frac{A^b\, r^2\, |\vec{k}|}{t^2}
\; t\, r^3 \label{te1} \\
B^\varepsilon_{\mbox{\tiny{\rm{vect}}}} &\sim&
\frac{A^b\, r}{t^4}\; t\, r^3\:+\: \frac{A^b\, r\: \Omega\,|\vec{k}|}{t^2}\;
t\, r^3\:. \label{te2}
\end{eqnarray}
These contributions actually diverge as~$\varepsilon \searrow 0$,
but we can compensate for them using the mechanism explained after~(\ref{traceless5}).
More precisely, we introduce the additional tails
\begin{eqnarray}
\sum_{\beta=1}^3 h_\beta(\omega) &\asymp&
\kappa_1\: \hRf^{p, q}\!\left(\varepsilon, \varepsilon^{\frac{45}{192}}, \frac{13}{8}, -\omega \right) \:+\:
\kappa_2\: \hRf^{p, q}\!\left(\varepsilon^{\frac{45}{192}}, \varepsilon^{\frac{43}{192}}, \frac{13}{8}, -\omega \right)
\label{lTailA} \\
\sum_{\beta=1}^3 g_\beta(\omega) &\asymp&
\delta_1\: \hRf^{p, q}\!\left(\varepsilon, \varepsilon^{\frac{45}{192}},
\frac{9}{8}, -\omega \right) \:+\:
\delta_2\: \hRf^{p, q}\!\left(\varepsilon^{\frac{45}{192}}, \varepsilon^{\frac{43}{192}},
\frac{9}{8}, -\omega \right) , \label{lTailZ}
\end{eqnarray}
and also introduce log-tails, with the relative prefactor chosen exactly as in
the tails leading to the singularities~(\ref{te1}, \ref{te2}). Since the log-tails
are straightforward, we do not consider them here.
By choosing the parameters~$\kappa_i$ and~$\delta_i$ appropriately,
these tails give rise to additional vector dominated cusps
at~$t \sim \varepsilon^{\frac{11}{48}}$ and~$t \sim
\varepsilon^{\frac{23}{96}}$, which do not contribute to~(\ref{116}),
but compensate for the
leading terms in~(\ref{te1}, \ref{te2}).
After choosing~$p$ and~$q$ sufficiently large,
all the correction terms can be treated by perturbing these two
vector dominated cusps.
The parameters~$\kappa_i$ and~$\delta_i$ can be chosen to have
the following scaling in~$\varepsilon$,
\[ \kappa_1 \;\sim\; \varepsilon^\frac{3977}{2688}\:,\quad
\delta_1 \;\sim\; \varepsilon^\frac{1625}{896}\:,\qquad
\kappa_2 \;\sim\; \varepsilon^\frac{1255}{768}\:,\quad
\delta_2 \;\sim\; \varepsilon^2\:.
 \]

Finally, we need to take into account that the contributions
of the innermost layer to the momentum cone conditions~(\ref{XYscal})
cease to exist if~$r \lesssim \varepsilon^\frac{5 (3+2 \sigma)}{832}$
(see Figure~\ref{figorigin}). However, this ``missing contribution''
to~$B^\varepsilon$ tends to zero as~$\varepsilon \searrow 0$.
This concludes the proof of~(3).

To prove~(4), we note that for~$q \not \in \overline{{\mathcal{C}}^\wedge}$,
the integration range in the convolution integral~(\ref{ci}) becomes
unbounded (cf.\ Figure~\ref{fig1}). For large~$p$, the
integrand in~(\ref{ci}) is easily computed to be a
{\em{positive}} definite matrix. The same is true for the
regularized integrand in~(\ref{convolv3}), showing that
the convolution integral~(\ref{convolv3}) diverges
as~$\varepsilon \searrow 0$ to~$+\infty$.
\QED

\section{Going Beyond the Distributional~$\M  \!\cdot\! P$-Product} \label{appB}
\setcounter{equation}{0}
We saw in Section~\ref{sec6a} that the regularization tails give us a lot of freedom
to modify~$\M[A^\varepsilon_{xy}]$ near the origin. We now use this freedom to
go beyond the distributional $\M P$-product. \\[1em]
{\mbox{\em{Proof of Theorem~\ref{thmbeyond}. }}}
Our task is to arrange additional contributions near the origin which leave the
continuum limit of~$\M[A^\varepsilon_{xy}]$ unchanged but in the
product~(\ref{Qxydef}) give rise to the extra contribution
\beq \label{extra}
\frac{1}{2}\, \lim_{\varepsilon \searrow 0} \M[A^\varepsilon_{xy}]\, P^\varepsilon(x,y)
\;\asymp\; c_2\, \delta^4(\xi) - c_3\, i \Pdd \delta^4(\xi) - c_4\, \Box \delta^4(\xi)\:.
\eeq
Our method is similar to that discussed in Section~\ref{sec6a} after~(\ref{traceless5}).
The main difference is that instead of working with the tails~$g$ and~$h$,
it is preferable here to use the tails~$f$ and~$h$. More precisely, we consider a
scalar regularization tail of the form~(\ref{hreg}) as well as the vector tail
\beq \label{freg} f(\omega) \;\asymp\; e^{\varepsilon \omega}\,
\frac{\nu}{\Gamma(\alpha-\frac{1}{2})}\: |\omega|^{\alpha-\frac{1}{2}-1}\; \Theta(-\omega)\:.
\eeq
Choosing
\beq \label{condkn}
\kappa \ll \nu\:,
\eeq
the region near the light cone is bilinear dominated, and thus
adding the above tails to the tails already considered earlier, in such a way that
the new tails are active
for even smaller values of~$s$, the multilayer structure near the light cone is not affected.
Near the origin, we get a vector dominated cusp of the form
\beq \label{rscale}
r \;\sim\; \frac{\kappa}{\nu}\: \sqrt{|t|}\:.
\eeq
We arrange the tails to be active on the scale~$|t| \sim \tau$. 
Then for the regularization tails
to be small, we must satisfy the conditions
\beq \label{optimal}
\kappa \;\ll\; \tau^{-1+\alpha} \:,\qquad \nu \;\ll\; \tau^{-\frac{3}{2}+\alpha}\:.
\eeq
Choosing~$\kappa$ and~$\tau$ according to this optimal scaling, we clearly satisfy~(\ref{condkn}),
whereas~(\ref{rscale}) yields~$r \sim \tau$. This shows that we can arrange that
the vector dominated cusp has a nonzero opening angle on the scale~$\tau$.
On this scale, we have the following asymptotic forms for composite expressions
\begin{eqnarray*}
A^\varepsilon_{xy} - \frac{1}{4} \, \Tr(A^\varepsilon_{xy}) &\asymp&
\xi\slsh\:r_0\:\kappa\:\tau^{-5-\alpha}\:\epsilon(t)
\:+\: \gamma^0 \:r_1\:\nu\:\tau^{-\frac{5}{2}-\alpha} \\
P^\varepsilon(x,y) &\asymp& i \xi\slsh\:r_2\:\tau^{-4} + \gamma^0\: c_0\: \nu\:
\tau^{-\frac{1}{2}-\alpha}
+r_3\: \tau^{-2}  + \kappa\:c_1\:\tau^{-1-\alpha}
\end{eqnarray*}
with real coefficients~$r_j$ and complex coefficients~$c_j$.
By direct inspection one sees that we can arrange terms which have
the same symmetry as the terms in~(\ref{extra}) and the correct scaling;
these terms are of the form~$\sim \xi\slsh \kappa^2$ and~$\sim \nu \kappa$.
Moreover, we have the freedom to choose an arbitrarily large number of pairs
of tails in~$f$ and~$h$ active on different scales~$\tau$.
Using all this freedom, it is straightforward to
verify that there are regularizations having all the required properties.
\QED
We remark that there is an alternative method for proving Theorem~\ref{thmbeyond}.
Namely, instead of considering the regularization tails, one can work with additional
high-energy states, which are not close to the mass cone or are no surface states
(for definitions and a discussion see~\cite[Chapter~4]{PFP}). Actually, working with
the high-energy states gives us more flexibility in modifying~$\M[A^\varepsilon_{xy}]$
near the origin. However, at least for regularizations which have the natural scaling
(see for example~\cite[eqn~(4.3.12)]{PFP}), we get the same results as with the
regularization tails. For this reason, here we shall not enter the construction of the
high-energy states. 

It is an obvious question whether one can arrange additional {\em{contributions}}
to~$Q$ supported {\em{on the light cone}}. More specifically, from the scaling the
following contributions seem possible,
\begin{eqnarray}
\sim \:\xi\slsh\: \delta(\xi^2)\, \epsilon(\xi^0) \:&,&\qquad
\sim \:\delta(\xi^2)\, \epsilon(\xi^0) \label{con2} \\
\sim i \xi\slsh\: \delta(\xi^2) \qquad\, \:&,&\qquad \sim \:\delta(\xi^2) \:. \label{con1}
\end{eqnarray}
Computing the Fourier transform, the contributions~(\ref{con2}) are supported on the
mass cone~$\{k^2=0\}$ and are thus not of interest. The terms~(\ref{con1}), however,
do contribute in the lower masse cone; more precisely, the corresponding contribution
to~$\hat{Q}$ can be written as
\[ \hat{Q}(k) \;\asymp\; c_5\: \frac{k\slsh}{k^4} \:+\: c_6\: \frac{1}{k^2} \qquad\qquad
(k \in {\mathcal{C}}^\wedge) \]
with real constants~$c_5$ and~$c_6$. A potential method for arranging these extra contributions
would be to construct additional layers whose contribution to~$\M[A^\varepsilon_{xy}]$
vanishes as~$\varepsilon \searrow 0$, but which, when multiplied by~$P^\varepsilon(x,y)$,
give rise to the terms~(\ref{con1}). Although we have no proof, we conjecture that this
cannot be accomplished. In any case, arranging the extra terms on the light cone~(\ref{con1})
is much more difficult than arranging the terms at the origin~(\ref{extra}).
One difficulty is that for contributions on the light cone one must get the scaling in the
radius right. Furthermore, a symmetry argument shows that the product~$\M[A^\varepsilon_{xy}]
P(x,y)$ can only give rise to the uninteresting contributions~(\ref{con2}). Thus to
obtain~(\ref{con1}) one needs to take into account the regularization tails of the
factor~$P^\varepsilon(x,y)$ in the product~$\M[A^\varepsilon_{xy}]\, P^\varepsilon(x,y)$.
This goes beyond our considerations used for the momentum cone conditions and does not seem easy.

Finally, it is a good question why at all we want to implement the
distributional $\M P$-product. This property is very useful because it makes it possible
to analyze the Euler-Lagrange equations in the continuum, but this is certainly not a compelling reason
for imposing it. Another fair approach would be to consider right away the regularized product
\beq \label{QE}
Q^\varepsilon(x,y) \;=\; \frac{1}{2}\: \M[A^\varepsilon_{xy}]\, P^\varepsilon(x,y)\:,
\eeq
and to try to choose the regularization tails such as to give the singularities of~(\ref{QE})
on the light cone a meaning in the distributional sense. We strongly conjecture that this
procedure does not work, although we again have no proof. Namely, giving~(\ref{QE})
a distributional meaning leads to the following serious difficulties. First, the pole
of~(\ref{QE}) on the light cone is of higher order than that of~$\M$, and thus we would have
to introduce more counterterms. Second, for fixed~$x$ and~$y$, (\ref{QE}) is a
non-selfadjoint matrix which also involves a scalar component. This again increases the
number of counterterms by more than a factor of two. In view of the difficulties
encountered in Sections~\ref{sec4}--\ref{sec6z}, it seems hopeless to construct all
these counterterms. This argument gives
a justification for the assumption of a distributional $\M P$-product which does not
refer to aesthetics nor usefulness: this assumption drastically simplifies the structure
of the singularities on the light cone which need to be given a distributional meaning.

\section{General Remarks} \label{sec8}
\setcounter{equation}{0}
In the preceding constructions we worked with a special class of
regularizations. This raises the following questions:
\begin{description}
\item[(1)] Considering only our restrictive class of regularizations
satisfying Theorem~\ref{thm1}, is there still enough freedom to
fulfill additional constraints? More specifically, is it possible to
satisfy the condition of half-occupied surface states? Are the
regularization parameters still linearly independent? Can the
fermionic states still be properly normalized?
\item[(2)] To what extent was our choice of regularizations only a
matter of convenience, and to what extent a matter of necessity?
Clearly, many details of the regularization could be modified,
but are there general properties which are canonical?
In more physical terms (assuming that a suitably regularized fermionic
projector describes nature), what do we learn about the microscopic
structure of space-time?
\end{description}
In this section we answer or at least discuss these questions.

\begin{Remark} (half occupied surface states and the normalization
of the fermionic states) \label{remhalfnorm} \em
In the constructions of Sections~\ref{sec4}--\ref{sec6}, we imposed
conditions on the following combinations of regularization functions,
\[ \sum_{\beta=1}^3 h_\beta\:,\qquad
\sum_{\beta=1}^3 \alpha_\beta\, h_\beta\:,\qquad
\sum_{\beta=1}^3 g_\beta\:,\qquad
\sum_{\beta=1}^3 \alpha_\beta\, g_\beta\:,\qquad
\sum_{\beta=1}^3 f_\beta \]
(see~(\ref{TZ}--\ref{TJ}), (\ref{Th2}--\ref{Tf2}), (\ref{Th3}--\ref{Tf3}),
(\ref{TailB}), (\ref{TailC}),
(\ref{h4tail}--\ref{h4tailz}), (\ref{TailA}--\ref{TailZ}),
(\ref{finf}),
(\ref{lTailA}--\ref{lTailZ})).
Moreover, we implicitly assumed that
\[ \sum_{\beta=1}^3 \alpha_\beta\, f_\beta \;=\; 0 \:. \]
The higher orders in the mass expansion are so small near the light
cone that they do not contribute in the limit~$\varepsilon \searrow 0$.
Hence in the construction of the tails we prescribed six functions
of~$\omega$. On the other hand, the regularization of each of the three
Dirac seas involves the four free functions~$f_\beta$, $g_\beta$,
$h_\beta$ and~$\alpha_\beta$. Thus for every~$\omega$, we have twelve
free parameters to satisfy six conditions. This leaves us with six
free parameters, two for each Dirac sea. This is precisely what we
need to satisfy the condition of half occupied surface states~(\ref{hoss})
and to normalize the fermionic states according to~\cite[\S~2.6]{PFP}:
\begin{itemize}
\item In a system with three generations, there is a family of
regularizations~$(P_\varepsilon)_{\varepsilon>0}$ having all the
properties in Theorem~\ref{thm1}, and furthermore the surface
states are half occupied with properly normalized fermions.
\end{itemize}
It is remarkable that the situation would be completely different if
we had worked with less than three generations. More precisely,
counting the number of free parameters, one immediately gets the
following results:
\begin{itemize}
\item In a system with two generations, there is a family of
regularizations~$(P_\varepsilon)_{\varepsilon>0}$ having all the
properties in Theorem~\ref{thm1}, and furthermore the surface
states are half occupied {\em{or}} properly normalized.
\item In a system with one generation, it does not seem
possible to regularize the fermionic projector in such a way
that Theorem~\ref{thm1} holds.
\end{itemize}
This consideration gives a natural explanation why three generations of elementary particles appear in nature.
\QEDrem
\end{Remark}

\begin{Remark} (linear independence of the regularization parameters) \label{rem7} \em \\
In the continuum limit~\cite[Chapter~4]{PFP}, the unknown microscopic
structure of space-time is taken into account by a small number
of free parameters, the so-called regularization parameters.
In~\cite[Appendix~E]{PFP} it was shown by an explicit construction
of suitable regularizations that the regularization parameters are
linearly independent. It is an important question whether this
result remains true for the more restrictive class of regularizations
considered here.
To answer this question, we first note that in~\cite[Appendix~E]{PFP}
we worked with the {\em{moments}} of the regularization functions,
such as the quantities
\[ \int_0^\infty \omega^n\: h_\alpha(\omega)\, d\omega\:, \]
whereas in the preceding sections we analyzed the {\em{tails}}
of the regularization. The tails and the moments describe
complementary properties of the regularization functions,
which can can be prescribed independently. Therefore,
the considerations of the tails of the present paper
are not in conflict with the considerations
in~\cite[Appendix~E]{PFP}. Writing down families of
regularization functions which satisfy both Theorem~\ref{thm1} and~\cite[Theorem~E.1]{PFP}
seems straightforward (although the explicit formulas would clearly become
rather complicated).

Alternatively, this result can be understood in position space as follows.
The regularization tails are active in the region~$s \gg E_P$.
The regularization parameters, on the other hand, model the behavior
of the fermionic projector with interaction near the light cone~$s \sim E_P$.
Since the range of~$s$ is different in both cases, it is clear
that the tails do not affect the regularization parameters.

To avoid misunderstandings, we repeat the warning from~\cite[Appendix~E]{PFP}
that linear independence of the regularization parameters
does not mean that, by choosing suitable regularization functions, the
regularization parameters can be given arbitrary values.
The regularization parameters might well be constrained by
certain inequalities. For this reason, in applications one should
always verify that the values for the
regularization parameters needed in the effective continuum
theory can actually be realized by suitable regularization
functions.
\QEDrem
\end{Remark}

\begin{Remark} (universal properties of the regularizations) \label{remmls}
\em \\
Clearly, many construction steps in this paper were arbitrary or merely
a matter of convenience. Nevertheless, based on the experience of
numerous calculations involving different kinds of regularization
functions, it seems that a few properties of our regularizations are
universal in the sense that every admissible family of regularizations
should have these properties. More precisely, we can make the following
general considerations.
\begin{itemize}
\item The structure of our variational principle leads us to
distinguish between vector and bilinear dominated regions.
In order to get correspondence to Minkowski space, the boundary between
these regions must be close to the light cone, and it might involve
a transition region with one or several layers. The property
of a distributional~$\M P$-product gives so many conditions that it
seems impossible to satisfy all these conditions with only one layer.
This means that every admissible regularization should have a
{\em{multilayer structure}} near the light cone.
\item At first sight, it might seem a promising strategy to make
the regularization ``as Lorentz invariant as possible'' by arranging
that the boundary between the vector and bilinear dominated regions
lies on hyperbolas~$\xi^2={\mbox{const}}$. However,
this strategy does not seem to work. Instead, the radial scaling argument
on page~\pageref{radscal} yields boundaries~$s \sim r^{-\gamma}$
with~$\gamma<1$, which break Lorentz symmetry. Clearly, all the
contributions which are not Lorentz invariant must cancel each other
as~$\varepsilon \searrow 0$. This seems possible only by working with
several layers which have a different scaling in~$r$. As a consequence,
these layers must also scale differently in~$\varepsilon$. In short, we
can say that the regularization should involve {\em{several length
scales}}.
\item As explained in Section~\ref{sec6a}, the effect of the regularization
tails on~$\M[A^\varepsilon_{xy}]$ changes drastically near the origin.
This seems to imply that as~$\varepsilon \searrow 0$, the operator $\M[A^\varepsilon_{xy}]$ will develop singularities at the origin, unless
these singularities are compensated by additional regularization tails
which are active near~$\xi=0$. In other words, the admissible
regularizations should involve additional {\em{cusplike regions
close to the origin}}.
\item In Section~\ref{sec6b} we saw that the behavior of~$\M[A^\varepsilon_{xy}]$ near infinity requires special attention. Our method of introducing
a shear of the surface states changes the behavior of the surface states
considerably if their energy is large. This suggests that the fermionic
projector of the vacuum should have a {\em{nontrivial high-energy
structure}}.
\end{itemize}
In any case, these considerations suggest that
the fermionic projector of the vacuum should have a
complicated structure on the Planck scale.
At present, getting more detailed information than what we just
discussed seems difficult. Also, in view of the fact that direct
measurements on the Planck scale are out reach, it might seem
rather speculative to further discuss possible regularizations.
The good news coming out of our analysis is that we do not need to
know the detailed structure of the fermionic projector on the
Planck scale. Since we know that~$P^\varepsilon$ and~$\M[A^\varepsilon_{xy}]$
can be arranged to have a well defined continuum limit,
it is consistent to work with~$P$ and~$\M$ as Lorentz invariant
distributions in Minkowski space.
\QEDrem
\end{Remark}

\appendix
\section{The Weight Factors~$\rho_\beta$} \label{appA}
\setcounter{equation}{0}
In the ansatz for the fermionic projector of the vacuum~(\ref{A}) we introduced
weight factors~$\rho_\beta$. We now discuss the physical meaning of these weight
factors and mention their implication for the analysis of the continuum limit.

First of all, we point out that the weights are of no importance for the constructions
in this paper. By choosing~$\rho_\beta=1$, one gets back to the setting of~\cite{PFP}.
The motivation for introducing general weights is that this gives more freedom for
choosing the vacuum Dirac sea configuration. This additional freedom is of
advantage in the stability analysis~\cite{FH}.

At first sight, the weights~$\rho_\beta$ might seem to contradict physical observations.
However, a careful consideration shows that the weights do not lead to obvious contradictions,
as we now explain. First, the weights are not in conflict with the proper normalization of
the fermionic states (see~\cite[\S2.6 and Appendix~C]{PFP}). Namely, thinking
of the fermionic projector as being the continuum limit of suitable discretizations,
one can arrange Dirac seas with smaller weights for example by occupying the states on the
lower mass shell more sparsely.
Second, it is not clear how to measure the weights in experiments. For example, one
might expect that the weights have an influence on the scattering amplitudes; for instance,
it might seem more likely for pair creation to occur if the weights are larger.
To see that this argument is incorrect, one has to keep in mind that the scattering amplitudes
are computed from the Dirac propagator (and possibly the propagators of the gauge fields),
where the weights do not enter. Thus scattering experiments do not seem appropriate for
measuring the weights. Indeed, even after discussing the issue with several high energy
physicists, the author cannot think of any experiment to measure the weights.
It would clearly be desirable to measure the weights~$\rho_\beta$ or to have a
theoretical argument which determines the weights. However, as long as these measurements
have not been carried out and no such argument has been given, it seems most honest
to treat the weights~$\rho_\beta$ as free positive parameters.

However, the weights have implications on the analysis of the continuum
limit (see~\cite[Chapter~4]{PFP}). More precisely, the factors~$\rho_\beta$ enter the
chiral asymmetry matrix~$X$, which in~\cite[\S2.3]{PFP} was introduced by
\[ X \;=\; \bigoplus_{a=1}^N \bigoplus_{\beta=1}^3 \:X_{a \beta}\:, \]
where the direct summands~$X_{a \beta}$ are the identity matrix in the massive sectors and
equal the chiral projector~$\chi_L$ in the neutrino sector.
To take into account the weights, we introduce for every Dirac sea a parameter~$\rho_{a \beta}>0$
and set
\[ X_{a \beta} \;=\; \left\{ \begin{array}{cl}
\rho_{a \beta} \,\1 & {\mbox{in the massive sectors}} \\
\rho_{a \beta}\, \chi_L & {\mbox{in the neutrino sectors}} \:.
\end{array} \right.  \]
When introducing an interaction, we must satisfy the {\em{causality compatibility condition}}
(see~\cite[Def.~2.3.2]{PFP})
\beq \label{cacc}
X^* \: (i \Pdd + {\mathcal{B}} - m Y) \;=\; (i \Pdd + {\mathcal{B}} - m Y) \: X \:.
\eeq
This condition makes it impossible to introduce gauge fields which describe an
interaction of Dirac seas with different weights. However, in~\cite{PFP}
a similar problem occurred with the neutrino sector,
and this problem was overcome by introducing the dynamical mass matrices
and by transforming to the effective gauge fields
(see~\cite[Chapters~7 and~8]{PFP}). Furthermore, the condition~(\ref{cacc}) was
weakened in~\cite[Def.~7.1.1]{PFP}. In view of these constructions, the
weight factors~$\rho_\beta$ do not lead to obvious problems.
However, the implications of the weight factors should be analyzed carefully
when working out the continuum limit in more detail. \\

\noindent
{\em{Acknowledgments:}} I want to thank Joel Smoller and the referee for
helpful comments on the manuscript.


\noindent
NWF I -- Mathematik,
Universit{\"a}t Regensburg, 93040 Regensburg, Germany, \\
{\tt{Felix.Finster@mathematik.uni-regensburg.de}}


\begin{thebibliography}{99}
\bibitem{Baez} J.C.\ Baez, ``An Introduction to Spin Foam Models of
Quantum Gravity and BF Theory,'' {\em{Lect.Notes Phys.}}\ {\bf{543}} (2000) 25-94
\bibitem{BLMS} L.\ Bombelli, J.\ Lee, D.\ Meyer, R.\ Sorkin, ``Space-time as
a causal set,'' {\em{Phys.\ Rev.\ Lett.}}\ {\bf{59}} (1987) 521-524
\bibitem{CC} A.H.\ Chamseddine, A.\ Connes, ``The spectral action principle,''
{\em{Commun.\ Math.\ Phys.}}\ {\bf{186}} (1997) 731-750
\bibitem{C} A.\ Connes, ``Noncommutative Geometry,'' {\em{Academic Press}} (1994)
\bibitem{DFS} A.\ Diethert, F.\ Finster, D.\ Schiefeneder, ``Fermion systems in discrete
space-time exemplifying the spontaneous generation of a causal structure,''
arXiv:0710.4420 [math-ph] (2007)
\bibitem{PFP} F.\ Finster, ``The Principle of the Fermionic Projector,''
{\em{AMS/IP Studies in Advanced Mathematics}} {\bf{35}} (2006)
\bibitem{F1} F.\ Finster, ``A variational principle in discrete space-time -- existence of minimizers,'' math-ph/0503069, {\em{Calc.\ Var.\ and Partial Diff.\ Eq.}}\ {\bf{29}} (2007) 431-453
\bibitem{F2} F.\ Finster, ``Fermion systems in discrete space-time,''
hep-th/0601140, {\em{J.\ Phys.: Conf.\ Ser.}}\ {\bf{67}} (2007) 012048
\bibitem{F3} F.\ Finster, ``Fermion systems in discrete space-time -- outer symmetries and spontaneous symmetry breaking,''  math-ph/0601039,
{\em{Adv.\ Theor.\ Math.\ Phys.}}\ {\bf{11}} (2007) 91-146
\bibitem{F4} F.\ Finster, ``From discrete space-time to Minkowski space:
Basic mechanisms, methods and perspectives,'' arXiv:0712.0685 [math-ph] (2007)
\bibitem{FH} F.\ Finster, S.\ Hoch, ``An action principle for the masses of Dirac particles,''
arXiv:0712.0678 [math-ph] (2007)
\bibitem{H} S.W.\ Hawking, ``Space-time Foam,'' {\em{Nucl.\ Phys}}.\ {\bf{B144}}
(1978) 349-362
\bibitem{V} A.\ Iqbal, N.\ Nekrasov, A.\ Okounkov, C.\ Vafa, ``Quantum foam and topological strings,'' arXiv:hep-th/0312022v2 (2003)
\bibitem{Oeckl} R.\ Oeckl, ``Discrete Gauge Theory: From Lattices to TQFT,''
{\em{Imperial College Press}} (2005)
\bibitem{O} D.\ Oriti, ``A quantum field theory of simplicial geometry and the emergence of spacetime,'' {\em{J.\ Phys.: Conf.\ Ser.}}\  {\bf{67}} (2007) 012052
\bibitem{RW} T.\ Regge, R.M.\ Williams, ``Discrete structures in gravity,''
arXiv:gr-qc/0012035, {\em{J.Math.Phys.}}\ {\bf{41}} (2000) 3964-3984
\bibitem{W} J. A.\ Wheeler, in ``Relativity, Groups and Topology,'' edited by B.S.\ and C.M.\ DeWitt,
{\em{Gordon and Breach}}, New York (1964)
\bibitem{R} C.\ Rovelli, ``Quantum Gravity,'' {\em{Cambridge University Press}} (2004)
\end{thebibliography}
\end{document}